\documentclass[aps,twocolumn,superscriptaddress,floatfix,letterpaper]{revtex4-1}

\pdfoutput=1

\usepackage[all]{xy}

\newcommand\sA{\EuScript{A}}

\newcommand\T{\mathbb{T}}

\newcommand\Rep{\EuScript{R}\mathrm{ep}}
\newcommand\sRep{\mathrm{s}\EuScript{R}\mathrm{ep}}

\newcommand{\w}{{\rm w}}
\newcommand{\Tor}{\text{Tor}}

\newcommand{\Hom}{\text{Hom}}

\newcommand{\Sq}{\text{Sq}}
\newcommand{\Bs}{{\cal B}}
\newcommand{\RZ}{{\mathbb{R}/\mathbb{Z}}}
\newcommand\se[1]{\overset{\scriptscriptstyle #1}{=}}
\newcommand\hcup[1]{\underset{{\scriptscriptstyle #1}}{\smile}}

\newcommand\toZ[1]{\lfloor #1 \rfloor}

\newcommand {\emptycomment}[1]{}

\newcommand{\tDm}[6]{{\bmm #1\emm}_{ 
b_{#2#3#4} 
b_{#2#3#5} 
b_{#2#3#6} 
b_{#2#4#5} 
b_{#2#4#6} 
b_{#2#5#6} 
b_{#3#4#5} 
b_{#3#4#6} 
b_{#3#5#6} 
b_{#4#5#6};
c_{#2#3#4#6} 
c_{#2#3#4#5} 
}^{ 
\hat a_{#2#3} 
\hat a_{#2#4} 
\hat a_{#2#5}
\hat a_{#2#6} 
\hat a_{#3#4} 
\hat a_{#3#5}  
\hat a_{#3#6} 
\hat a_{#4#5}  
\hat a_{#4#6} 
\hat a_{#5#6};
c_{#3#4#5#6} 
c_{#2#4#5#6} 
c_{#2#3#5#6} 
}}

\newcommand{\tAw}[3]{{#1}_{l_{#2#3} }^{v_{#2}v_{#3}}}

\newcommand{\tC}[5]{{#1}_{v_{#2}v_{#3}v_{#4}v_{#5};
t_{#2#4#5}
t_{#2#3#5} 
t_{#3#4#5}
}^{
l_{#2#3} l_{#2#4} l_{#2#5} l_{#3#4} l_{#3#5} l_{#4#5};
t_{#2#3#4} 
}}

\newcommand{\tAwR}[3]{{#1}_{a_{#2#3} }^{v_{#2}v_{#3}}}

\newcommand{\tCR}[5]{{#1}_{v_{#2}v_{#3}v_{#4}v_{#5};
b_{#2#4#5}
b_{#2#3#5} 
b_{#3#4#5}
}^{
a_{#2#3} a_{#2#4} a_{#2#5} a_{#3#4} a_{#3#5} a_{#4#5};
b_{#2#3#4} 
}}

\usepackage{tikz}
\usetikzlibrary{matrix,arrows,calc}
\tikzset{cd/.style={matrix of math nodes,row sep=2em,column sep=2em, text height=1.5ex, text depth=0.5ex}}
\tikzset{cdar/.style={->,auto}} 

\tikzset{triar/.style={anchor=mid,->}} 
\tikzset{tridar/.style={anchor=mid,double,double equal sign distance,-implies}}

\theoremstyle{plain}
\newtheorem{theorem}{Theorem}[section]
\newtheorem{statement}{Statement}[section]

\newtheorem{lemma}[theorem]{Lemma}

\theoremstyle{definition}

\theoremstyle{remark}

\begin{document}

\begin{titlepage}

\title{
Topological non-linear $\sigma$-model, higher gauge theory,\\
 and a realization of all 3+1D topological orders for boson systems
}

\author{Chenchang Zhu} 
\affiliation{Mathematics Institute, Georg-August-University of
  G\"ottingen, G\"ottingen 37073, Germany}
\affiliation{Center for Mathematical
  Sciences, Huazhong University of Science and Technology, Wuhan, China}

\author{Tian Lan} 
\affiliation{Institute for Quantum Computing,
  University of Waterloo, Waterloo, Ontario N2L 3G1, Canada}

\author{Xiao-Gang Wen}
\affiliation{Department of Physics, Massachusetts Institute of
Technology, Cambridge, Massachusetts 02139, USA}

\begin{abstract} 
A discrete non-linear $\sigma$-model is obtained by triangulate both the
space-time $M^{d+1}$ and the target space $K$.  If the path integral is given
by the sum of all the complex homomorphisms $\phi: M^{d+1} \to K$, with an
partition function that is independent of space-time triangulation, then the
corresponding non-linear $\sigma$-model will be called topological non-linear
$\sigma$-model which is exactly soluble.  Those exactly soluble models suggest
that phase transitions induced by fluctuations with no topological defects
({\it i.e.} fluctuations described by homomorphisms $\phi$) usually produce a
topologically ordered state and are topological phase transitions, while phase
transitions induced by fluctuations with all the topological defects give rise
to trivial product states and are not topological phase transitions.  If $K$ is
a space with only non-trivial first homotopy group $G$ which is finite,
those topological non-linear $\sigma$-models can realize all 3+1D bosonic
topological orders without emergent fermions, which are
described by Dijkgraaf-Witten theory with gauge group $\pi_1(K)=G$.  Here, we
show that the 3+1D bosonic topological orders with emergent fermions can be
realized by topological non-linear $\sigma$-models with $\pi_1(K)=$ finite
groups, $\pi_2(K)=Z_2$, and $\pi_{n>2}(K)=0$.  A subset of those topological
non-linear $\sigma$-models corresponds to 2-gauge theories, which realize and
classify bosonic topological orders with emergent fermions that have no
emergent Majorana zero modes at triple string intersections.  The
classification of 3+1D bosonic topological orders may correspond to a
classification of unitary fully dualizable fully extended \emph{topological
quantum field theories} in 4-dimensions.  

\end{abstract}


\maketitle

\end{titlepage}

{\small \setcounter{tocdepth}{1} \tableofcontents }

\section{Introduction}

\subsection{Background}

The study of topological phase of matter has become a very active field of
research in condensed matter physics, quantum computation, as well as in part
of quantum field theory and mathematics.  However, ``topological'' may have
very different meanings, even in the same context of topological phase of
matter.  

\begin{figure}[tb] 
\centering 
\includegraphics[height=1.3in]{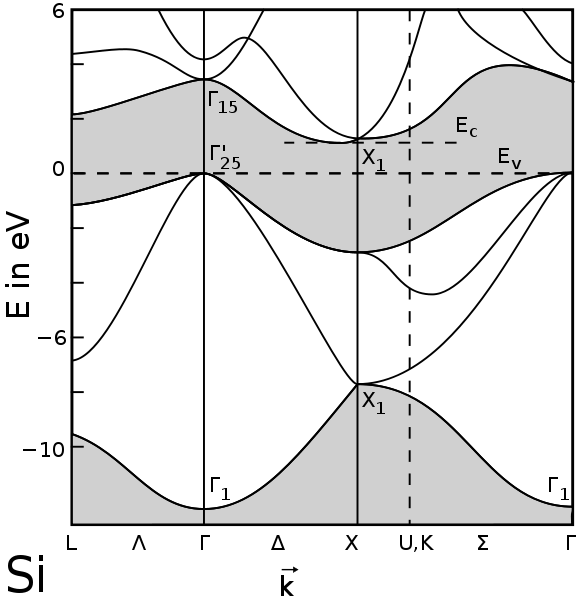} ~
\includegraphics[height=0.9in]{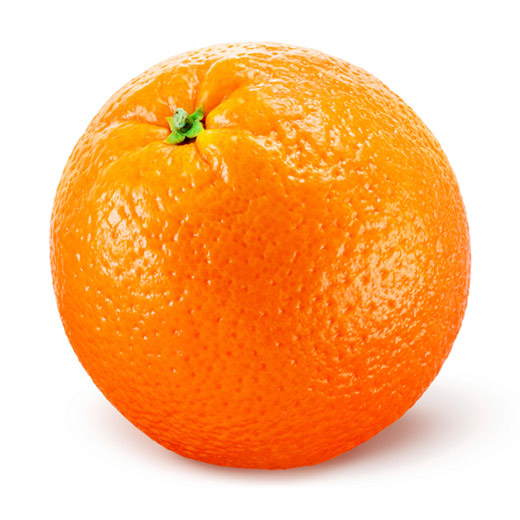} 
\includegraphics[height=0.9in]{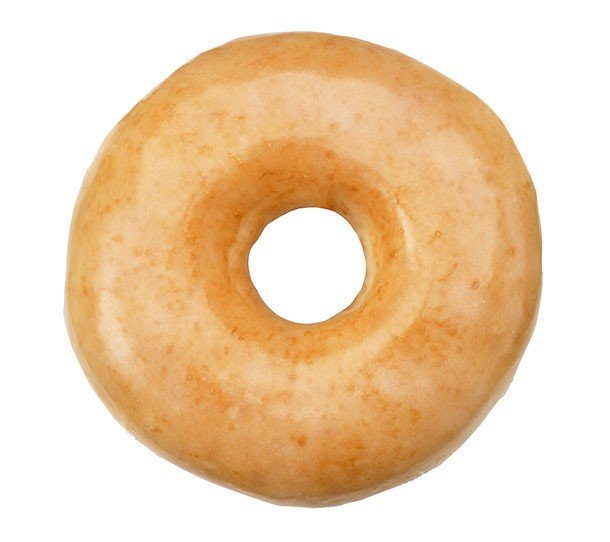} 
\caption{ 
``Topology'' in topological insulator/superconductor (2005) corresponds
to the twist in the band structure of orbitals, which is similar to
the topological structure that distinguishes  a sphere from a torus.
This kind of topology is \emph{classical  topology}.
}
\label{cTop} 
\end{figure}

In \emph{topological} insulator/superconductor 
\cite{KM0502,BZ0611399,MB0706,FKM0703,QHZ0824,R0664}, ``topological'' means the
twist in the band structure of orbitals (see Fig. \ref{cTop}), which is
described by curvature, Chern number, finite dimensional fiber bundle, \etc
\cite{TKN8205,ASS8351,ASS8829,K0986}.  Such ``topological'' properties can be
defined even without any particles.

\begin{figure}[tb] 
\centering 
\includegraphics[height=1.2in]{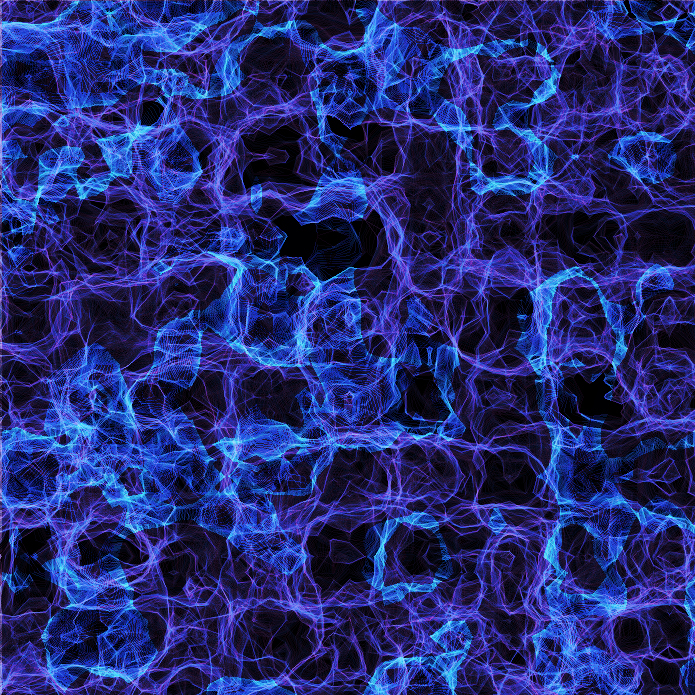} ~
\includegraphics[height=1.0in]{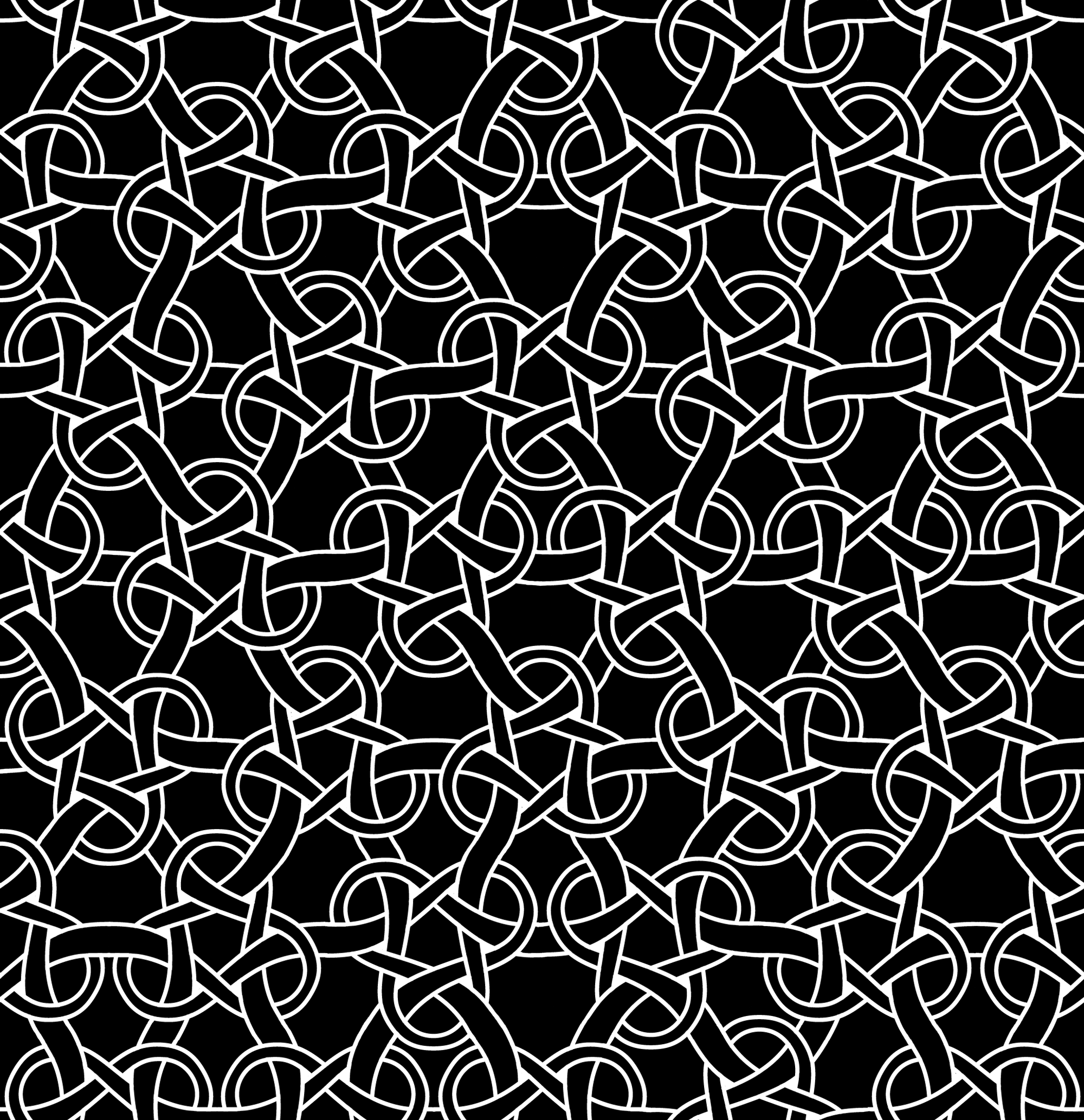} 
\includegraphics[height=1.0in]{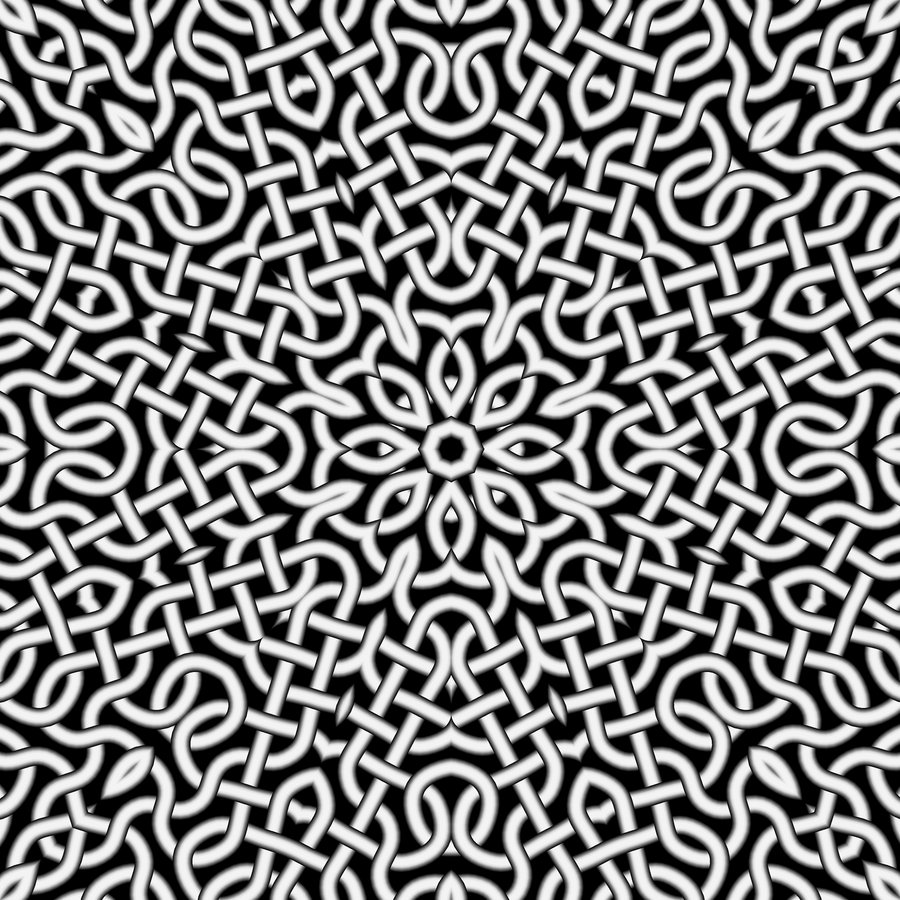} 
\caption{ 
``Topology'' in topological order (1989) corresponds to pattern of many-body
entanglement in many-body wave function $\Psi(m_1,m_2,\cdots,m_N)$, that is
robust against any local perturbations that can break any symmetry.  Such
robustness is the meaning of ``topological'' in topological order.  This kind
of topology is \emph{quantum topology}.
}
\label{qTop} 
\end{figure}

However, in \emph{topological} order \cite{Wtop,Wrig,WNtop},
``topological'' means the pattern of quantum entanglement
\cite{KP0604,LW0605,CGW1038} in many-body wave functions of $N\sim 10^{20}$
variables:
\begin{align}
\Psi(m_1,m_2,\cdots,m_N) .
\end{align}
It is hard to visualize the patterns of many-body entanglement in such
complicated  many-body systems.  We may use Celtic knots to help us to get some
spirit of topological order or pattern of many-body entanglement (see Fig.
\ref{qTop}).  


So the ``topology'' in topological order is very different from the classical
topology that distinguishes a sphere from a torus. We will refer this new kind
of ``topology'' as quantum topology.  It turns out that the mathematical
foundation for quantum topology is related to topological quantum field theory,
braided fusion category, cohomology, \etc
\cite{W8951,LW0510,K062,RSW0777,CGL1314,BBC1440,K1467,LW160205946,KW170200673}.

\begin{figure}[tb] 
\centering 
\includegraphics[height=0.8in]{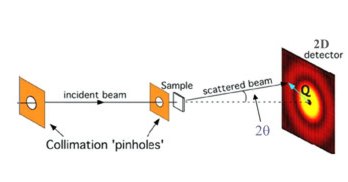} ~
\includegraphics[height=1.1in]{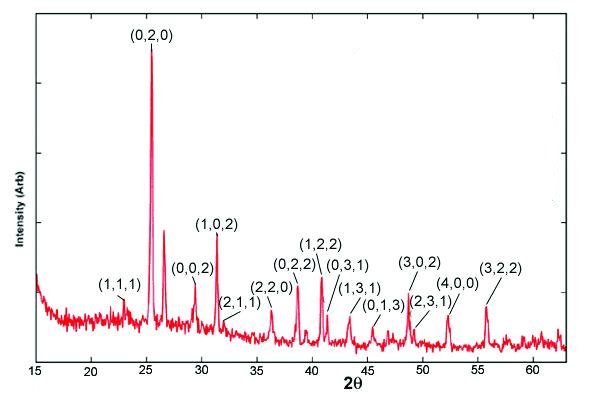} 
\caption{ 
X-ray scattering is a universal probe for all crystal orders.
}
\label{XRay} 
\end{figure}

To develop a quantitative theory for topological order and the related pattern
of many-body entanglement, we need to identify physical probes that can measure
topological order \cite{Wtop,Wrig,WNtop}, \ie identify topological invariants
that can characterize topological order.  We know that, for crystal order,
X-ray scattering is a universal probe that can measure all crystal orders (see
Fig.  \ref{XRay}).  So we like to ask: do we have a single universal probe that
can measure all topological orders?

One potential universal probe (topological invariant) for topological orders is
the partition function $Z$.  Let us consider bosonic systems described by the
path integral of non-linear $\si$-models:
\begin{align}
\label{ZTL}
 Z(M^{d+1}; K,\cL) = \sum_{\phi(x)} \ee^{- \int_{M^{d+1}} \dd^{d+1} x\; \cL(\phi(x), \prt \phi(x), \cdots)}.
\end{align}
Here $M^{d+1}$ is a $d+1$D space-time manifold and $K$ a target manifold.
$\sum_{\phi(x)}$ sum over all the maps $\phi: M^{d+1}\to K$, $x\in M^{d+1}$ and
$\phi(x) \in K$.  $\dd^{d+1} x\; \cL(\phi(x), \prt \phi(x), \cdots)$ is a
$(d+1)$-form at $x$ that depends on $\phi(x), \prt \phi(x)$ \etc.
$\cL(\phi(x), \prt \phi(x), \cdots)$ is also called the Lagrangian density in
physics.

The pair $(K,\cL)$ labels the  bosonic systems, and the partition function $Z$
is a map from space-time manifolds to complex numbers
\begin{align}
  Z(-; K,\cL): \{ M^{d+1} \} \to \C .
\end{align}
So the partition function  $Z$ is a physical probe that measure the bosonic
system.  However, $Z(-; K,\cL)$ does not measure topological order, since two
systems $(K,\cL)$ and $(K',\cL')$ that are in the same topologically ordered
phase can have different partition functions: $Z(-; K,\cL) \neq Z(-; K',\cL')$.
In other words, the  partition function $Z(-; K,\cL)$ is not a topological
invariant. 

We know that the leading term in the partition function
comes from the energy density $\veps(x)$:
\begin{align}
\label{ZMTL}
 Z(M^{d+1}; K,\cL)) = \ee^{-\int_{M^{d+1}} \dd^{d+1} x\; \veps(x)}
 Z^\text{top}(M^{d+1}; K,\cL)) ,
\end{align}
where the subleading term $Z^\text{top}(M^{d+1}; K,\cL)$ is of order 1 in large
space-time volume limit.  The leading term $\ee^{-\int_{M^{d+1}} \dd^{d+1} x\;
\veps(x)}$ is not topological, since even when two systems $(K,\cL)$ and
$(K',\cL')$ are in the same topologically ordered phase, their energy densities
$\veps(x)$ and $\veps'(x)$ can be different.  

However, the idea of using partition function to characterize topological order
is not totally wrong.  In particular, the subleading term is believed to be
topological.\cite{KW1458} So $Z^\text{top}(M^{d+1}; K,\cL)$ are topological
invariants that can be used to measure/define topological order.
\Ref{WW180109938} describes ways to extract topological invariant
$Z^\text{top}(M^{d+1}; K,\cL)$ from non-topological partition function
$Z(M^{d+1}; K,\cL)$ via surgery operations.

After identifying the topological invariants that characterize and define
topological orders, the next issue is to systematically construct bosonic
systems $(K,\cL)$ that realize all kinds of topological orders, which is the
topic of this paper:  
\begin{enumerate}
\item
We will describe in details a general way to construct exactly soluble bosonic
models: topological non-linear $\si$-models, and their special cases -- higher
gauge theories.   We believe that topological non-linear $\si$-models can
realize all bosonic topological orders with gappable boundary.  In particular,
higher gauge theories realize and classify all bosonic topological orders with
the following property: the topological orders have a gapped boundary that all
pointlike, stringlike and other higher dimensional excitations on the boundary
have a unit quantum dimension.
\item 
We use exactly soluble 2-gauge theories to systematically realize and classify
EF1 topological orders --   3+1D bosonic topological orders with emergent
bosons and fermions where triple string intersections carry no Majorana zero
modes.  The rest of 3+1D bosonic topological orders  with emergent bosons and
fermions are EF2 topological orders where some triple string intersections must
carry Majorana zero modes.\cite{LW180108530} EF2 topological orders can be
realized by topological non-linear $\si$-models which are beyond 2-gauge
theories.  
\end{enumerate}

Recently, there are many works
\cite{KT1321,BM160606639,BM170200868,CT171104186,P180201139,DT180210104,BT180300529,NW180400677}
on higher gauge theories and their connection to topological phases of matter.
In this paper, we present a detailed description of ``lattice higher gauge
theories'', in a way to make their connection to non-linear $\si$-model
explicit.  In our presentation, we do not require higher gauge symmetry and
higher gauge holonomy. We even do not mod out higher gauge transformations.
Our ``lattice higher gauge theories'' are just lattice non-linear $\si$-models
with only lattice scalar fields (\ie lattice qubits).  However, lattice
non-linear $\si$-models (without higher gauge symmetry) can realize topological
orders whose low energy effective theories are higher gauge theories with
emergent  higher gauge symmetry. In other words, we describe how higher gauge
theories can emerge from lattice qubit models (\ie quantum spin models in
condensed matter).  In this paper, we also apply 2-gauge theories to classify a
subclass of 3+1D bosonic topological orders with emergent fermions.  We point
out that the rest of 3+1D bosonic topological orders with emergent fermions are
beyond 2-gauge theories and can be realized by more general topological
non-linear $\si$-models.

\subsection{Realize topological orders via disordered symmetry breaking states
without topological defects}

\label{topdef}

In this paper, we show that all the higher gauge theories can be viewed as
non-linear $\si$-models with some complicated target space and carefully
designed action.  Such a duality relation between non-linear $\si$-models and
higher gauge theories suggests that we may be able to use disordered symmetry
breaking states (which are described by non-linear $\si$-models) to realize a
large class of topological orders.  In other words, starting with a symmetry
breaking state and letting the order parameter have a strong quantum
fluctuation, we may get a symmetric disordered ground state with topological
order.  

However, this picture seems to contradict with many previous results that a
symmetric disordered ground state is usually just a trivial product state
rather than a topological state.  The study in this paper suggests that the
reason that we get a trivial disordered state is because the strongly
fluctuating order parameter in the disordered state contains a lot of
topological defects, such as vortex lines, monopoles, etc.  

The importance of the topological defects \cite{LD8851} in producing
short-range correlated disordered states have been emphasized by Kosterlitz and
Thouless in \Ref{KT7381}, which shared 2016 Nobel prize ``for theoretical
discoveries of topological phase transitions and topological phases of
matter''.  

In this paper, we show that the phase transitions driven by
fluctuations with all possible topological defects produce disordered states
that have no topological order, and correspond to non-topological phase
transitions.  While transitions driven by fluctuations without any topological
defects usually produce disordered states that have non-trivial topological
orders, and correspond to topological phase transitions.
Thus, it may be confusing to refer the transition driven by topological defects
as a topological phase transitions, since the appearance of topological defects
decrease the chance to produce topological phases of matter. 

More precisely, if the fluctuating order parameter in a disordered state has no
topological defects, then the corresponding disordered state will usually have
a non-trivial topological order.  The type of the topological order depends on
the topology of the degenerate manifold $K$ of the order parameter (\ie the
target space of the non-linear $\si$-model).  For example, if $\pi_1(K)$ is a
finite group and $\pi_{n>1}(K)=0$, then the disordered phase may have a
topological order described by a gauge theory of gauge group $G=\pi_1(K)$.  If
$\pi_1(K),\pi_2(K)$ are finite groups and $\pi_{n>2}(K)=0$, then the disordered
phase may have a topological order described by a 2-gauge theory of
2-gauge-group $\cB(\pi_1(K),\pi_2(K))$.  

It is the absence of topological defects that enable the symmetric disordered
state to have a non-trivial topological order.  When there are a lot of
topological defects, they will destroy the topology of the degenerate manifold
of the order parameter (\ie the degenerate manifold effectively becomes a
discrete set with trivial topology). In this case the symmetric disordered
state becomes a product state with no topological order.  Certainly, if the
fluctuating order parameter contains only a subclass of topological defects,
then only part of the topological structure of the degenerate manifold is
destroyed by the defects.  The corresponding symmetric disordered state may
still have a topological order.

\subsection{Realizations of all 3+1D bosonic topological orders}

It was shown \cite{LW170404221,LW180108530} that all 3+1D bosonic topological
orders belong to two classes: AB topological orders
where all pointlike excitations are bosonic and  EF topological orders where
some pointlike excitations are fermionic.  \Ref{LW180108530} shows that 
all EF
topological orders have a unique gapped boundary with the following properties:
\begin{enumerate}
\item
All stringlike boundary excitations have a unit quantum dimension.  Those
boundary strings form a finite group $\hat G_b$ under string fusion.  The group
$\hat G_b$ is an extension of a finite group $G_b$ by $Z_2^m$: $\hat G_b =
Z_2^m \gext G_b$. (See Section \ref{notation} for the definition of $ Z_2^m
\gext G_b$.)
\item
There is one non-trivial type of pointlike boundary excitations
which is fermionic and has a unit quantum dimension.
\item
There are on-string pointlike excitations -- Majorana zero modes of quantum
dimension $\sqrt 2$.  The Majorana zero mode always lives at the pointlike domain wall
where a string labeled by $g$ joins a string  labeled by $g m $. Here
$g\in \hat G_b$ and $m$ is the non-trivial element in $Z_2^m$.
\end{enumerate}

We note that the boundary fermions can form a topological $p$-wave
superconducting (pSC) chain.\cite{K0131} The boundary strings labeled by $\hat
G_b$ can be viewed as the boundary strings labeled by $G_b$ plus the pSC chain.
In particular, a string labeled by $g$ and a string  labeled by $g m $ differ
by a pSC chain.

If $\hat G_b$ is the trivial extension of $G_b$ by $Z_2^m$: $\hat G_b = Z_2^m
\times G_b$, the corresponding bulk topological order is called a EF1
topological order.  If $\hat G_b$ is a non-trivial extension of $G_b$ by
$Z_2^m$: $\hat G_b = Z_2^m \gext_{\rho_2} G_b$ where $\rho_2 \in H^2(\cB
G_b;\Z_2^m)$, the corresponding bulk topological order is called a EF2
topological order.  Here, we have used a conjecture -- a holographic
principle\cite{LW1384,KW1458,KW170200673} -- that the boundary topological
order completely determines the bulk topological order.

When $\hat G_b$ is the trivial extension: $\hat G_b = Z_2^m \times G_b$, we can
drop boundary strings that come from the pSC chain (by regarding the pSC chain
as a kind of trivial strings).  Thus, the  EF1 topological order has a simpler
gapped boundary: In addition to the boundary strings of unit quantum dimension
labeled by a finite group $G_b$, there is one and only one non-trivial type of
pointlike boundary excitations which is fermionic and has a unit quantum
dimension.\cite{LW180108530} 

\begin{figure}[tb] 
\centering \includegraphics[scale=0.8]{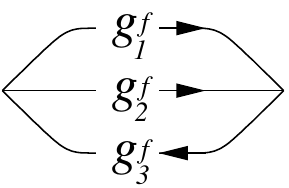} 
\caption{
A string configuration in the bulk described by a triple
$(\chi_{g_1^f},\chi_{g_2^f},\chi_{g_3^f})$, where $\chi_{g^f}$ is a conjugacy class in $G_f$
containing $g^f\in G_f$ and the triple satisfy $g_1^f g_2^f=g_3^f$.
}
\label{3strings} 
\end{figure}

In the above, we have defined EF1 and EF2 topological orders via their boundary
properties.  To distinguish EF1 and EF2 topological order through their bulk
properties, we consider a stringlike excitation in the bulk that has triple
string intersections (see Fig. \ref{3strings}).  Note that a triple string
intersection is described by the conjugacy classes $\chi_{g_1^f}, \chi_{g_2^f},\chi_{g_3^f}
\subset G_f$ that satisfy $ g_1^fg_2^f =g_3^f$.  By measuring the appearance of
Majorana zero mode at triple string intersections for different triples
$\chi_{g_1^f}, \chi_{g_2^f},\chi_{g_3^f}$, we can determine the cohomology class of
$\rho_2$.\cite{LW180108530} If the measured $\rho_2$ is a coboundary, the
bulk topological order is an EF1 or an AB topological order.  Otherwise, the
bulk topological order is an EF2 topological order.

It has been shown that all 3+1D AB topological orders are classified and
realized by 1-gauge theories (\ie Dijkgraaf-Witten gauge
theories).\cite{LW170404221} In this paper, we show that all 3+1D EF1
topological orders are classified and realized by 2-gauge theories with
2-gauge-group $\cB(G_b,Z_2^f)$.  The pointlike topological excitations
(including emergent fermions) are described by symmetric fusion category
$\sRep(Z_2^f \gext G_b)$, where $Z_2^f \gext G_b$ is an extension of $G_b$ by
$Z_2^f$.

We will also discuss how to systematically realize 3+1D EF2 topological orders
through topological non-linear $\si$-models whose target space $K$ satisfies
$\pi_1(K)=G_b$ and $\pi_2(K)=\Z_2$.  Those topological non-linear $\si$-models
are beyond 2-gauge theories.  The resulting EF2 topological orders have
pointlike topological excitations described by $\sRep(Z_2^f \gext G_b)$.

Our results suggest the following more general picture: 
\begin{statement}
Exactly soluble $n$-gauge theories can realize all bosonic topological orders
in $n+1$ spatial dimensions that have a gaped boundary where all boundary
excitations (including on $d$-brane excitations) have a unit quantum
dimension.
\end{statement}
This is because higher groups can be viewed as higher monoidal categories where
all objects and higher morphisms are invertible.  For more general bosonic
topological orders whose gapped boundary excitations have non-unit quantum
dimensions, we need to use more general exactly soluble models, such as
topological non-linear $\si$-model or even more general tensor network models,
to realize them.\cite{KW1458}

Combining the above realization results and the boundary results in
\Ref{LW180108530}, we obtain the following classification of EF topological
orders:
\begin{statement}\label{2cat}
3+1D EF topological orders 
are classified by unitary  fusion
2-categories that have the following properties:\\
(1) The simple objects are labeled by $\hat G_b=Z_2^m \gext_{\rho_2} G_b$, and their fusion is described by the group $\hat G_b$. \\
(2) For each simple object $g$ there is one nontrivial invertible
1-morphism corresponding to a fermion $f_g$. \\
(3) In addition, there are quantum-dimension-$\sqrt 2$ 1-morphisms $\si_{g,gm}$ that connect
two objects $g$ and $gm$, where $g \in \hat G_b$ and $m$ is the generator of
$Z_2^m$.\\
(4) The fusion of 1-morphisms is given by $f_g f_g =\one$ and
$\si_{g,gm}\si_{gm,g}=\one\oplus f_g$.
\end{statement}

\subsection{Notations and conventions}
\label{notation}

Let us first explain some notations used in this paper.  We will use
extensively the mathematical formalism of cochains, coboundaries, and cocycles,
as well as their higher cup product $\hcup{k}$, Steenrod square $\Sq^k$, and
the Bockstrin homomorphism $\Bs_n$. A brief introduction can be found in
Appendix \ref{cochain}.  We will abbreviate the cup product $a\smile b$ as $ab$
by dropping $\smile$.  We will use a symbol with bar, such as $\bar a$ to
denote a cochain on the target complex $\cK$.  We will use $a$ to denote the
corresponding pullback cochain on space-time $\cM^{d+1}$: $a = \phi^* \bar a$,
where $\phi$ is a homomorphism of complexes $\phi: \cM^{d+1} \to \cK$.

We will use $\se{n}$ to mean equal up to a multiple of $n$, and use $\se{\dd }$
to mean equal up to $\dd f$ (\ie up to a coboundary).  We will use 
$\toZ{x}$ to denote the greatest integer less than or equal to $x$,
and
$\<l,m\>$ for the greatest common divisor of $l$ and $m$ ($\<0,m\>\equiv
m$).  

Also, we will use $Z_n=\{1,\ee^{\ii \frac{2\pi}{n}},\ee^{\ii 2
\frac{2\pi}{n}},\cdots,\ee^{\ii (n-1) \frac{2\pi}{n}} \}$ to denote an Abelian
group, where the group multiplication is ``$*$''.  We use
$\Z_n=\{\toZ{-\frac{n}2+1},\toZ{-\frac{n}2+1}+1,\cdots,\toZ{\frac n 2}\}$ to denote an
integer lifting of $Z_n$, where ``+'' is done without mod-$n$.  In this sense,
$\Z_n$ is not a group under ``+''.  But under a modified equality $\se{n}$,
$\Z_n$ is the $Z_n$ group under ``+''.  Similarly, we will use
$\RZ=(-\frac12,\frac12]$ to denote an $\R$-lifting of $U_1$ group.  Under a
modified equality $\se{1}$, $\RZ$ is the $U_1$ group under ``+''.  In this
paper, there are many expressions containing the addition ``+'' of
$\Z_n$-valued or $\RZ$-valued, such as $a^{\Z_n}_1+a^{\Z_n}_2$ where
$a^{\Z_n}_1$ and $a^{\Z_n}_2$ are $\Z_n$-valued.  Those  additions ``+'' are
done without mod $n$ or mod 1.  In this paper, we also have expressions like
$\frac1n a^{\Z_n}_1$.  Such an expression convert a $\Z_n$-valued $a^{\Z_n}_1$
to a $\RZ$-valued $\frac1n a^{\Z_n}_1$, by viewing the $\Z_n$-value as a
$\Z$-value. (In fact, $\Z_n$ is a $\Z$ lifting of $Z_n$.)

We introduced a symbol $\gext$ to construct fiber bundle $X$ from the fiber $F$
and the base space $B$:
\begin{align}
pt\to  F \to X=F\gext B \to B\to pt .
\end{align}
We will also use $\gext$ to construct group extension of $H$ by $N$
\cite{Mor97}:
\begin{align}
1 \to  N \to N\gext_{e_2,\al} H \to H\to 1 .
\end{align}
Here $e_2 \in H^2[H;Z(N)]$ and $Z(N)$ is the center of $N$.  Also $H$ may have
a non-trivial action on $Z(N)$ via $\al: H \to \text{Aut}(N)$.
$e_2$ and $\al$ characterize different group extensions.

We will use $K(\Pi_1,\Pi_2,\cdots,\Pi_n)$ to denote a connected topological
space with homotopy group $\pi_i(K(\Pi_1,\Pi_2,\cdots,\Pi_n)) =\Pi_i$ for
$1\leq i\leq n$, and $\pi_i(K(\Pi_1,\Pi_2,\cdots,\Pi_n)) =0$ for $i>n$.  In
this paper, \emph{we assume that all $\Pi_n$'s are finite.} We note that
$\pi_i$ is abelian for $i>1$.  If only one of the homotopy groups, say $\Pi_d$,
is non-trivial, then $K(\Pi_1,\Pi_2,\cdots,\Pi_n)$ is the Eilenberg-MacLane
space, which is denoted as $K(\Pi_d,d)$.  If only two of the homotopy groups,
say $\Pi_d$, $\Pi_{d'}$, is non-trivial, then we denote the space as
$K(\Pi_d,d; \Pi_{d'},d')$, \etc.  We will use $\cK(\Pi_1;\Pi_2;\cdots;\Pi_n)$,
$\cK(\Pi_d,d)$, and $\cK(\Pi_d,d; \Pi_{d'},d')$ to denote the simplicial
complexes that describe a triangulation of $K(\Pi_1,\Pi_2,\cdots,\Pi_n)$,
$K(\Pi_d,d)$, and $K(\Pi_d,d; \Pi_{d'},d')$ respectively.  We will use
$\cB(\Pi_1;\Pi_2;\cdots;\Pi_n)$, $\cB(\Pi_d,d)$, and $\cB(\Pi_d,d;
\Pi_{d'},d')$ to denote the simplicial sets with only one vertex
satisfying Kan conditions that describe a special triangulation of
$K(\Pi_1,\Pi_2,\cdots,\Pi_n)$, $K(\Pi_d,d)$, and $K(\Pi_d,d; \Pi_{d'},d')$
respectively. Since simplicial sets satisfying Kan conditions are viewed as
higher groupoids in higher category theory,  the  simplicial sets
$\cB(\Pi_1;\Pi_2;\cdots;\Pi_n)$, $\cB(\Pi_d,d)$, and $\cB(\Pi_d,d;
\Pi_{d'},d')$, with only one vertex (unit),  can be viewed as higher groups.
In this paper, higher groups are treated therefore as this sort of special
simplicial sets.

\section{Topological non-linear $\si$-models and topological tensor network models}

\subsection{Discrete defectless non-linear $\si$-models}

The non-linear $\si$-model \eq{ZTL} is widely used in field theory to describe
a bosonic system.  If we require the map $\phi(x)$ to be continuous, then the
non-linear $\si$-model will be defectless, \ie the fluctuations contain no
defects.  But the corresponding path integral \eq{ZTL} is not well defined
since the summation $\sum_{\phi(x)}$ over $\infty^\infty$ number of the
continuous maps is not well defined.  To obtain a well defined theory, we
discretize both the space-time $M^{d+1}$ and the target space $K$.  We replace
them by simplicial complexes $\cM^{d+1}$ and $\cK$.  

\subsubsection{A detailed description of simplicial complex}

Let us first describe the simplicial complexes systematically. We introduce
$M_0,M_1,M_2,\cdots$ as the sets of vertices, links, triangles, \etc that form
the space-time complex $\cM^{d+1}$.  The complex $\cM^{d+1}$ is formally
described by
\begin{equation}\label{eq:nerveM}
\xymatrix{ 
M_0 & 
M_1 \ar@<-1ex>[l]_{d_0, d_1}\ar[l] & 
M_2 \ar@<-1ex>_{d_0, d_1 , d_2}[l] \ar@<1ex>[l] \ar[l] & 
M_3 \ar@<-1ex>[l]_{d_0, ..., d_3} \ar@<1ex>[l]_{\cdot} & 
M_4 \ar@<-1ex>[l]_{d_0, ..., d_4} \ar@<1ex>[l]_{\cdot} \cdots ,
}
\end{equation}
where $d_i$ are the face maps, describing how the $(n-1)$-simplices are
attached to a $n$-simplex.  Similarly, the complex $\cK$ is formally described
by
\begin{equation}\label{eq:nerveT}
\xymatrix{ 
K_0 & 
K_1 \ar@<-1ex>[l]_{d_0, d_1}\ar[l] & 
K_2 \ar@<-1ex>_{d_0, d_1 , d_2}[l] \ar@<1ex>[l] \ar[l] & 
K_3 \ar@<-1ex>[l]_{d_0, ..., d_3} \ar@<1ex>[l]_{\cdot} & 
K_4 \ar@<-1ex>[l]_{d_0, ..., d_4} \ar@<1ex>[l]_{\cdot} \cdots ,
}
\end{equation}
where $K_0$, $K_1$, $K_2$, $\cdots$ are the sets of vertices, links, triangles,
\etc that form the target complex $\cK$.  

In this paper, we will use $v_1,v_2,\cdots \in K_0$ to label different vertices
in the complex $\cK$.  We will use $l_1,l_2,\cdots \in K_1$ to label different
links in the complex $\cK$, and $t_1,t_2,\cdots \in K_2$ different triangles,
\etc.  We choose a fine triangulation on $\cM^{d+1}$ such that the links,
triangles, \etc can be be labeled by their vertices.  In other words, we will
use $i$ to label vertices in $M_0$.  We will use $(ij)$ to label links in
$M_1$, and $(ijk)$ to label triangles in $M_2$, \etc.

The continuous maps between manifolds $\phi(x): M^{d+1}\to K$ is replaced by
homomorphisms between complexes $\phi: \cM^{d+1}\to \cK$.  The homomorphism
$\phi$ is a set of maps $\phi^{(0)}: M_0\to K_0$, $\phi^{(1)}: M_1\to K_1$,
$\phi^{(2)}: M_2\to K_2$, \etc that preserve the attachment structure of
simplices described by the face maps $d_i$.  For example, if $(ij)$ is
attached to $(ijk)$ by the face map $d_3$ in space-time complex $\cM^{d+1}$,
then $\phi^{(1)}((ij))$ is attached to $\phi^{(2)}((ijk))$ by the face map
$d_3$ in target space complex $\cK$.  The homomorphism is the discrete version
of continuous map. Physically, the continuous map or the homomorphism describes
fluctuations without any topological defects and any kind of ``tears''.

\subsubsection{A simple definition of discrete non-linear $\si$-model} 

Now, a \emph{discrete non-linear $\si$-model} is defined via the following path integral 
\begin{align} 
\label{ZMKom} 
Z(\cM^{d+1};\cK,\bar\om_{d+1})=\sum_{\phi} \ee^{2\pi \ii \int_{\cM^{d+1}} \phi^*\bar\om_{d+1}}
\end{align} 
where $\sum_{\phi}$ sums over all the homomorphisms $\phi: \cM^{d+1}\to \cK$.
It is clear that the map $\phi$ assign a label $v_i$ to each vertex $i\in M_0$,
a label $l_{ij}$ to each link $(ij) \in M_1$, a label $t_{ijk}$ to each
triangle $(ijk) \in M_2$, \etc.  Thus we can view the map $\phi$ as a
collection of fields on the space-time complex $\cM$: a field $v_i$ on the
vertices $M_0$, a field $e_{ij}$ on the links $M_1$, a field $t_{ijk}$ on the
triangles $M_2$, \etc.  We can rewrite the path integral as a integration of
those fields:
\begin{align}
\label{Zgab} 
Z(\cM^{d+1};\cK,\bar\om_{d+1})=\sum_{v_i,l_{ij},t_{ijk},\cdots} 
\ee^{2\pi \ii \int_{\cM^{d+1}} \om_{d+1}(v,l,t,\cdots)  } .
\end{align}
Although those fields $v_i,l_{ij},t_{ijk},\cdots$ satisfy certain local
constraints described by the face maps $d_i$,  we can impose those local
constraints by energy penalty: The field configurations that do not satisfy
attachment conditions will cost a large energy.  Thus we can view those fields
as independent fields.

The term $\ee^{2\pi \ii \int_{\cM^{d+1}} \phi^*\bar\om_{d+1}}$ in the path
integral is the action amplitude. Here $\phi^* \bar\om_{d+1}\equiv \om_{d+1}$
is a real-valued $(d+1)$-cochain on $\cM^{d+1}$ which is a pull back of a
real-valued $(d+1)$-cochain $ \bar\om_{d+1}$ on $\cK$.  
The resulting path integral defines a discrete non-linear $\si$-model 
whose fluctuations have no defects.

However, the above definition of discrete non-linear $\si$-model has an
inconvenience: different choices of space-time triangulation may lead to
different phases of the bosonic systems.  To avoid this problem, we like to
choose some special triangulation $\cK$ of the target space $K$, and some
special $\bar\om_{d+1}$'s on $\cK$ such that, for a given pair $(\cK,
\bar\om_{d+1})$, the corresponding discrete defectless non-linear $\si$-model
will realize the same phase for any space-time triangulations, as long as they
are very fine triangulations (\ie in the thermodynamic limit).  Such kind of
choice of $(\cK, \bar\om_{d+1})$ turns out to give rise exactly soluble models.
To describe how we choose $(\cK, \bar\om_{d+1})$, we will first discuss a more
general class of discrete bosonic discrete non-linear $\si$-models -- tensor
network models.

In the above definition of discrete non-linear $\si$-models, we assign each
$d+1$-simplex $\Del^{d+1}$ a field-dependent complex number $\ee^{\ii 2\pi
\int_{\Del^{d+1}} \om_{d+1}}$, and multiply all those numbers together to get
an action amplitude.  In the more general tensor network models, we also assign
each $n$-simplex $\Del^{n}$, $n<d+1$, a field-dependent real positive number,
and multiply all those numbers together to get additional contributions to the
action amplitude.  In the following, we will describe tensor network models in
details.

\subsection{Exactly soluble tensor network models}
\label{ETensor}

\begin{figure}[t]
\begin{center}
\includegraphics[scale=0.6]{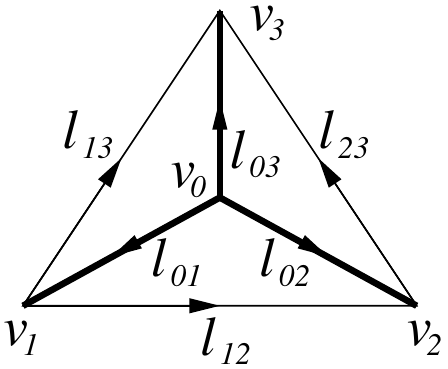}
\end{center}
\caption{
The tensor $\tC{C}0123$ is associated with a tetrahedron, which has a branching
structure.  If the vertex-0 is above the triangle-123, the tetrahedron
has an orientation $s_{0123}=*$.  If the vertex-0 is below the
triangle-123, the tetrahedron has an orientation $s_{0123}=1$. The
branching structure gives the vertices a local order: the $i^{th}$ vertex has
$i$ incoming links.  
}
\label{tetr}
\end{figure}

Let us describe a tensor network model in 2+1D space-time complex $\cM^3$ as an
example. The tensor network model is constructed from a tensor set $\T$ of two
real and one complex tensors: $\T=(w_{v_0}, \tAw{w}01,\tC{C}0123)$.  We will
call $\tC{C}0123)$ the top tensor and $w_{v_0}, \tAw{w}01$ the weight tensors.
The complex tensor $\tC{C}0123$ can be associated with a tetrahedron $(0123)$,
which has a branching structure (see Fig.  \ref{tetr}).  A branching structure
is a choice of orientation of each link in the complex so that there is no
oriented loop on any triangle (see Fig.  \ref{tetr}).  Here the $v_0$ index is
associated with the vertex-0, the $l_{01}$ index is associated with the
link-$01$, and the $t_{012}$ index is associated with the triangle-$012$.  They
represents the degrees of freedom on the vertices, links, and the triangles.
Similarly, the real tensor $\tAw{w}01$ is associated with a link $(01)$, and
$w_{v_0}$ with a vertex $0$.  

Using the tensors, we can define a path integral on any 3-complex that has no
boundary:\cite{KW1458}
\begin{align}
\label{Z3d}
 Z(\cM^3;\T)=
\hskip -3em
\sum_{ v_i,\cdots; l_{ij},\cdots; t_{ijk},\cdots}
&\prod_{i} w_{v_{i}} 
\prod_{(ij)} \tAw{w}ij\times
\\
&
\prod_{(ijkm)} [\tC{C}ijkm ]^{s_{ijkm}}
\nonumber 
\end{align}
where $\sum_{v_i; l_{ij}; t_{ijk}}$ sums over all the
vertex indices, the link indices, and the triangle indices, 
$s_{ijkm}=1$ or $*$
depending on the orientation of tetrahedron $(ijkm)$ (see Fig.  \ref{tetr}).

\begin{figure}[t]
\begin{center}
\includegraphics[scale=0.5]{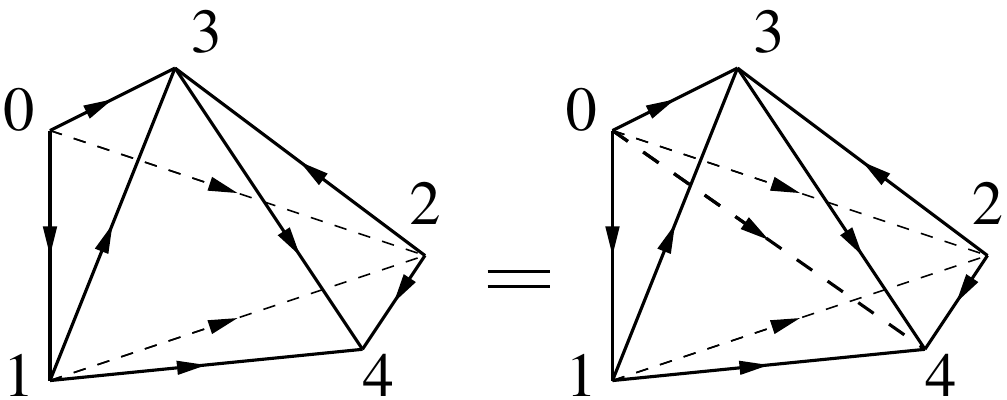}
\end{center}
\caption{
A retriangulation of a 3D complex, obtained by dividing the five 3-simplices on
the boundary of the 4-simplex $(01234)$ into $[(0123),(1234)]$ and
$[(0124),(0134),(0234)]$.
}
\label{2to3}
\end{figure}
\begin{figure}[t]
\begin{center}
\includegraphics[scale=0.5]{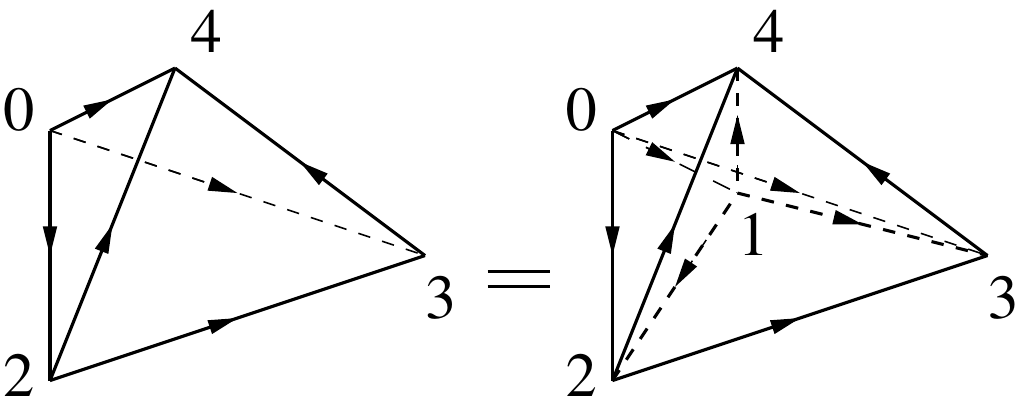}
\end{center}
\caption{
A retriangulation of another 3D complex, obtained dividing the five 3-simplices
on the boundary of the 4-simplex $(01234)$ into $[(0234)]$ and $[(0123),
(0124),(0134),(1234)]$.
}
\label{1to4}
\end{figure}

On the complex $\cM^3$ with boundary: $\cB^2= \prt \cM^3$,
the partition function is defined differently:
\begin{align}
\label{Z3dB}
 Z=\sum_{ \{ v_i; l_{ij}; t_{ijk} \} }
&\prod_{i \notin \cB^2} w_{v_{i}} 
\prod_{(ij) \notin \cB^2} \tAw{w}ij\times
\\
&
\prod_{(ijkm)} [\tC{C}0123 ]^{s_{ijkm}}
\nonumber 
\end{align}
where $\sum_{v_i; l_{ij}; t_{ijk}}$ only sums over the vertex indices, the link
indices, and the triangle indices that are not on the boundary.  
The resulting
$Z$ is actually a complex function of $v_{i}$'s, $l_{ij}$'s, and $t_{ijk}$'s on
the boundary $\cB^2$: $Z=Z(\{v_{i};l_{ij};t_{ijk}\})$.  Such a function is a
vector in a Hilbert space $\cH_{\cB^2}$.  We will denote such a vector by
$|\Psi(\cM^3)\>$.

Consider two complexes $\cM^3_1$ and $\cM^3_2$ with the same boundary $\cB=\prt
\cM^3_1=-\prt \cM^3_2$, the inner product between $|\Psi(\cM^3_1)\>$ and
$|\Psi(\cM^3_2)\>$ can be obtained by gluing $\cM^3_1$ and $\cM^3_2$ together
$\cM^3= \cM^3_1 \cup \cM^3_2$ and perform the path integral on $\cM^3$
\begin{align}
\< \Psi(\cM^3_2)|\Psi(\cM^3_1)\> = Z(\cM^3;\T).
\end{align}
We note that, in the definition of $|\Psi(\cM^3_1)\>$ and $|\Psi(\cM^3_2)\>$,
the tensors $w_{v_{i}}$ and $\tAw{w}ij$ are absent for the vertices and the
links on the boundary.  When we glue two boundaries together, those tensors
$w_{v_{i}}$ and $\tAw{w}ij$ need to be added back.  So the tensors $w_{v_{i}}$
and $\tAw{w}ij$ defines the inner product in the boundary Hilbert space
$\cH_{\cB^2}$.  Therefore, we require $w_{v_{i}}$ and $\tAw{w}ij$ to satisfy
the following unitary condition (or the reflection positivity condition)
\begin{align}
 w_{v_{i}} > 0, \ \ \ \tAw{w}ij >0.
\end{align}

The tensor network model \eq{Z3d} are also inconvenient since for a fixed
tensor set $\T$, different choices of the triangulations of the space-time
$\cM^3$ may lead to different phases.  To solve this problem, we want to choose
the tensors $(w_{v_0}$, $\tAw{w}01$, $\tC{C}0123)$ such that the path integral
is re-triangulation invariant.  The corresponding models will be called a
topological tensor network model, which can realize the same phase for any
triangulations of the space-time $\cM^3$. In general such a phase has a
non-trivial topological order that has gappable boundary.  

The invariance of $Z$ under the re-triangulation in Fig.
\ref{2to3} requires that
\begin{align}
\label{CC23}
&\ \ \ \
\sum_{\phi_{123}} 
\tC{C}0123 
\tC{C}1234
\nonumber\\
&=
\sum_{l_{04}} \tAw{w}04
\sum_{ t_{014} t_{024} t_{034} }
\tC{C}0124 \times
\nonumber\\ & \ \ \ \ \ \ \
\tC{C^*}0134
\tC{C}0234 .
\end{align}
The invariance of $Z$ under the re-triangulation in Fig. \ref{1to4}
requires that
\begin{widetext}
\begin{align}
\label{CC14}
&
\tC{C}0234
=
\sum_{l_{01}l_{12}l_{13}l_{14},v_1} w_{v_1} 
\tAw{w}01 \tAw{w}12 \tAw{w}13 \tAw{w}14 
\sum_{ 
t_{012} 
t_{013} 
t_{014} 
t_{123} 
t_{124} 
t_{134} 
}
\\
&\ \ \ \ \ \ \ \ \ \ \ \ 
\tC{C}0123
\tC{C^*}0124 
\tC{C}0134
\tC{C}1234
\nonumber 
\end{align}
\end{widetext}
There are other similar conditions for different choices of the branching
structures.  To obtain those conditions, we start with a 4-simplex $(01234)$,
and divide the five 3-simplices on the boundary of the 4-simplex $(01234)$ into
two groups.  Then the partition function \eq{Z3dB} on one group of the
3-simplices must equal to the partition function on the other group of the
3-simplices, after a complex conjugation.  

The above two types of the conditions are sufficient to determine the tensor
set $\T$ that produces a topologically invariant partition function $Z$ for any
triangulated space-time $\cM^3$.  For such a tensor set, its  partition
function $Z=Z^\text{top}$ (\ie the energy density in \eqn{ZMTL} $\veps(x) =
0$).  Such topological partition function $Z^\text{top}(\cM^3)$ is nothing but
the topological invariant for three manifolds introduced by Turaev and
Viro.\cite{TV9265} 

\subsection{Topological non-linear $\si$-models}

A subclass of topological tensor network models happen to have a form of
discrete defectless non-linear $\si$-models.  Such topological
tensor network models (\ie exactly soluble discrete non-linear $\si$-models)
are called topological non-linear $\si$-models.

In the following, we will explain why a subclass of topological tensor network
models can be viewed as discrete defectless non-linear $\si$-models.  Again we
will use a 2+1D non-linear $\si$-model as example.  The target complex $\cK$
has a set of vertices labeled by $v$, a set of links labeled by $l$, a set of
triangles labeled by $t$, \etc.  We assume that each tetrahedron in $\cK$ is
uniquely determined by its vertices $v_0,v_1,v_2,v_3$, its links $ l_{01},
l_{02}, l_{03}, l_{12}, l_{13}, l_{23} $, and its triangles $ t_{012}, t_{023},
t_{013} $.

We first assign a complex number to each tetrahedron in $\cK$, which can be
written as $\tC{C}0123$.  When the indices  $v_0,v_1,v_2,v_3$, $ l_{01},
l_{02}, l_{03}, l_{12}, l_{13}, l_{23} $, $ t_{012}, t_{023}, t_{013} $ are not
vertices, links, and triangles of a tetrahedron in $\cK$, then the
corresponding $\tC{C}0123 =0$.  Similarly, we also choose a real tensor
$\tAw{w}01$ whose value is positive when $v_0,v_1,l_{01}$, are the vertices and
the link of a triangle in $\cK$.  Otherwise $\tAw{w}01=0$.  We also assign a
real positive value $w_v$ to each vertex $v$ in $\cK$.  For such a choice of
tensor set $\T$, the partition function \eq{Z3d} actually describes a discrete
defectless non-linear $\si$-model.

To see this we note that a homomorphism $\phi: \cM^3 \to \cK$ assigns a value
$v_i$ (a vertex in $\cK$) to each vertex $i$ in $\cM^3$.  $\phi$ also assigns a
value $l_{ij}$ to each link $(ij)$ and assigns a value $t_{ijk}$ to each
triangle $(ijk)$ in $\cM^3$.  The terms in the summation in \eqn{Z3d} are
non-zero only when the fields $v_i$, $l_{ij}$, $t_{ijk}$ correspond to a
homomorphism $\phi: \cM^3 \to \cK$.  Thus, the summation $\sum_{ \{ v_i;
l_{ij}; t_{ijk} \} }$ in \eqn{Z3d} corresponds to a summation $\sum_{\phi}$
over all the homomorphisms $\phi: \cM^3 \to \cK$. In this case, \eqn{Z3d} can
be viewed as a discrete defectless non-linear $\si$-model.  If the tensors
$w_{v_0}$ $\tAw{w}01$, $\tC{C}0123$ also satisfy the conditions \eqn{CC23} and
\eqn{CC14}, then the corresponding discrete defectless non-linear $\si$-model
will be a topological non-linear $\si$-model.


\subsection{Labeling simplices in a complex}

In the above example, most components  of the tensor $\tC{C}0123$ are zero.
This is because most combinations of $v_0,v_1,v_2,v_3$, $ l_{01}, l_{02},
l_{03}, l_{12}, l_{13}, l_{23} $, $ t_{012}, t_{023}, t_{013} $ are not
vertices, links, and triangles of a tetrahedron in $\cK$.  In the following, we
will describe a more economical way to label simplices in a complex, such that
each label will have a smaller range and a larger fraction of the tensor
elements will be non-zero.

We still use $v$ to label different vertices in the complex $\cK$.  Thus
$K_0=\{v\}$.  To label links in $\cK$, we will first try to use two vertices
$v_0,v_1$ on the two ends of the link to label it.  If there are many links
with the same end points $v_0,v_1$, we will introduce additional label $a_{01}$
to label those links with the same set of end points.  Thus, different links in
$\cK$ are labeled by $(v_0,v_1,a_{01})$, and $K_1=\{ (v_0,v_1,a_{01})\}$.  We
see that the new link label $a_{01}$ has a smaller range than the original link
label $l_{01} \sim (v_0,v_1,a_{01}) $.

In general, the set of the extra labels, $\{a_{01}\}$, depends on the end
points $v_0,v_1$. In this paper, we will only consider a special type of
complex $\cK$ such that the set of the extra labels, $\{a_{01}\}$, does not
depend on the end points $v_0,v_1$.  In this case $a_{01}$ can be treated as a
new label that is independent of vertex label $v_i$.  

Similarly, different triangles $t_{012}$ in $\cK$ are labeled by $t_{012}\sim
(v_0,v_1,v_2,a_{01},a_{12},a_{02},b_{012})$, and
$K_2=\{(v_0,v_1,v_2,a_{01},a_{12},a_{02},b_{012})\}$.  Again the complex $\cK$
has a property that $b_{012}$ is a new label independent of vertex and link
labels $v_i,a_{jk}$.  We like to stress that not all combinations
$\{(v_0,v_1,v_2,a_{01},a_{12},a_{02},b_{012})\}$ correspond to valid triangles
in $\cK$.  Only when $v_0,v_1,v_2,a_{01},a_{12},a_{02},b_{012}$'s satisfy
certain conditions, can they label the triangles in $\cK$.  Using the new set
of labels, the tensors that define topological non-linear $\si$-model can be rewritten as
$w_{v_0}$, $\tAwR{w}01$, and $\tCR{C}0123$, where the indices have a smaller
range.

\section{Dijkgraaf-Witten gauge theories from topological non-linear $\si$-models}

In this section, we will introduce 1-gauge theories (\ie Dijkgraaf-Witten gauge
theories), as topological non-linear $\si$-models.  We will show that 1-gauge
theories are nothing but a special kind of topological non-linear $\si$-models
whose target space $K$ is modeled by a special one-vertex complex $\cK$ and
satisfy $\pi_1(K)=G$, $\pi_{k>1}(K)=0$. Such a one-vertex complex $\cK$ is a
simplicial set and is denoted by $\cB G$.  Similarly $n$-gauge theories are
nothing but a special kind of topological non-linear $\si$-models whose target
spaces $K$ is modeled by a simplicial set $\cB(\pi_1(K),\pi_2(K),\cdots)$ and
satisfy $\pi_{k>n}(K)=0$.

\subsection{Lattice gauge theories from topological non-linear $\si$-models}
\label{labelSimp}

The simplest class of topological non-linear $\si$-models has a simple target
space $K(G)$, the Eilenberg-MacLane spaces with only non-trivial
$\pi_1(K(G))=G$.  For a finite $G$, $K(G)$ is the classifying space $BG$.  
To construct a discrete non-linear $\si$-model from the classifying space
$BG=K(G)$, we need to choose a triangulation of $BG=K(G)$ which is a simplicial
complex.  Here we will choose a triangulation that contains only one vertex.
The corresponding triangulation is a simplicial set denoted by $\cB G$ or
$\cB(G)$.  We will show that for such a one-vertex triangulation, the
topological non-linear $\si$-model becomes a (Dijkfraaf-Witten) lattice gauge
theory, which is also called 1-gauge theory.

The triangulation $\cB G =\cB(G)$ is obtained in the following way: 
\begin{enumerate}
\item
 There is only
one vertex $(\cB G)_0=\{pt\}$ (called the base point) in $BG$.
\item
 The links are the loops starting and ending at the base point. We pick one
loop in each homotopic class of loops: $(\cB G)_1=\pi_1(BG)$.  Thus the links
are labeled by the group elements $a_{ij}\in G$: $(\cB G)_1=G$.  
\item
For arbitrary three links $a_{01},\ a_{12},\ a_{02}$ they may not form the
links around a triangle.  Only when they satisfy $a_{01}a_{12}=a_{02}$, the
composition of the three links is a contractible loop. In this case, there is a
triangle $t_{012}$ bounded by the links $a_{01},a_{12},a_{02}$.  Note that, for
a finite $G$, $\pi_n(BG)=0$ for $n\geq 2$.  Thus all different choices of
triangles are homotopy equivalent.  Here we just pick a particular one.  
This gives rise to the set of 2-simplices labeled
by the three links  $a_{01},a_{12},a_{02}$ that satisfy  $a_{01}a_{12}=a_{02}$.
Thus the set of 2-simplices is $(\cB G)_2 = G^{\times 2}$, labeled by
$a_{01},a_{12}$.  
\item
The set of 3-simplices $(\cB G)_3$ is obtained by filling all four triangles
that form a 2-cycle.  Using a similar consideration, we find the set of
3-simplices to be $(\cB G)_3= G^{\times 3}$, labeled by
$a_{01},a_{12},a_{23}$..  
\end{enumerate}

The sets of higher simplices $(\cB G)_n= G^{\times n}$ are obtained in the same
way.  To summarize, the complex $\cB G$ has the following nerve 
\begin{equation}\label{eq:nerveBG}
\xymatrix{ 
pt & 
G \ar@<-1ex>[l]_{d_0, d_1}\ar[l] & 
G^{\times 2} \ar@<-1ex>_{d_0, d_1 , d_2}[l] \ar@<1ex>[l] \ar[l] & 
G^{\times 3} \ar@<-1ex>[l]_{d_0, ..., d_3} \ar@<1ex>[l]_{\cdot} & 
G^{\times 4} \ar@<-1ex>[l]_{d_0, ..., d_4} \ar@<1ex>[l]_{\cdot} \cdots
}
\end{equation}

Next, let us determine the set of tensors that satisfy the retriangulation
invariance conditions like \eq{CC23} and \eq{CC14}.  We assume the space-time
dimension to be $d+1$.  For each $d+1$-simplex labeled by
$(a_{01},a_{12},\cdots,a_{d,d+1})$ in $\cB G$, we assign a complex number
\begin{align}
 T_{d+1}(a_{ij}) = w_{d+1} \ee^{\ii 2\pi \bar\om_{d+1}(a_{01},a_{12},\cdots,a_{d,d+1})}
\end{align}
where $\bar\om_{d+1}(a_{01},a_{12},\cdots,a_{d,d+1})$ is a $\R/\Z$-valued
cocycle on $\cB G$: $\bar\om_{d+1} \in H^{d+1}(\cB G;\R/\Z)$.  $T$ is the top
tensor in the tensor set $\T$, like the tensor $\tCR{C}0123$ in Section
\ref{labelSimp}.  For each $n$-simplex, $n\leq d$,  we assign a positive number
$w_n$.  $w_n$'s correspond to the weight tensors $w_{v_0}$ and $\tAwR{w}01$ in
Section \ref{labelSimp}.  The partition function of the corresponding
topological non-linear $\si$-model is then given by
\begin{align}
\label{ZBG}
 Z & = \sum_{\phi} 
\Big[ \prod_{n=0}^{d+1} (w_n)^{N_n} \Big]
\ee^{\ii 2\pi \int_{\cM^{d+1}} \phi^* \bar \om_{d+1}}
\end{align}
where $N_n$ is the number of $n$-simplices in $\cM^{d+1}$ and $ \sum_{\phi}$
sums over all the homomorphisms $\phi: \cM^{d+1}\to \cB G$.  Because $\bar
\om_{d+1}$ is a cocycle on $\cB G$, the term $\ee^{\ii 2\pi \int_{\cM^{d+1}}
\phi^* \bar \om_{d+1}}$ is independent on how we triangulate the space-time
$\cM^{d+1}$.  But the term $\sum_{\phi} \prod_{n=0}^{d+1} (w_n)^{N_n}$ does
dependent on the triangulation of $\cM^{d+1}$.  The idea is to choose the
weight tensors $w_n$ to cancel such triangulation dependence.

Let us define two homomorphisms $\phi$ and $\phi'$ to
be homotopic if there exist a homomorphism $\Phi: I\times \cM^{d+1} \to \cB G$
such that, when restricted to the two boundaries of $I\times \cM^{d+1}$, $\Phi$
becomes $\phi$ and $\phi'$.  For such two homomorphisms, we have 
\begin{align}
\label{gaugephi}
 \ee^{2\pi \ii \int_{\cM^{d+1}} \phi^* \bar\om_{d+1}}
=\ee^{2\pi \ii \int_{\cM^{d+1}} \phi^{\prime *} \bar\om_{d+1}}
\end{align}
if the space-time $\cM^{d+1}$ has no boundary. Such a property is called gauge
invariance.  Since the phase factor $\ee^{2\pi \ii \int_{\cM^{d+1}}
\phi^*\bar\om_{d+1}}$ only depends on the homotopic classes $[\phi]$, we can
rewrite it as $\ee^{2\pi \ii \int_{\cM^{d+1}} [\phi]^*\bar\om_{d+1}}$.  For two
homotopic  homomorphisms $\phi$ and $\phi'$, their corresponding field
configurations $a$ and $a'$ are said to be gauge equivalent.  The action
amplitude that satisfies \eqn{gaugephi} is also said to have a ``generalized
global symmetry'' or a higher symmetry.\cite{GW14125148} 

Let us describe the homotopic classes $[\phi]$ in more detail.
First, there is a surjective map
\begin{align}
 \phi \twoheadrightarrow \Hom(\pi_1(\cM^{d+1}),G)
\end{align}
where $\Hom(\pi_1(\cM^{d+1}),G)$ is the set of group homomorphisms.
There is another  surjective map
\begin{align}
\Hom(\pi_1(\cM^{d+1}),G)
\twoheadrightarrow 
\{ [\phi] \} . 
\end{align}
where $\{[\phi]\}$ is the set of homotopic classes of the complex homomorphisms
$\cM^{d+1} \xrightarrow{\phi} \cB G$.  Two group homomorphisms $\ga,\ga' \in
\Hom(\pi_1(\cM^{d+1}),G)$ are said to be equivalent if their are related by
\begin{align}
 \ga = g \ga' g^{-1},\ \ \ g \in G.
\end{align}
Let $[\ga]$ be an equivalent class of the group homomorphisms
$\Hom(\pi_1(\cM^{d+1}),G)$.  It turns out that
\begin{align}
 \{[\ga]\} = \{[\phi]\}
\end{align}
where $\{[\ga]\}$ is the set of equivalent classes of the group homomorphisms.  

Now, $\sum_{\phi}$ is reduced to a summation over the homotopic classes of the
homomorphisms $\phi$, $\sum_{[\phi]}$, which is a sum with only a few terms:
\begin{align} 
Z=
\sum_{[\phi]}  &
\Big[ \prod_{n=0}^{d+1} (w_n)^{N_n} \Big]
N([\phi],\cM^{d+1},\cB G) 
\ee^{2\pi \ii \int_{\cM^{d+1}} [\phi]^*\bar\om_{d+1}}
\end{align} 
where $N([\phi],\cM^{d+1},\cB G)$ is the number of the homomorphisms $\phi:
\cM^{d+1} \to \cB G$ in the homotopic class $[\phi]$.  Due to the one-to-one
correspondence between $[\phi]$ and $[\ga]$, we can also write
$N([\phi],\cM^{d+1},\cB G)$ as $N([\ga],\cM^{d+1},\cB G)$.  The total number of
the homomorphisms $\phi$ is given by
\begin{align}
\label{Nphi}
 N(\cM^{d+1},\cB G) = \sum_{[\phi]} N([\phi],\cM^{d+1},\cB G).
\end{align}

To count $N([\phi],\cM^{d+1},\cB G)$, we note that, in the above discrete
non-linear $\si$-models, the map $\phi$ sends all vertices in $\cM^{d+1}$
(labeled by $i=0,\cdots,N_v-1$) to the base point $pt$ in $\cB G$.  The map
$\phi$ sends an link $ (ij) \in \cM^{d+1}$ to an link  $a_{ij} \in \cB G$.
Thus on each link $(ij)$ of space-time complex $\cM^{d+1}$, we have a degree of
freedom $a_{ij}$.  Note that if three links in space-time complex, $(01)$,
$(12)$, and $(02)$, form the boundary of a triangle $(012)$, then the map
$\phi$ will sends such a triangle to the triangle $t_{012}\in \cB G$ bounded by
$a_{01},a_{12},a_{02}$.  This implies that there is no extra degrees of freedom
on the triangles except those come from the links $a_{01},a_{12},a_{02}$. It
also implies that $a_{ij}$ on the three links $(ij)$ satisfy a flat condition:
\begin{align}
\label{flat}
 a_{ij}a_{jk}=a_{ik}.
\end{align}
This is an example of the conditions discussed above.  Using similar
considerations, we see that there are no extra degrees of freedom on the
3-simplices and higher simplices.  Thus the summation $\sum_{\phi}$
 can be rewritten as $\sum_{a_{ij}}$
where $\sum_{a_{ij}}$ sum over all $a_{ij} \in G$ on link $(ij) \in \cM^{d+1}$,
so that $a_{ij}$ satisfy the flat condition \eq{flat}.

Since the set of $a_{ij}$ describes a flat $G$-gauge connection, we see that
$N([\phi],\cM^{d+1},\cB G)$ is the number gauge equivalent flat $G$-gauge
connections on $\cM^{d+1}$.
We find that 
\begin{align}
N([\phi],\cM^{d+1},\cB G) &= N([\ga],\cM^{d+1},\cB G)
\\
&= |G|^{N_0}  W_\text{top}([\ga],M^{d+1},\cB G)
\nonumber\\
W_\text{top}([\ga],M^{d+1},\cB G) &=W_\text{top}([\phi],M^{d+1},\cB G)=
|[\ga]|/|G| .
\nonumber
\end{align}
where $|G|$ is the number of the elements in the group $G$ and $ |[\ga]|$ is
the number of the elements in the equivalent class $[\ga]$.  Here the factor
$|G|^{N_0}$ comes from the numbers of gauge transformations
\begin{align}
 a_{ij}\to g^i a_{ij} g_j^{-1}
\end{align}
generated by $g_i\in G$ on each vertex $i$ in $\cM^{d+1}$.  Also
$1/W_\text{top}([\phi],M^{d+1},\cB G)$ is the number of gauge transformations
that leave a gauge field $a$ (or $\phi$) invariant.  So
$1/W_\text{top}([\ga],M^{d+1},\cB G)$ is given by the number of the elements in
the subgroup of $G$ thats leave $\ga$ invariant, which is $|G|/|[\ga]|$.  Thus
$W_\text{top}([\phi],M^{d+1},\cB G)$ is independent of the triangulation on
$\cM^{d+1}$.  

$N_0$ in $|G|^{N_0} W_\text{top}([\phi],M^{d+1},\cB G)$ depends on the
triangulation of $M^{d+1}$.  We want to choose $w_n$ to cancel the $N_0$
dependence, which turns out to be
\begin{align}
w_0=|G|^{-1},\ \ \  
\text{ other } w_n=1.
\end{align}
In this case, the partition function \eq{ZBG} becomes
\begin{align}
\label{Za}
 Z & = \sum_{a_{ij}}
\Big( \prod_i |G|^{-1} \Big)
\ee^{\ii 2\pi \int_{\cM^{d+1}} \om_{d+1}(a_{01},a_{12},\cdots,a_{d,d+1}) }
\nonumber\\
& = \sum_{[\phi]}  W_\text{top}([\phi],M^{d+1},\cB G)
\ee^{\ii 2\pi \int_{\cM^{d+1}} [\phi]^* \bar \om_{d+1}}
\end{align}
which is invariant under the retriangulation of space-time $M^{d+1}$.
Such choice of tensors give us a topological non-linear $\si$-model.

We see that the topological non-linear $\si$-models with $\cB G$ as the target complex are
classified by the $(d+1)$-cohomology classes $H^{d+1}(\cB G,\R/\Z)$.  When
$\om_{d+1}=0$, the partition function is given by the equal weight summation of
all flat connections $a_{ij}$ on the links of space-time complex, which give
rise to a $G$-gauge theory in the deconfined phase.  If we choose a non-trivial
cocycle $\om_{d+1}\in H^{d+1}(\cB G,\R/\Z)$, then the path integral \eq{Za}
will gives rise to a Dijkgraaf-Witten lattice gauge theory.

\subsection{Classification of exactly soluble 1-gauge theories}

We have seen that by choosing a classifying space $BG=K(G)$ as the target space
and choosing a particular triangulation of $K(G)$, $\cB(G)$, as the target
complex, we obtain the Dijkgraaf-Witten gauge theories for a finite gauge group
$G$.  For each finite gauge group $G$, we only have one corresponding $K(G)$.
The different $(d+1)$-cohomology classes $\om_{d+1} \in H^{d+1}(K(G), \R/\Z)$
give rise to different Dijkgraaf-Witten gauge theories.  Thus Dijkgraaf-Witten
gauge theories (or 1-gauge theories) are classified by pairs $(G,\om_{d+1})$.  

We have seen that Dijkgraaf-Witten gauge theories are topological non-linear $\si$-models.
It is natural to ask if topological non-linear $\si$-models with target complex $\cB G$
are Dijkgraaf-Witten gauge theories.
In other words, we have shown that the tensor set
\begin{align}
\label{Tw0}
 T_{d+1}(a_{ij}) =  \ee^{\ii 2\pi \bar\om_{d+1}(a_{01},a_{12},\cdots,a_{d,d+1})}
,\ \ \ \
w_0= |G|^{-1}
\end{align}
satisfy the retriangulation invariance conditions, such as \eqn{CC23} and
\eq{CC14}.  The question is that if all the solutions of the retriangulation
invariance conditions (such as \eqn{CC23} and \eq{CC14}) have the form
\eqn{Tw0} as described by a cocycle $\bar\om_{d+1}$.
There is another related question: given a triagulation $\cK$ of the
classifying space $BG$ ($\cK$ may not be a simplicial set), are all the
topological non-linear $\si$-models with target complex $\cK$ equivalent to
Dijkgraaf-Witten gauge theories (\ie produce the same topological invariant
$Z^\text{top}$ or produce the same topological order)? We left the questions
for future work.

\section{2-gauge theories from topological non-linear $\si$-models}

In this section, we are going to discuss exactly soluble 2-gauge theories and
their classification, from a point of view of topological non-linear $\si$-model.  We have
seen that if the target space $K$ has only non-trivial $\pi_1(K)$, we can
get a 1-gauge theory from the topological non-linear $\si$-model.  If the target space
$K$ has only non-trivial $\pi_1(K)$ and $\pi_2(K)$, then we can get a 2-gauge
theory.

\subsection{2-groups}
\label{2group}

To obtain a 2-gauge theory via a topological non-linear $\si$-model, we choose a special
triangulation of $K(G,\Pi_2)$, the simplicial set $\cB(G,\Pi_2)$, as the target
complex.  The simplicial set $\cB(G,\Pi_2)$ is called a 2-group.  The
corresponding topological non-linear $\si$-model can be a 2-gauge theory. 
In this section, we concentrate on 2-groups $\cB(G,\Pi_2)$, where
 $G$ is a finite group and $\Pi_2$ a finite abelian group.

The simplicial set $\cB(G;\Pi_2)$ (the 2-group) can be viewed as a fiber bundle
with $\cB(0;\Pi_2)=\cB(\Pi_2,2)$ as the fiber and $\cB(G)$ as the base space:
\begin{align} \label{eq:2gp}
 \cB(\Pi_2,2) \to \cB(G;\Pi_2) \to \cB(G)  .
\end{align}
Thus a classification of $\cB(G;\Pi_2)$ can be obtain using the following
general result:
\begin{lemma}\label{lem:fibration}
The simplicial set 
$\cB(\pi_1;\cdots;\pi_n)$ has the following fibration
\[
\cB(\pi_n,n) \to \cB(\pi_1;\cdots;\pi_n) \to  \cB(\pi_1;\cdots;\pi_{n-1}), 
\] 
Thus $\cB(\pi_1;\cdots;\pi_n)$ for fixed $\pi_i$'s are classified by $H^{n+1}
[\cB(\pi_1;\cdots;\pi_{n-1}); \pi_n]$ with local coefficient $\pi_n$.
\end{lemma}
\noindent
The $n=2$ case was discussed in \Ref{BLm0307200}, Theorem 43. 

Using the above result, we find that, for a fixed pair $(G,\Pi_2)$, the
2-groups $\cB(G;\Pi_2)$ are classified by $H^{3} (\cB(G), \Pi_2)= H^{3} (\cB G,
\Pi_2^{\al_2})$.  The local coefficient $\Pi_2$ in topological cohomology
classes $H^{3} (\cB G, \Pi_2^{\al_2})$ means that $G$ may have a non-trivial
action on $\Pi_2$, which is described by $\al_2: G\to \text{Aut}(\Pi_2)$.  Such
an action is indicated by the superscript $\al_2$ in $\Pi_2^{\al_2}$.  

To summarize, 2-groups $\cB(G;\Pi_2)$  are classified by the following data
\begin{align}
\label{2grp}
  G;\ \Pi_2, \al_2, \bar n_3
\end{align}
where $G$ is a finite group, $\Pi_2$ a finite abelian group, $\al_2$ a group
action $\al_2: G\to \text{Aut}(\Pi_2)$, and $\bar n_3 \in H^3(\cB(G),
\Pi_2^{\al_2})$.  The cocycle condition that determines
$\bar n_3(a_{01},a_{12},a_{23})$ is given by
\begin{widetext}
\begin{align}
0 &= \al_2(a_{01})\cdot 
\bar n_3(a_{12},a_{23},a_{34})
-\bar n_3(a_{02},a_{23},a_{34})
+\bar n_3(a_{01},a_{13},a_{34})
-\bar n_3(a_{01},a_{12},a_{24})
+\bar n_3(a_{01},a_{12},a_{23})
\\
&= \al_2(a_{01})\cdot 
\bar n_3(a_{12},a_{23},a_{34})
-\bar n_3(a_{01}a_{12},a_{23},a_{34})
+\bar n_3(a_{01},a_{12}a_{23},a_{34})
-\bar n_3(a_{01},a_{12},a_{23}a_{34})
+\bar n_3(a_{01},a_{12},a_{23})
\nonumber 
\end{align}
for all $a_{01}$, $a_{12}$, $a_{23}$, and $a_{34}$.  

After knowing how to label all the 2-groups $\cB(G;\Pi_2)$ using the data
\eq{2grp}, the next important question is to obtain a detailed description
of the simplicial set $\cB(G;\Pi_2)$ from the classifying data \eq{2grp}.
The simplicial set $\cB(G;\Pi_2)$ has the following sets of simplices:
\begin{equation}\label{eq:nerve2a}
\xymatrix{ pt & 
G \ar@<-1ex>[l]_{d_0, d_1}\ar[l] & 
G^{\times 2} \times \Pi_2 \ar@<-1ex>_{d_0, ..., d_2}[l] \ar@<1ex>[l] \ar[l] & 
G^{\times 3} \times \Pi_2^{\times 3} \ar@<-1ex>[l]_{d_0, ..., d_3} \ar@<1ex>[l]_{\cdot} & 
G^{\times 4} \times \Pi_2^{\times 6} \ar@<-1ex>[l]_{d_0, ..., d_4} \ar@<1ex>[l]_{\cdot} \cdots& 
G^{\times 5} \times \Pi_2^{\times 10} \dots \ar@<-1ex>[l]_{d_0, ..., d_5} \ar@<1ex>[l]_{\cdot} 
}
\end{equation}
\end{widetext}
Let us describe the sets of simplices and the face map $\dd_m$ in more details.
First there is only one vertex $pt$ in $\cB(G;\Pi_2)$.  The links in
$\cB(G;\Pi_2)$ are labeled by elements $a_{ij}$ in $G$.  All the links connect
$pt$ to $pt$, and correspond to non-contractable loops in $\pi(\cB(G;\Pi_2))=G$.
Thus the face maps are give by
\begin{align}
 \dd_0 (a_{01}) = pt, \ \ \ \
 \dd_1 (a_{01}) = pt .
\end{align}
The boundary map is given by $\prt=\dd_0-\dd_1$. We see that $\prt (a_{01}) =0$
and the link $(a_{01})$ is a 1-cycle, for all $a_{01} \in G$.

The composition of two links $a_{01}$ and $a_{12}$
can be deformed into the link $a_{02}$ if and only if
\begin{align}
\label{ggg}
a_{01}a_{12}=a_{02}.
\end{align}
Thus $a_{01}$, $a_{12}$, and $a_{02}$ are boundary of a triangle if and only if
$a_{01}a_{12}=a_{02}$.  

The $G$-valued $a_{ij}$ on each link of $\cB(G,\Pi_2)$ define a $G$-valued
$1$-cochain $\bar a$, which is called a canonical $1$-cochain.  Using $\bar
a$, the above condition can be written as
\begin{align}
 \del \bar a \equiv a_{01}a_{12}a_{02}^{-1} =\one .
\end{align}
This implies that the canonical $1$-cochain $\bar a$ is a $G$-valued
$1$-cocycle.

When $a_{01}a_{12}=a_{02}$, there may be many
triangles with the same boundaries $a_{01}$, $a_{12}$, and $a_{02}$. Those
triangles are labeled by elements in $\Pi_2$.  Thus all the triangles are
labeled by $(a_{01}, a_{12}, a_{02}; b_{012})$ where $a_{ij}$ satisfy
\eqn{ggg}.  If we use independent $a_{ij}$, we find all the triangles are
labeled by $[a_{01}, a_{12}; b_{012}]$, which leads to the set of triangles
$G^{\times 2} \times \Pi_2$.
The face maps are given by
\begin{align}
 \dd_0 (a_{01}, a_{12}, a_{02}; b_{012}) = (a_{12}),
\nonumber\\
 \dd_1 (a_{01}, a_{12}, a_{02}; b_{012}) = (a_{02}),
\nonumber\\
 \dd_2 (a_{01}, a_{12}, a_{02}; b_{012}) = (a_{01}),
\end{align}
which map the traingle to one of its links.
From the face maps $\dd_m$, we obtain
the boundary map $\prt$:
\begin{align}
 \prt =\dd_0-\dd_1+\dd_2.
\end{align}
Thus the boundary of triangle  $(a_{01}, a_{12}, a_{02}; b_{012})$
is given by
\begin{align}
 \prt (a_{01}, a_{12}, a_{02}; b_{012}) = 
(a_{12}) -(a_{02}) +(a_{01}).
\end{align}

Using the above boundary map, we find that four triangles $-(a_{01}, a_{12},
a_{02}; b_{012})$, $(a_{01}, a_{13}, a_{03}; b_{013})$, $-(a_{02}, a_{23},
a_{03}; b_{023})$, $(a_{12}, a_{23}, a_{13}; b_{123})$, form a 2-cycle since
their boundaries cancel each other.  Note that $a_{ij}$ in each triangle must
satisfy \eqn{ggg}. Otherwise, they will not form triangles.  But the 2-cycle
formed by the four  triangles may not be the boundary of a tetrahedron in
$\cB(G;\Pi_2)$.  In order to have a tetrahedron in $\cB(G;\Pi_2)$ that fill the
2-cycle, $b_{ijk}$'s must satisfy a condition.  In other words $a_{ij}$'s and
$b_{ijk}$'s that label the links and triangles in a tetrahedron in
$\cB(G;\Pi_2)$ must satisfy a condition.  Such a condition can be described
using the cochain language (see Appendix \ref{cochain}) if we introduce a
$\Pi_2$-valued canonical $2$-cochain $\bar b$, as defined by the values
$b_{ijk}$ on all the triangles of $\cB(G;\Pi_2)$.  Using $\bar b$, the
condition on $b_{ijk}$ can be written as
\begin{align}
\label{bn3a}
 \dd \bar b = \bar n_3(\bar a) ,
\end{align}
So, the  canonical $2$-cochain $\bar b$ may not be a cocycle.  Its derivative
is given by a function of canonical $1$-cocycle $\bar a$.  When $\al_2$ is
trivial, the above have the following explicit expression:
$a_{ij}$'s and
$b_{ijk}$'s that label the links and triangles in a tetrahedron
satisfy
\begin{align}
\label{bn3b}
 b_{123}- b_{023}+ b_{013}- b_{012} = \bar n_3(a_{01},a_{12},a_{23}) .
\end{align}
When $\al_2$ is non-trivial, $\dd \bar b = \bar n_3(\bar a)$ becomes
\begin{align}
\label{bn3}
 \al_2(a_{01})\cdot b_{123}- b_{023}+ b_{013}- b_{012} = \bar n_3(a_{01},a_{12},a_{23}) .
\end{align}

We see that the tetrahedrons in $\cB(G;\Pi_2)$ are labeled by $(a_{01}$,
$a_{12}$, $a_{23}$, $a_{02}$, $a_{13}$, $a_{03}$; $ b_{012}$, $ b_{023}$, $
b_{013}$, $ b_{123})$ that satisfy \eqn{ggg} and \eqn{bn3}.  In other
words, the tetrahedrons in $\cB(G;\Pi_2)$ are labeled by independent indices
$[a_{01}$, $a_{12}$, $a_{23}$; $b_{012}$, $b_{023}$, $b_{013}]$. 
Those tetrahedrons form the set $G^{\times 3}\times \Pi_2^{\times 3}$
in \eqn{eq:nerve2a}.

The face maps $\dd_m$'s on tetrahedrons are given by
\begin{align}
\label{dmgb1}
\dd_0&
(a_{01},a_{12},a_{23},a_{02},a_{13},a_{03};b_{012},b_{023},b_{013},b_{123})
\nonumber\\
&=(a_{12},a_{23},a_{13};b_{123})
\nonumber\\
\dd_1&
(a_{01},a_{12},a_{23},a_{02},a_{13},a_{03};b_{012},b_{023},b_{013},b_{123})
\nonumber\\
&=(a_{02},a_{23},a_{03};b_{023})
\nonumber\\
\dd_2&
(a_{01},a_{12},a_{23},a_{02},a_{13},a_{03};b_{012},b_{023},b_{013},b_{123})
\nonumber\\
&=(a_{01},a_{13},a_{03};b_{013})
\nonumber\\
\dd_3&
(a_{01},a_{12},a_{23},a_{02},a_{13},a_{03};b_{012},b_{023},b_{013},b_{123})
\nonumber\\
&=(a_{01},a_{12},a_{02};b_{012})
\end{align}
Let us introduce $s[01]$ to describe the link $(a_{01})$, $s[012]$ the triangle
$(a_{01},a_{12},a_{02};b_{012})$, $s[0123]$ the tetrahedron
$(a_{01},a_{12},a_{23},a_{02},a_{13},a_{03};b_{012},b_{023},b_{013},b_{123})$,
\etc.  Then, the above expression can
be put in a more compact form
\begin{align}
\label{dms0123}
 \dd_0 s[0123]=s[123], \ \ \ 
 \dd_1 s[0123]=s[023], 
\nonumber\\
 \dd_2 s[0123]=s[013], \ \ \ 
 \dd_3 s[0123]=s[012] .
\end{align}
Using independent labels, \eqn{dmgb1} can be rewritten as
\begin{align}
\label{dms0123i}
& \dd_0 [a_{01},a_{12},a_{23};b_{012},b_{023},b_{013}] =[a_{12},a_{23};b_{123}]
=[a_{12},a_{23};
\nonumber\\
 & \ \ \ \ \al_2^{-1}(a_{01})\cdot (b_{023} -b_{013} +b_{012}
+\bar n_3(a_{01},a_{12},a_{23})) ],
\nonumber\\
& \dd_1 [a_{01},a_{12},a_{23};b_{012},b_{023},b_{013}] =[a_{02},a_{23};b_{023}],
\nonumber\\
& \dd_2 [a_{01},a_{12},a_{23};b_{012},b_{023},b_{013}] =[a_{01},a_{13};b_{013}],
\nonumber\\
& \dd_3 [a_{01},a_{12},a_{23};b_{012},b_{023},b_{013}] =[a_{01},a_{12};b_{012}] .
\end{align}
The boundary map $\prt$ for tetrahedron is given by
\begin{align}
 \prt=\dd_0-\dd_1+\dd_2-\dd_3 .
\end{align}
Thus
\begin{align}
& \prt [a_{01},a_{12},a_{23};b_{012},b_{023},b_{013}]
=[a_{12},a_{23};
\\
 & \ \ \ \ \al_2^{-1}(a_{01})\cdot (b_{023} -b_{013} +b_{012}
+\bar n_3(a_{01},a_{12},a_{23})) ]
\nonumber\\
& \ \ \ \ 
-[a_{02},a_{23};b_{023}]
+[a_{01},a_{13};b_{013}]
-[a_{01},a_{12};b_{012}] .
\nonumber 
\end{align}

In general, the $n$-simplices in $G^{\times n}\times \Pi_2^{(^n_2)}$ are
labeled by $(a_{ij}, b_{klm})$, $i<j$, $k<l<m$ $i,j,k,l,m=0,1,\cdots,n$, that
satisfy the conditions \eq{ggg} (after replacing $012$ by $i<j<k$) and
\eqn{bn3} (after replacing $0123$ by $i<j<k<l$).  We see that all the
$a_{ij}$'s are determined by the independent $a_{01}, a_{12}, \cdots,
a_{n-1,n}$.  Similarly, all the $b_{ijk}$'s are given by an independent subset
of $b_{ijk}$'s.  Such independent subset is obtained 
by picking $i=0$, and $j<k$.

Using the labeling scheme $(a_{ij}, b_{ijk})$, $i,j,k=0,1,\cdots,n$, where
$a_{ij}, b_{ijk}$ satisfy \eqn{ggg} and \eqn{bn3}, we can obtain a simple
description of the face map $\dd_m$ in \eqn{eq:nerve2a} that sends a
$n$-simplex to a $(n-1)$-simplex.  To describe the action of $\dd_m$, we start
with a $n$-simplex $(a_{ij}, b_{ijk})$.  The resulting $n-1$-simplex is
obtained by dropping all in $a_{ij}, b_{ijk}$ in the set $(a_{ij}, b_{ijk})$
that contain the vertex $m$.  This changes $(a_{ij}, b_{ijk})$ to its subset
which is written as
\begin{align}
\label{dmgb}
\dd_m (a_{ij}, b_{ijk}| 0\leq i,j,k \leq n )
=
(a_{ij}, b_{ijk}|i,j,k \neq m )
.
\end{align}
$a_{ij}, b_{ijk}$ in the subset also satisfy  \eqn{ggg} and \eqn{bn3}.  The
subset $\dd_m (a_{ij}, b_{ijk})$ describes the resulting $n-1$-simplex after
the $\dd_m$ map.  We see that the explicit expression for $\dd_m (a_{ij},
b_{ijk}| 0\leq i,j,k \leq n)$ is simple to construct using non-independent
$a_{ij}, b_{ijk}$'s. 

\subsection{2-gauge theories}

To define a $d+1$D topological non-linear $\si$-model (we will assume $d \geq 2$  since
there is no 2-gauge theory in $1+1$D), we need to specify the tensor set $\T$.
To do so, for each $d+1$-simplex labeled by $(a_{ij},b_{ijk})$ in
$\cB(G,\Pi_2)$ we assign a complex number
\begin{align}
\label{Tab}
 T_{d+1}(a_{ij},b_{ijk}) = w_{d+1} \ee^{\ii 2\pi \bar\om_{d+1}(a_{ij},b_{ijk})}
\end{align}
where $\bar\om_{d+1}(a_{ij},b_{ijk})$ is a $\R/\Z$-valued cocycle on $\cB
(G,\Pi_2)$: $\bar\om_{d+1} \in H^{d+1}(\cB(G,\Pi_2),\R/\Z)$.  $T$ is the top
tensor in the tensor set $\T$.  For each $n$-simplex in $\cB(G,\Pi_2)$, $n\leq
d$,  we assign a positive number $w_n$, which correspond to the weight tensors
in the tensor set.  The partition function of the corresponding topological non-linear
$\si$-model is then given by
\begin{align}
\label{ZBGPi2}
 Z & = \sum_{\phi} 
\Big[ \prod_{n=0}^{d+1} (w_n)^{N_n} \Big]
\ee^{\ii 2\pi \int_{\cM^{d+1}} \phi^* \bar \om_{d+1}}
\end{align}
where $N_n$ is the number of $n$-simplices in $\cM^{d+1}$ and $ \sum_{\phi}$
sums over all the homomorphisms $\phi: \cM^{d+1}\to \cB( G,\Pi_2)$.  

The pullbacks of the canonical cochains $\bar a$ and $\bar b$ on $\cB(
G,\Pi_2)$ by the homomorphisms $\phi$ give rise to cochains $a$ and $b$ on
$\cM^{d+1}$:
\begin{align}
 a=\phi^* \bar a,\ \ \ 
 b=\phi^* \bar b.
\end{align}
$a$ and $b$ are referred as gauge field and rank-2 gauge field in physics,
which satisfy
\begin{align}
 \del a =\one,\ \ \ \dd b =n_3(a).
\end{align}
In fact there is a one-to-one correspondence between the allowed field
configurations $a$ and $b$ and the homomorphisms.  Thus we can replace
$\sum_{\phi}$ and $\sum_{a,b}$:
\begin{align}
\label{ZBGPi2ab}
 Z & = \sum_{a,b} 
\Big[ \prod_{n=0}^{d+1} (w_n)^{N_n} \Big]
\ee^{\ii 2\pi \int_{\cM^{d+1}} \om_{d+1}(a,b)}
\end{align}

As shown in \eqn{gaugephi}, homotopic homomorphisms  $\phi$'s give rise to the
same action amplitude $\ee^{2\pi \ii \int_{\cM^{4}} \phi^*\bar\om_{d+1} }$.
Thus the partition function
can be written as
\begin{align}
\label{ZBGPi2class}
 Z  = \sum_{[\phi]} &
\Big[ \prod_{n=0}^{d+1} (w_n)^{N_n} \Big]
N([\phi],\cM^{d+1}, \cB( G,\Pi_2)) \times
\nonumber\\ & 
\ee^{\ii 2\pi \int_{\cM^{d+1}} [\phi]^* \bar \om_{d+1}}
\end{align}
where $N([\phi],\cM^{d+1}, \cB( G,\Pi_2))$ is the number of homomorphisms
$\phi: \cM^{d+1}\to \cB( G,\Pi_2)$ in the homotopic class $[\phi]$.

Let two field configurations $a_{ij},b_{ijk},\cdots$ and
$a_{ij}',b_{ijk}',\cdots$ on $\cM^{d+1}$ come from two homotopic homomorphisms
$\phi$ and $\phi'$.  Thus the two field configurations have the same the action
amplitude $\ee^{2\pi \ii \int_{\cM^{4}} \phi^*\bar\om_{d+1} }$.  We say that
the two configurations differ by a gauge transformation.

The gauge equivalent  field configurations are generated by two kinds of gauge
transformations: The first one is generated by $g_i$ on each vertex
\begin{align}
\label{gauge2a}
a_{ij}  &\to a_{ij}'=g_i a_{ij} g_j^{-1} , \ \
\nonumber\\
b_{ijk} &\to b_{ijk}' = b_{ijk} + \zeta_2(a_{ij},a_{jk},g_i,g_j,g_k)
\end{align}
where $\zeta_2(a_{ij},a_{jk},g_i,g_j,g_k)$ is a $\Pi_2$-valued function
that satisfy
\begin{align}
&\ \ \ \ (\dd \zeta_2)(a_{ij},a_{jk},a_{kl},g_i,g_j,g_k,g_l) 
\nonumber \\
&= 
-\zeta_2(a_{ij},a_{jk},g_i,g_j,g_k) 
+\zeta_2(a_{ik},a_{kl},g_i,g_k,g_l) 
\\
&\ \ \ 
-\zeta_2(a_{ij},a_{jl},g_i,g_j,g_l) 
+ \al_2(g_{ij})\cdot \zeta_2(a_{jk},a_{kl},g_j,g_k,g_l)
\nonumber\\
&= 
n_3(g_i a_{ij} g_j^{-1},g_j a_{jk} g_k^{-1},g_k a_{kl} g_l^{-1})
-n_3(a_{ij},a_{jk},a_{ik})
\nonumber 
\end{align}
Since $n_3$ is a cocycle, the above equation always has a solution.
The second one is generated by $\Pi_2$-valued $\la_{ij}$ on each link
\begin{align}
\label{gauge2b}
a_{ij}  & \to a_{ij}' =a_{ij}  , 
\nonumber\\
b_{ijk} & \to b_{ijk}'  = b_{ijk} + \la_{ij}-\la_{ik}+\al_2(g_{ij})\cdot \la_{jk} .
\end{align}
\eqn{gauge2a} and \eqn{gauge2b} generate the 2-gauge transformations.
The action amplitude $\ee^{2\pi \ii \int_{\cM^{4}} \om_{d+1}(a,b) }$ is
invariant under the 2-gauge transformations.

Since $N([\phi],\cM^{d+1}, \cB( G,\Pi_2))$ counts 2-gauge equivalent field
configurations, from the above form of 2-gauge transformations, we see that
\begin{align}
&\ \ \ \
 N([\phi],\cM^{d+1}, \cB( G,\Pi_2)) 
\nonumber\\
&= |G|^{N_0} |\Pi_2|^{N_1} 
W_\text{top}([\phi],M^{d+1}, \cB( G,\Pi_2)) .
\end{align}
To cancel the triangulation dependence $N_0$ and $N_1$, we choose the weight
tensors to be
\begin{align}
\label{w01}
 w_0=|G|^{-1},\ \ \
 w_1=|\Pi_2|^{-1},\ \ \
\text{ other } w_n=1.
\end{align}
Such choice of top and weight tensors, \eq{Tab} and \eq{w01}, give rise to a
topological non-linear $\si$-model which is a 2-gauge theory.

We like to remark that \eqn{Tab} and \eqn{w01} represent one class of the
solutions to the retriangulation invariance conditions (like \eqn{CC23} and
\eqn{CC14}).  It is not clear if \eqn{Tab} and \eqn{w01} represent all the
solutions to the retriangulation invariance conditions.
In other words, it is not clear if
topological non-linear $\si$-models with target complex 
$\cB( G,\Pi_2)$ are always 2-gauge theories described by (see \eqn{ZBGPi2ab})
\begin{align}
\label{Z2g}
 Z & = \sum_{a,b} 
\Big( \prod_i |G|^{-1} \prod_{(ij)}|\Pi_2|^{-1} \Big)
\ee^{\ii 2\pi \int_{\cM^{d+1}} \om_{d+1}(a,b)}
\nonumber\\
& = \sum_{[\phi]}  
W_\text{top}([\phi],M^{d+1}, \cB( G,\Pi_2)) 
\ee^{\ii 2\pi \int_{\cM^{d+1}} [\phi]^* \bar \om_{d+1}}
\end{align}

Since the data $(G;\ \Pi_2, \al_2, \bar n_3)$ classify the 2-groups, the
$d+1$D 2-gauge theories are then classified by the following data
\begin{align}
  G;\ \Pi_2, \al_2, \bar n_3;\ \bar \om_{d+1} 
\end{align}
where $\bar \om_{d+1} \in H^{d+1}(\cB(G,\Pi_2), \R/\Z)$.  Using the above data, we
can construct a 2-gauge theory \eqn{Z2g}.

\subsection{2-group cocycles}

$\bar\om_{d+1}$ in \eqn{Z2g} is called a 2-group cocycle.  In the following,
we give an explicit description of 2-group cocycles, based on the 
discussion in Section \ref{2group}.  First, a $d+1$D 2-group cochain $\bar\om_{d+1}$ with value $\M$ is
a function $\bar\om_{d+1}: G^{\times d}\times \Pi_2^{(^d_2)} \to \M$.  Then we
can define the differential operator $\dd$ acting on the 2-group cochains as
the following (see \eqn{dms0123} or \eqn{dms0123i}):
\begin{align}
(\dd \bar\om_{d+1})( s[0\cdots d+1] )
=\sum_{m=0}^{d+1} (-)^m \bar\om_{d+1}(s[1\cdots \hat m \cdots d+1]).
\end{align}
In each dimension, we obtain:
\begin{align}
&\ \ \ \ (\dd\bar\om_0)(a_{01}) = 0,
\end{align}
\begin{align}
&\ \ \ \ (\dd\bar\om_1)(a_{01}, a_{12}, b_{012})
= \bar\om_1(a_{01}) - \bar\om_1(a_{02}) + \bar\om_1(a_{12}) ,
\end{align}
\begin{align}
&\ \ \ \ (\dd\bar\om_2)(a_{01}, a_{12}, a_{23}, b_{012}, b_{013}, 
b_{023})
 \nonumber\\ 
&= -\bar\om_2(a_{01}, a_{12}, b_{012}) + \bar\om_2(a_{01}, a_{13}, 
b_{013}) 
\nonumber\\
& \ \ \ 
- \bar\om_2(a_{02}, a_{23}, b_{023}) + \bar\om_2(a_{12}, 
a_{23}, b_{123}) ,
\end{align}
\begin{align}
&\ \ \ \ (\dd\bar\om_3)(a_{01}, a_{12}, a_{23}, a_{34}, b_{012}, b_{013}, \
b_{014}, b_{023}, b_{024}, b_{034})
 \nonumber\\ 
&= + \bar\om_3(a_{01}, a_{12}, a_{23}, b_{012}, b_{013}, b_{023}) 
\nonumber\\
&\ \ \
- \bar\om_3(a_{01}, a_{12}, a_{24}, b_{012}, b_{014}, b_{024}) 
\nonumber\\
&\ \ \
+ \bar\om_3(a_{01}, a_{13}, a_{34}, b_{013}, b_{014}, b_{034}) 
\nonumber\\
&\ \ \
- \bar\om_3(a_{02}, a_{23}, a_{34}, b_{023}, b_{024}, b_{034}) 
\nonumber\\
&\ \ \
+ \bar\om_3(a_{12}, a_{23}, a_{34}, b_{123}, b_{124}, b_{134}) ,
\end{align}
\begin{widetext}
\begin{align}
 &\ \ \ \ (\dd\bar\om_4)(a_{01}, a_{12}, a_{23}, a_{34}, a_{45}, b_{012}, 
b_{013}, b_{014}, b_{015}, b_{023}, b_{024}, b_{025}, b_{034}, 
b_{035}, b_{045})
 \nonumber\\ 
&= -\bar\om_4(a_{01}, a_{12}, a_{23}, a_{34}, b_{012}, b_{013}, b_{014}, 
b_{023}, b_{024}, b_{034}) 
+ \bar\om_4(a_{01}, a_{12}, a_{23}, a_{35}, b_{012}, b_{013}, b_{015}, b_{023}, b_{025}, b_{035}) 
\nonumber\\
&\ \ \ - \bar\om_4(a_{01}, a_{12}, a_{24}, a_{45}, b_{012}, b_{014}, 
b_{015}, b_{024}, b_{025}, b_{045}) 
+ \bar\om_4(a_{01}, a_{13}, a_{34}, a_{45}, b_{013}, b_{014}, b_{015}, b_{034}, b_{035}, b_{045}) 
\nonumber\\
&\ \ \ - \bar\om_4(a_{02}, a_{23}, a_{34}, a_{45}, b_{023}, b_{024}, b_{025}, b_{034}, b_{035}, b_{045}) 
 + \bar\om_4(a_{12}, a_{23}, a_{34}, a_{45}, b_{123}, b_{124}, b_{125}, b_{134}, b_{135}, b_{145}),
\end{align}
\begin{align}
 &\ \ \ \ (\dd\bar\om_5)(a_{01}, a_{12}, a_{23}, a_{34}, a_{45}, a_{56}, 
b_{012}, b_{013}, b_{014}, b_{015}, b_{016}, b_{023}, b_{024}, 
b_{025}, b_{026}, b_{034}, b_{035}, b_{036}, b_{045}, b_{046}, b_{056})
 \nonumber\\ 
&= + \bar\om_5(a_{01}, a_{12}, a_{23}, a_{34}, a_{45}, b_{012}, b_{013}, b_{014}, b_{015}, b_{023}, b_{024}, b_{025}, b_{034}, b_{035}, b_{045}) 
\nonumber\\
&\ \ \ - \bar\om_5(a_{01}, a_{12}, a_{23}, a_{34}, a_{46}, b_{012}, b_{013}, b_{014}, b_{016}, b_{023}, b_{024}, b_{026}, b_{034}, b_{036}, b_{046}) 
\nonumber\\
&\ \ \ + \bar\om_5(a_{01}, a_{12}, a_{23}, a_{35}, a_{56}, b_{012}, b_{013}, b_{015}, b_{016}, b_{023}, b_{025}, b_{026}, b_{035}, b_{036}, b_{056}) 
\nonumber\\
&\ \ \ - \bar\om_5(a_{01}, a_{12}, a_{24}, a_{45}, a_{56}, b_{012}, b_{014}, b_{015}, b_{016}, b_{024}, b_{025}, b_{026}, b_{045}, b_{046}, b_{056}) 
\nonumber\\
&\ \ \ + \bar\om_5(a_{01}, a_{13}, a_{34}, a_{45}, a_{56}, b_{013}, b_{014}, b_{015}, b_{016}, b_{034}, b_{035}, b_{036}, b_{045}, b_{046}, b_{056}) 
\nonumber\\
&\ \ \ - \bar\om_5(a_{02}, a_{23}, a_{34}, a_{45}, a_{56}, b_{023}, b_{024}, b_{025}, b_{026}, b_{034}, b_{035}, b_{036}, b_{045}, b_{046}, 
b_{056}) 
\nonumber\\
&\ \ \ + \bar\om_5(a_{12}, a_{23}, a_{34}, a_{45}, a_{56}, b_{123}, b_{124}, b_{125}, b_{126}, b_{134}, b_{135}, b_{136}, b_{145}, b_{146}, b_{156})
\end{align}
\end{widetext}
In the above,  the variables $a_{ij}$ with $j-i>1$ and $b_{ijk}$ with $i\neq 0$
do not appear on the left-hand-side of the equation but appear  on the
right-hand-side of the equation.  In fact, those $a_{ij}$ and $b_{ijk}$ are
given by $a_{i,i+1}$'s and $b_{0mn}$'s that do appear on the left-hand-side of
the equation:
\begin{align}
 a_{ij}&=a_{i,i+1}\cdots a_{j-1,j},\ \ \ \text{if } j-i \geq 2,
\nonumber\\
 b_{ijk} &= \al_2^{-1}(a_{01})\cdot [ b_{0jk}-b_{0ik}+b_{0ij}+\bar n_3(a_{0i},a_{ij},a_{jk}) ].
\end{align}
So the above are conditions on the functions of
$a_{i,i+1}$'s and $b_{0mn}$'s.

With the above definition of $\dd$ operator, we can define the 2-group cocycles
as the 2-group cochains that satisfy $\dd\bar\om_{d+1} =0$.  This generalizes
the notion of group cocycle to 2-group cocycle.  Two different 2-group cocycles
$\bar\om_{d+1}$ and $\bar\om_{d+1}'$ are equivalent if they differ by a 2-group
coboundary $\dd \bar \nu_{d}$.  The set of equivalent classes of $d+1$D 2-group
cocycles is denoted as $H^{d+1}(\cB(G_b;\Pi_2), \M)$.

\subsection{Cohomology of 2-group}

One way to understand the structure of $H^{d+1}(\cB(G_b;\Pi_2), \M)$ is to use
the fibration $\cB(\Pi_2,2) \to \cB(G_b; \Pi_2) \to \cB G_b$
(see \eqn{eq:2gp}), and use spectral sequence to reduce the cohomology of
$\cB(G_b; \Pi_2)$ to cohomology groups of $G_b$ and $\cB(\Pi_2,2)$.  In particular,
from Appendix \ref{LHS}, we see that every element in $H^{d+1}(\cB(G_b;\Pi_2),
\R/\Z)$ can be labeled by $(k_0,k_1,\cdots,k_d)$ where $ k_l\in
H^l[\cB G_b,H^{d+1-l}(\cB(\Pi_2,2);\R/\Z)_{G}]$, although some
$(k_0,k_1,\cdots,k_d)$'s may not correspond to any elements in
$H^{d+1}(\cB(G_b;\Pi_2), \R/\Z)$, and some different
$(k_0,k_1,\cdots,k_d)$'s may correspond to the same element in
$H^{d+1}(\cB(G_b;\Pi_2), \R/\Z)$.  (When $\cB(G_b; \Pi_2) = \cB(\Pi_2,2)
\times \cB G_b$, $(k_0,k_1,\cdots,k_d)$ will be the one-to-one label of all the
elements in $H^{d+1}(\cB(G_b;\Pi_2), \R/\Z)$.)

Next, let us concentrate on a special case of $\Pi_2=\Z_2$, and try to compute
$H^{d+1}(\cB(G_b;\Z_2), \R/\Z)$.  Since $\Z_2$ group has no non-trivial
automorphism, $\al_2$ is always trivial.  But $\bar n_3 \in H^3(\cB G_b; \Z_2)$
is in general non-trivial.  Thus, a 2-group $\cB(G_b;\Z_2)$ is characterized by
a pair $G, \bar n_3$.  The cohomology $H^*(\cB(\Z_2,2),\Z)$ is given by
\cite{Clement02}
\begin{align}
\bmm
d:                   & 0  & 1 & 2 & 3    & 4 & 5    & 6    & 7    \\
H^d(\cB(\Z_2,2),\Z): & \Z & 0 & 0 & \Z_2 & 0 & \Z_4 & \Z_2 & \Z_2 \\
\emm
\end{align}
Using the universal coefficient theorem
\begin{align}
\label{ucf}
 H^n(X,\M) &\simeq 
H^n(X;\Z) \otimes_{\Z} \M \oplus \Tor(H^{n+1}(X;\Z),\M)  .
\end{align}
and $ \Z_n \otimes_\Z \R/\Z
=0$,  $\Tor(\Z_n,\R/\Z)=\Z_n$, we find that
$H^{n}(\cB(\Z_2,2),\R/\Z)=H^{n+1}(\cB(\Z_2,2),\Z)$:
\begin{align}
\label{HKZ22}
\bmm
d:                      & 0     & 1 & 2 & 3    & 4    & 5    & 6    \\
H^d(\cB(\Z_2,2),\R/\Z): & \R/\Z & 0 & \Z_2 & 0 & \Z_4 & \Z_2 & \Z_2 \\
\emm
\end{align}
Using the above result, we find that $H^{4}(\cB(G_b;\Z_2), \R/\Z)$ can be
labeled by
\begin{align}
\label{om4mmm}
H^4(\cB(\Z_2,2);\R/\Z) &= \Z_4=\{\bar k_0\},
\nonumber\\
H^1[\cB G_b;H^{3}(\cB(\Z_2,2);\R/\Z)] &= \{0\},
\nonumber\\
H^2[\cB G_b;H^{2}(\cB(\Z_2,2);\R/\Z)] &= H^2(\cB G_b;\Z_2) =\{k_2\},
\nonumber\\
H^3[\cB G_b;H^{1}(\cB(\Z_2,2);\R/\Z)] &= \{0\},
\nonumber\\
H^4[\cB G_b;\R/\Z)] &= \{k_4\} .
\end{align}

Since 2-gauge theories in 3+1D are classified  pairs $(\bar n_3, \bar \om_4)$, $\om_4\in
H^{4}(\cB(G_b;\Z_2), \R/\Z)\}$,  we find that each 3+1D 2-gauge theory 
corresponds to
one or more elements in a subset of
\begin{align}
& H^3[\cB G_b;\Z_2)] \times H^4(\cB(\Z_2,2);\R/\Z) \times 
\nonumber\\
& H^2[\cB G_b;\Z_2)] \times H^4[\cB G_b;\R/\Z)]
\end{align}
The first $H$ comes from $\bar n_3$ and the rest $H$'s from $\bar \om_4$.  

If the index $\bar k_0 \in H^4(\cB(\Z_2,2);\R/\Z)=\Z_4$ is $\bar
k_0=2$, the 2-gauge theory has emergent fermions.  The index $k_2$
in $H^2(\cB G_b;\Z_2)$ describes the extension of $G_b$ by $\Z_2$ to obtain
$G_f$.  $\sRep(G_f)$ describes the particlelike excitations in the 2-gauge
theory.  For details, see Section \ref{modelGbZ2}.

\section{Pure 2-gauge theory of 2-gauge-group $\cB(\Pi_2,2)$ }

In the last section, we discuss some general properties of
2-gauge theory. In this section, we are going to discuss a
special 2-gauge theory, \emph{pure 2-gauge theory}.

\subsection{Pure 2-group and pure 2-gauge theory}

If we choose the target complex of the topological non-linear $\si$-model to be
$\cB(0;\Pi_2)=\cB(\Pi_2,2)$, we will get a pure 2-gauge theory of 2-gauge-group
$\cB(\Pi_2,2)$, where $\Pi_2$ is a finite abelian group.  There is only one
complex of $\cB(\Pi_2,2)$-type.  The complex $\cB(\Pi_2,2)$ has a structure
\begin{align}
\label{eq:nerveK2}
\xymatrix{ pt & 
pt \ar@<-1ex>[l]_{d_0, d_1}\ar[l] & 
\Pi_2 \ar@<-1ex>_{d_0, d_1 , d_2}[l] \ar@<1ex>[l] \ar[l] & 
\Pi_2^{\times 3} \ar@<-1ex>[l]_{d_0, ..., d_3} \ar@<1ex>[l]_{\cdot} & 
\Pi_2^{\times 6}  \ar@<-1ex>[l]_{d_0, ..., d_4} \ar@<1ex>[l]_{\cdot}& 
\Pi_2^{\times 10}  \dots \ar@<-1ex>[l]_{d_0, ..., d_5} \ar@<1ex>[l]_{\cdot} }
\end{align}

In this case $\bar n_3=0$, $\al_2$ is trivial, and $b_{ijk}$ satisfy
\begin{align}
 b_{123}- b_{023}+ b_{013}- b_{012} = 0 .
\end{align}
We see that canonical 2-cochain $\bar b$ is a $\Pi_2$-valued 2-cocycle on
target complex $\cB(\Pi_2,2)$.  The action of $\dd$ on the cochains in
$\cB(\Pi_2,2)$ are given by
\begin{align}
&\ \ \ \ (\dd\om_0)() = 0,
\end{align}
\begin{align}
&\ \ \ \ (\dd\om_1)(b_{012})
= \om_1() ,
\end{align}
\begin{align}
\label{cocycle2}
&\ \ \ \ (\dd\om_2)(b_{012}, b_{013}, b_{023})
 \nonumber\\ 
&= -\om_2(b_{012}) + \om_2(b_{013}) 
   - \om_2(b_{023}) + \om_2(b_{123}) ,
\end{align}
\begin{align}
&\ \ \ \ (\dd\om_3)(b_{012}, b_{013}, b_{014}, b_{023}, b_{024}, b_{034})
 \nonumber\\ 
&= \om_3(b_{012}, b_{013}, b_{023}) 
- \om_3(b_{012}, b_{014}, b_{024}) 
\nonumber\\
&\ \ \
+ \om_3(b_{013}, b_{014}, b_{034}) 
- \om_3(b_{023}, b_{024}, b_{034}) 
\nonumber\\
&\ \ \
+ \om_3(b_{123}, b_{124}, b_{134}) ,
\end{align}
\begin{widetext}
\begin{align}
 &\ \ \ \ (\dd\om_4)(b_{012}, b_{013}, b_{014}, b_{015}, b_{023}, b_{024}, b_{025}, b_{034}, 
b_{035}, b_{045})
 \nonumber\\ 
&= -\om_4(b_{013}, b_{014}, b_{023}, b_{024}, b_{034}) 
+ \om_4(b_{012}, b_{013}, b_{015}, b_{023}, b_{025}, b_{035}) 
- \om_4(b_{012}, b_{014}, b_{015}, b_{024}, b_{025}, b_{045}) 
\nonumber\\
&\ \ \ 
+ \om_4(b_{013}, b_{014}, b_{015}, b_{034}, b_{035}, b_{045}) 
- \om_4(b_{023}, b_{024}, b_{025}, b_{034}, b_{035}, b_{045}) 
+ \om_4(b_{123}, b_{124}, b_{125}, b_{134}, b_{135}, b_{145}),
\end{align}
\end{widetext}
In the above, the variables $b_{ijk}$ for $i\neq 0$ do not appear on the
left-hand-side of the equation, but appear on the right-hand-side of the
equation.  In fact, those $b_{ijk}$ are given by $b_{0mn}$'s that do appear on
the left-hand-side of the equation:
\begin{align}
 b_{ijk} &=  b_{0jk}-b_{0ik}+b_{0ij}.
\end{align}
So the above are the conditions on functions of $b_{0mn}$'s.

Clearly, 
\begin{align}
 H^0(\cB(\Pi_2,2),\R/\Z)=\R/\Z, \ \ \
 H^1(\cB(\Pi_2,2),\R/\Z)=0 .
\end{align}
From \eqn{cocycle2}, we see that, for $\Pi_2=\Z_n$, a 2-group 2-cocycle has a
form
\begin{align}
 (\om_2)_{ijk} = \frac{m}{n} b_{ijk} + c,\ \ \ m=0,\cdots,n-1,
\end{align}
The constant term $c$ is a coboundary. Thus
$H^2(\cB(\Z_n,2),\R/\Z)=\Z_n$. 
This allows us to show that for a finite $\Pi_2$
\begin{align}
 H^2(\cB(\Pi_2,2),\R/\Z)=\Pi_2 .
\end{align}
which agrees with
$H^2(\cB(\Z_2,2),\R/\Z)=\Z_2$ (see \eqn{HKZ22})

\def\arraystretch{2.2} 
\setlength\tabcolsep{2pt}
\begin{table*}[t]
\caption{Volume independent partition function $Z^\text{top}(\cM^4;\cB,\om_4)$
for the constructed local bosonic models, on closed 4-dimensional space-time
manifolds. The space-time $\cM^4$ considered have vanishing Euler number and
Pontryagin number $\chi(\cM^4)=P_1(\cM^4)=0$, which makes $Z^\text{top}(M^4)$
to be a topological invariant.\cite{KW1458}  
Here $L^3(p)$ is the 3-dimensional lens space and
$F^4=(S^1\times S^3)\# (S^1\times S^3)\#\C P^2\# \overline{\C P}^2$.  $F^4$ is
not spin. 
} 
\label{tab:topinv}
 \centering
 \begin{tabular}{ |c|c|c|c|c|c| }
 \hline
Models $\backslash\ M^4$:  & $T^4 $ & $T^2\times S^2$ & $S^1\times L^3(p) $  & $F^4$ & Low energy effective theory \\
\hline
\parbox{1.9in}{$Z^\text{top}(\cM^4;\cB(\Z_n,2),\frac{m}{2n} \Sq^2 b)$  \eq{Zb4m}
\\ $n=$ even, $m=0,\cdots,2n-1$
}  & $\<m,n\>^3$ & $\<m,n\>$ & $\<m,n,p\>$  & 
\parbox{1.5in}{$\<m,n\>$ if $\frac{mn}{\<m,n\>^2}=$ even\\
0 if $\frac{mn}{\<m,n\>^2}=$ odd}
& \parbox{1.8in}{$Z_{\<m,n\>}$ gauge theory with fermions iff $\frac{mn}{\<m,n\>^2}=$ odd}\\
\hline
\parbox{1.8in}{$Z^\text{top}(\cM^4;\cB(\Z_n,2),\frac{k}{n} \Sq^2 b)$ \eq{ZbZnodd}
\\ $n=$ odd, $k=0,\cdots,n-1$
}  & $\<2k,n\>^3$ & $\<2m,n\>$ & $\<2k,n,p\>$  & 
$\<2k,n\>$
& Untwisted $Z_{\<2k,n\>}$ gauge theory \\
\hline
$Z^\text{top}(\cM^4;\cB\Z_n,0)$  & $n^3$ & $n$ & $\<n,p\>$ & $n$ & \parbox{1.8in}{Untwisted $Z_n$ gauge theory}\\
\hline
 \end{tabular}
\end{table*}
\def\arraystretch{1} 

To compute $H^4(\cB(\Pi_2,2),\R/\Z)$, let us first assume $\Pi_2=\Z_2$.  From
\eqn{HKZ22}, we see that $H^4(\cB(\Z_2,2),\R/\Z)=\Z_4$.  One of the
4-dimensional 2-group cocycle is given by
\begin{align}
 \om_4(b) = \frac12 b^2.
\end{align}
We note that $2 \om_4 \se{1} 0$.  Thus $\om_4$ only generate $\Z_2$ subgroup of
$\Z_4=H^4(\cB(\Z_2,2),\R/\Z)$.

To obtain the generator of $H^4(\cB(\Z_2,2),\R/\Z)$, we note that, if we view
$b$ as $\Z$-valued 2-cochain, we have $\dd b =2 c$ where $c$ is a $\Z$-valued
3-cochain. Then, from \eqn{Sqdef} and \eqn{Sqd1}, we see that
\begin{align}
\dd \Sq^2 b
=  \Sq^2 \dd b + 2 \Sq^3 b
= 4 (c \hcup{1} c + bc).
\end{align}
Thus
\begin{align}
\om_4(b)= 
\frac14 \Sq^2 b 
\end{align}
is a $\R/\Z$-valued 4-cocycle: $ \dd \om_4(b) \se{1} 0 $.  Such a $\om_4$
generates the full group $\Z_4=H^4(\cB(\Z_2,2),\R/\Z)$.

In general, if $b$ is a $\Z_n$-valued 2-cocycle, we have $\dd b =n c$ where $c$
is a $\Z$-valued 3-cochain.  From \eqn{cupkrel1}, we see that
\begin{align}
\dd \Sq^2 b
= \Sq^2 \dd b + 2 \Sq^3 b
= n^2 c \hcup{1} c + 2n bc.
\end{align}
This result tells us that when $n=$ odd,
\begin{align}
\om_4(b)= 
\frac1n \Sq^2 b
\end{align}
is a $\R/\Z$-valued 4-cocycle, while
when $n=$ even
\begin{align}
\om_4(b)= 
\frac1 {2n} \Sq^2 b
\end{align}
is a $\R/\Z$-valued 4-cocycle.  $\om_4(b)$ generates a $\Z_n$ group
when $n=$ odd, and a $\Z_{2n}$ group when $n=$ even.  This suggests that
$H^4(\cB(\Z_n,2),\R/\Z)=\Z_n$  when $n=$ odd, and
$H^4(\cB(\Z_n,2),\R/\Z)=\Z_{2n}$  when $n=$ even.

\subsection{Pure 2-gauge theory in 3+1D}

\subsubsection{$n=$ odd case}

We see that, when $n=$ odd, we have $n$ different 3+1D $\cB(\Z_n,2)$ 2-gauge
theories, described by partition function
\begin{align}
\label{ZbZnodd}
 Z(\cM^4; \cB(\Z_n,2),k) = 
\sum_{\dd b \se{n} 0} \ee^{2\pi \ii \int_{\cM^4} \frac{k}{n} b^2 }
\end{align} 
where $k=0,1,\cdots,n-1$.  Clearly, the action amplitude $\ee^{2\pi \ii
\int_{\cM^4} \frac{k}{n} b^2 }$ is invariant under the 2-gauge transformation
$b \to b+\dd \la$.  The above 2-gauge theory was studied in \Ref{W161201418}.
It was found that the theory realizes a 3+1D $Z_{\<2k,n\>}$-gauge theory.  It
is an untwist $Z_{\<2k,n\>}$-gauge theory since $2kn/\<2k,n\>^2$ is always
even.

\subsubsection{$n=$ even case}

When $n=$ even, we have $2n$ different
3+1D $\cB(Z_n,2)$ 2-gauge theories,
described by partition function
\begin{align}
\label{Zb4m}
 Z(\cM^4; \cB(\Z_n,2),m) = 
\sum_{\dd b \se{n} 0} \ee^{2\pi \ii \int_{\cM^4} \frac{m}{2n} 
\Sq^2 b }
\end{align} 
where $m=0,1,\cdots,2n-1$.  Noticing that the $\Z_n$-valued 2-cocycle $b$
satisfies $\dd b = n c$.  Under the 2-gauge transformation $b \to b+\dd \la$
generated by  $\Z_n$-valued 1-cochain $\la$, we see that,
from \eqn{Sqgauge} and using $\dd b = n c$
\begin{align}
\Sq^2 (b +\dd \la) - \Sq^2 b  \se{2n,\dd} 0 .
\end{align} 
This implies the 2-gauge invariance of the action amplitude $\ee^{2\pi \ii
\int_{\cM^4} \frac{k}{2n} \Sq^2 b}$ for the $n=$ even case.

\subsubsection{Properties and duality relations}

The pure 2-gauge theories \eq{ZbZnodd} and \eq{Zb4m} were studied for
$n=$ odd cases and for $n=$ even and $m= 2k$ cases in \Ref{W161201418}.  In
those cases, it was found that the theory realizes a 3+1D $Z_{\<2k,n\>}$-gauge
theory.  The $Z_{\<2k,n\>}$-gauge theory has emergent fermions if
$2kn/\<2k,n\>^2 = $ odd, and it is  a untwist $Z_{\<2k,n\>}$-gauge theory if
$2kn/\<2k,n\>^2 = $ even.  
To understand the properties of the model \eq{Zb4m} for $n=$ even and $m=$
odd cases, we compute the partition function \eq{Zb4m} in Appendix
\ref{Z2gauge}.  The result is summarized in Table \ref{tab:topinv}.  We see
that, for $n=$ even, the 3+1D pure 2-gauge theory is equivalent to
$Z_{\<m,n\>}$-gauge theory.  The theory has emergent fermion iff $mn/\<m,n\>^2
= $ odd.

The higher gauge theories are labeled by a pair $(K,\om_{d+1})$: a target space
$K$ and a cocycle $\om_{d+1}$ on it.  Some times two different  higher gauge
theories may realize the same topologically ordered phase.  In this case, we
say that the two theories are equivalent or dual to each other.  The results in
Table \ref{tab:topinv} suggest the following duality relations, where we use
$[\cB(\Pi_1.\Pi_2,\cdots), \bar \om_{d+1}]$ to label different higher gauge
theories:\\ 
(1) for $n=$
even and $\frac{mn}{\<m,n\>^2}=$ even
\begin{align}
 [\cB(\Z_n,2), \frac{m}{2n} \Sq^2 b ]
\ \sim \ [\cB(\Z_{\<m,n\>}), 0] .
\end{align}
(2) for $n=$ odd 
\begin{align}
 [\cB(\Z_n,2), \frac{k}{n} \Sq^2 b ]
\ \sim \ [\cB(\Z_{\<2k,n\>}), 0] .
\end{align}
We note that $[\cB(\Z_{n}), 0]$ is an untwisted $Z_n$-gauge theory.

\section{3+1D 2-gauge theory of 2-gauge-group $\cB(G_b,Z^f_2)$}
\label{modelGbZ2}

In this section, we are going to consider more general 3+1D 2-gauge
theories which have 2-gauge-group $\cB(G_b,Z^f_2)$. 

\subsection{The Lagrangian and space-time path integral}

Since $\Z^f_2$ has no non-trivial
automorphism, so $\al_2$ is trivial.  As a result, such 2-gauge theories are
classified by
\begin{align}
 G_b; \bar n_3; \bar\om_4
\end{align}
where $\bar n_3 \in H^3(\cB G_b;\Z^f_2)$ and $\bar\om_4 \in H^4(\cB(G_b;\Z^f_2),\R/\Z)$.

To write down the  Lagrangian and space-time path integral for the 2-gauge
theories, the key is to find $\bar\om_4$. To do so, we note that the links  in
$\cB(G_b;\Z^f_2)$ are labeled by $(a)$, $a \in G_b$.  The triangles in
$\cB(G_b;\Z^f_2)$ are labeled by $(a_{ij},a_{jk},a_{ik},b_{ijk})$ that satisfy
\eqn{ggg} and \eqn{bn3}.  We see that on each link of $\cB(G_b;\Z^f_2)$, we
have a label $a_{ij}$, and on each triangle we have a label $b_{ijk}$.  We may
view $a_{ij}$ as the canonical $G_b$-valued 1-cocycle $\bar a$ (due to
\eqn{ggg}), and $b_{ijk}$ as the canonical $\Z^f_2$-valued 2-cochain $\bar b$
on $\cB(G_b;\Z^f_2)$.  The canonical 1-cocycle and the $2$-cochain are related
\begin{align}
 \dd \bar b = \bar n_3 (\bar a).
\end{align}
We may use the 1-cocycle $\bar a$
and the 2-cochain $\bar b$ to write down $\bar \om_4$.

We note that each $\bar\om_4 \in  H^4(\cB(G_b;\Z^f_2),\R/\Z)$ corresponds (see
\eqn{om4mmm}) to one or more elements in a subset of
\begin{align}
\label{HHH2gauge}
H^4(\cB(\Z^f_2,2);\R/\Z) \times
H^2(\cB G_b;\Z^f_2) \times
H^4(\cB G_b;\R/\Z) .
\end{align}
To construct a $\bar\om_4$, we may guess $\bar\om_4 =\frac{\bar k_0} 4 \Sq^2
\bar b$. Using \eqn{Sqd1}, we find that
\begin{align}
\dd  \Sq^2 \bar b &= \Sq^2 \bar n_3(\bar a) + 2\Sq^3 \bar b
\nonumber\\
&= \Sq^2 \bar n_3(\bar a) + 2\bar b \bar n_3(\bar a) 
.
\end{align}
So $\bar \om_4 =\frac{\bar k_0} 4 \Sq^2 \bar b$ is not a cocycle.  But the error is only
a function of 1-cocycle $\bar a$ if $\bar k_0=2$.  In this case, we can fix the
error by adding a function of $\bar a$, $\bar \nu_4(\bar a)$.  Similarly, we can try $\bar \om_4
=\frac12 \bar b \bar e_2(\bar a)$, where $\bar e_2(\bar a) \in Z^2(\cB G_b;\Z^f_2)$.  But $\dd [\bar b
\bar e_2(\bar a)] = \bar n_3(\bar a) \bar e_2(\bar a)$.  Again $\bar \om_4 =\frac12  \bar b \bar e_2(\bar a)$ is not a cocycle.
Again we can fix it by adding a function $\bar \nu_4(\bar a)$. Thus, we come up with the
following general expression of $\bar\om_4$:
\begin{align}
\label{om4exp}
\bar \om_4 (\bar a,\bar b) &=
 \frac{k_0}{2} \Sq^2 \bar b + \frac{1}{2} \bar b \bar e_2(\bar a)
+ \bar \nu_4(\bar a),
\end{align}
where $ \bar \nu_4(\bar a)$ is a $\R/\Z$-valued
cochain in $C^4(\cB G_b,\R/\Z)$ that satisfy
\begin{align}
 - \dd\bar  \nu_4(\bar a) = \frac{k_0}{2} \Sq^2 \bar n_3(\bar a)  
+\frac{1}{2} \bar n_3(\bar a) \bar e_2(\bar a).
\end{align}
In this case, $\bar\om_4(\bar a,\bar b)$ will be a cocycle $\dd \bar \om_4 \se{1} 0$.  The three
terms in \eqn{om4exp} correspond to the three cohomology classes in
\eqn{HHH2gauge}. Thus our construction of $\bar \om_4$ is complete (for $\bar n_3\neq
0$).

Using the expression \eq{om4exp} for $\bar \om_4$, we can construct 
a topological non-linear $\si$-model (\ie a 2-gauge theory):
\begin{align}
\label{Z2gaugeZ}
&\ \ \ 
 Z(\cM^4; \cB(G_b;\Z^f_2),\bar\om_4) 
\\
& = \sum_{\phi} 
\Big( \prod_i |G_b|^{-1} \prod_{(ij)}2^{-1} \Big)
\ee^{2\pi \ii \int_{\cM^{4}} \phi^*\bar\om_4 } 
\nonumber\\
&= |G_b|^{-N_0} 2^{-N_1} \sum_{\del a=\one,\dd b=n_3} \ee^{2\pi \ii \int_{\cM^{4}}
\nu_4(a) + \frac{k_0}{2} \Sq^2 b + \frac{1}{2} b e_2(a) }
,
\nonumber 
\end{align}
where $ \sum_{\del a=\one,\dd b=n_3}$ sum over the $G_b$-valued 1-cochains
$a_{ij}$ and the $\Z^f_2$-valued 2-cochains $b_{ijk}$ on the space-time complex
$\cM^4$, that satisfy 
\begin{align}
(\del a)_{ijk}\equiv a_{ij}a_{jk}a_{ik}^{-1} =\one,\ \ \ \ \ \dd b = n_3(a).  
\end{align}

In the above $k_0=0,1$ labels the elements of the $\Z_2$ subgroup of
$H^4(\cB(\Z^f_2,2);\R/\Z)=\Z_4$, $\bar e_2(a)$ labels the elements in $H^2(\cB
G_b;\Z^f_2)$, and different $\bar \nu_4(a)$ differ by the elements in $H^4(\cB
G_b;\R/\Z)$.  Plus $\bar n_3 \in H^3(\cB G_b;\Z^f_2)$, the four pieces of data,
$(k_0, \bar e_2, \bar n_3, \bar \nu_4)$, classify 2-gauge theories of 2-gauge-group
$\cB(G_b;\Z^f_2)$. 

\subsection{The equivalence between  $[k_0,\bar e_2(\bar a), \bar n_3(\bar a), \bar \nu_4(\bar a)]$'s }

The Lagrangian of the 2-gauge theory \eq{Z2gaugeZ} is labeled by the data
$[k_0, \bar e_2(\bar a), \bar n_3(\bar a), \bar \nu_4(\bar a)]$:
\begin{align}
 \bar e_2(a_{01},a_{12}) & \in Z^2(\cB G_b;\Z_2) ,
\nonumber\\
 \bar n_3(a_{01},a_{12},a_{23}) & \in Z^3(\cB G_b;\Z_2) ,
\nonumber\\
 \bar \nu_4(a_{01},a_{12},a_{23},a_{34})  & \in C^{d+1}(\cB G_b;\RZ) ,
\end{align}
that satisfy
\begin{align}
 \dd \bar \nu_4(\bar a) & \se{1} \frac12 [ \Sq^2 \bar n_3(\bar a) +\bar n_3(\bar a) \bar e_2(\bar
 a) ] .\label{consis}
\end{align}
As local bosonic systems, the different
2-gauge theories labeled by different data may realize the same bosonic
topological phase.  We say that those 2-gauge theories or those data are
equivalent.

Note that the Lagrangian is a
2-group cocycle, and two Lagrangians differing by a
2-group coboundary should be equivalent.
This kind of equivalent relation is generated by 
the following three kinds of transformations:\\
(1) a transformation generated by a $1$-cochain $\bar l_1 \in C^1(\cB G_b;\Z_2)$
\begin{align}
  \bar e_2 &\to \bar e_2 + \dd \bar l_1 ,
\\
  \bar n_3&\to \bar n_3 ,
\nonumber\\
  \bar \nu_4&\to \bar \nu_4 + \frac{1}{2} \bar n_3\bar l_1 .
\nonumber 
\end{align}
(2) a transformation generated by
 a $2$-cochain $\bar u_2 \in C^2(\cB G_b;\Z_2)$ 
\begin{align}
  \bar e_2 &\to \bar e_2  ,
\\
  \bar n_3&\to \bar n_3+\dd \bar u_{2},
\nonumber\\
  \bar \nu_4&\to \bar \nu_4+
  \frac{k_0}{2}\left(
   \dd \bar u_2 \hcup{2} \bar n_3+ \Sq^2 \bar u_2 + \bar u_2\bar e_2
\right),
\nonumber 
\end{align}
(3) a transformation generated by a $3$-cochain $\bar \eta_3\in C^3(\cB G_b;\RZ)$:
\begin{align}
  \bar e_2 &\to \bar e_2  ,
\\
  \bar n_3&\to \bar n_3,
\nonumber\\
  \bar \nu_{4}&\to \bar \nu_{4}+\dd \bar \eta_3  .
\nonumber 
\end{align}
Under those transformations, the Lagrangian $ \nu_4(a) + \frac{k_0}{2} \Sq^2
b + \frac{1}{2} b e_2(a) $ only changes by a coboundary. Those
transformations do not change the topological partition function and do not
change the topological order in the ground state. 

We like to point out that the different transformations of the second type
do not commute. Those transformation may generate changes $(\bar e_2,\bar
n_3,\bar \nu_4) \to (\bar e_2,\bar n_3,\bar \nu_4 +\Del \bar \om_4)$ where
$\Del \bar \om_4$ is a cocycle in $Z^4(\cB G_b;\RZ)$.

We also want to mention that the above transformations can not
generate all possible equivalent relations.
In particular, an isomorphism of the target space
$\cB(G_b,Z_2^f)\to\cB(G_b,Z_2^f)$ (2-group isomorphism) may relate two
Lagrangians whose difference is not a 2-group coboundary. We are not sure if
there are more general ``duality'' equivalent relations between 2-gauge
theories. This will be left for future work.

\subsection{2-gauge transformations in the cocycle $\si$-model}

As a local bosonic model, the discrete non-linear $\si$-model \eq{Z2gaugeZ}
do not have to have any symmetry.  However, in \eqn{Z2gaugeZ} we choose a very
special Lagrangian, the pullback of a cocycle on the target space.  For such a
special Lagrangian, the model is exactly soluble.  Such a special Lagrangian
has a large set of accidental symmetries: invariant under 2-gauge
transformations.  We may also say that the model has accidental higher
symmetries \cite{GW14125148}.  We note that breaking all those symmetries in the
Lagrangian by a small but arbitrary perturbation will not change the
topological order in the ground state.  Thus we may say that all those
arbitrary perturbations are irrelevant, and the cocycle $\si$-model is the
fixed point theory of the given topological order.  In other words,
topological order has emergent  higher symmetries.

In this section, we are going to discuss those accidental emergent higher
symmetries in the cocycle $\si$-model \eq{Z2gaugeZ}.  In this case, the
emergent higher symmetries is called 2-gauge symmetries.

The first type of 2-gauge transformation is given by 1-cochain $\la_1  \in
C^1(M^{d+1};\Z_2)$:
\begin{align}
 b &\to b + \dd \la_1, \ \ \ \ a \to a;
\end{align}
We find that, using \eqn{Sqplus} and \eqn{Sqd}
\begin{align}
&\ \ \ \
  k_0 \Sq^2 (b+\dd \la_1) +  (b+\dd \la_1) e_2(a)
-  k_0 \Sq^2 b -  b e_2(a)
\nonumber\\
&
\se{2,\dd} k_0 \Sq^2 \dd \la_1 
\se{2,\dd} 0 .
\end{align}
Therefore, the Lagrangian changes by only a total derivative term under the
first type of 2-gauge transformation.

The second type of 2-gauge transformation is given by 0-cochain $g_i  \in
C^0(M^{d+1};G_b)$:
\begin{align}
 b &\to b + \zeta_2(a,g), \ \ \ \ a_{ij} \to a^g = g_i a_{ij} g_j^{-1}.
\end{align}
Under the above transformation
\begin{align}
 n_3(a) &\to n_3(a^g) \se{2} n_3(a) +\dd  \zeta_2(a,g).
\nonumber\\
 e_2(a) &\to e_2(a^g) \se{2} e_2(a) +\dd  \xi_1(a,g).
\end{align}
which defines $ \zeta_2(a,g)$.
Thus the condition $\dd b \se{2} n_3(a)$ is maintained under the 2-gauge
transformation.  We find that, using \eqn{Sqplus1} and \eqn{Sqd}
\begin{align}
&\ \ \ \
 k_0 \Sq^2 (b+\zeta_2) + (b+\zeta_2) (e_2 + \dd  \xi_1) - k_0 \Sq^2 b - b e_2
\nonumber\\
&\se{2,\dd} k_0 \dd b \hcup{2} \dd \zeta_2
+ b \dd  \xi_1 + \zeta_2 e_2 + \zeta_2  \dd  \xi_1 .
\nonumber\\
&\se{2,\dd} k_0 n_3 \hcup{2} \dd \zeta_2
+ n_3  \xi_1 + \zeta_2 e_2 + \zeta_2  \dd  \xi_1 .
\end{align}
We note that the above only depends on $a$ and $g$.  Thus, if $\nu(a)$
satisfies
\begin{align}
 \nu(a^g) - \nu(a)
\se{2,\dd}  k_0 n_3 \hcup{2} \dd \zeta_2
+ n_3  \xi_1 + \zeta_2 e_2 + \zeta_2  \dd  \xi_1,
\end{align}
 the Lagrangian changes by only a total derivative term under the
second type of 2-gauge transformation.

\subsection{The vanishing of the partition function}

We have seen that if we change $b$ by a coboundary, the action amplitude
$\ee^{2\pi \ii \int_{\cM^{4}} \nu_4(a) + \frac{k_0}{2} \Sq^2 b + \frac{1}{2}
b e_2(a) }$ does not change.  However, if we change $b$ by a cocycle $b_0$,
the action amplitude will change.
Using \eqn{Sqplus} and \eqn{Sqd}, we find that
\begin{align}
&\ \ \ \
  k_0 \Sq^2 (b+b_0) +  (b+b_0) e_2
-  k_0 \Sq^2 b -  b e_2
\nonumber\\
&
\se{2,\dd} k_0 \Sq^2 b_0  +b_0 e_2
\se{2,\dd} [k_0 (\w_2+\w_1^2)  + e_2] b_0 .
\end{align}
Thus the action amplitude depends on $b_0$ via $ \ee^{\pi \ii
\int_{\cM^{4}}[k_0 (\w_2+\w_1^2)  + e_2] b_0 } $.
When we integral over $b$ (\ie $b_0$) in the path integral,
such a term will cause to partition function to vanish
if 
\begin{align}
 k_0 (\w_2+\w_1^2)  + e_2 \neq \text{ $\Z_2$-valued coboundary}. 
\end{align}
This allows us to conclude that the local bosonic system
has emergent pointlike excitations that are described by representations of
$G_f = Z_2\gext_{e_2} G_b$.  If $k_0=1$, the local bosonic system has emergent
fermions.\cite{W161201418}

\subsection{The pointlike and stringlike excitations in the 2-gauge theory}

There are two types of pointlike excitations in the 2-gauge theory.  Let $S^1$
be the world line of a pointlike excitation  of the first type.  The presence
of the pointlike excitation modifies the path integral via a Wilson loop:
\begin{align}
&\ \ \ 
 Z(\cM^4; \cB(G_b;\Z^f_2))
\\
&= 
\hskip -1em \sum_{\del a=\one,\dd b=n_3} \hskip -1em
 [\Tr \prod_{S^1}R_{G_b}(a_{ij})]
\ee^{2\pi \ii \int_{\cM^{4}} \nu_4(a)
+ \frac{k_0}{2} \Sq^2 b  + \frac{1}{2} b e_2(a) }
,
\nonumber 
\end{align}
where $R_{G_b}(a),\ a\in G_b$, is a representation of $G_b$ and $\prod_{S^1}R_{G_b}(a_{ij})$
is a product $R_{G_b}(a_{ij})$ along the loop $S^1$.

To describe the second type of pointlike excitations, let $f_3$ be the
Poincar\'e dual of the worldline of the pointlike excitations.
Then the  second type of pointlike excitations
are created by modifying the
condition $\dd b =n_3(a)$ to
\begin{align}
 \dd b =n_3(a) + f_3.
\end{align}
Now the path integral with the  second type of pointlike excitations becomes
\begin{align}
&\ \ \ 
 Z(\cM^4; \cB(G_b;\Z^f_2))
\\
&= 
\hskip -1em \sum_{\del a=\one,\dd b=n_3+f_3} \hskip -1em
\ee^{2\pi \ii \int_{\cM^{4}} \nu_4(a)
+ \frac{k_0}{2} \Sq^2 b + \frac{1}{2} b e_2(a) }
,
\nonumber 
\end{align}
To understand the property of the second type of excitations, let us assume the
worldline $S^1$ to be the boundary of a disk $D^2$.  Let a $\Z_2$-valued
2-cochain $s_2$ to be the Poincar\'e dual of $D^2$.  Then we have $f_3=\dd
s_2$.  The above path integral can be rewritten as
\begin{align}
&\ \ \ 
 Z(\cM^4; \cB(G_b;\Z^f_2))
\\
&= 
\hskip -1em \sum_{\del a=\one,\dd b=n_3+\dd s_2} \hskip -1em
\ee^{2\pi \ii \int_{\cM^{4}} \nu_4(a)
+ \frac{k_0}{2} \Sq^2 b + \frac{1}{2} b e_2(a) }
\nonumber \\
&= 
\hskip -1em \sum_{\del a=\one,\dd b=n_3} \hskip -1em
\ee^{2\pi \ii \int_{\cM^{4}} \nu_4(a)
+ \frac{k_0}{2} \Sq^2 (b+ s_2) + \frac{1}{2} (b+ s_2) e_2(a) }
\nonumber \\
&= 
\ee^{k_0 \pi \ii \int_{\cM^{4}}  \Sq^2 s_2 }
\nonumber\\
&
\sum_{\del a=\one,\dd b=n_3} \hskip -1em
\ee^{2\pi \ii \int_{\cM^{4}} \nu_4(a)
+ \frac{k_0}{2} \Sq^2 b + \frac{1}{2} b e_2  }
\ee^{\pi \ii \int_{\cM^{4}} 
k_0 f_3 \hcup{2} n_3  + s_2 e_2  }
,
\nonumber 
\end{align}
where we have used \eqn{Sqplus1}.  We note that the term $ \ee^{\pi \ii
\int_{\cM^{4}} s_2 e_2(a) } $ is the only one on the disk $D^2$ that depends
on the 1-cocycle field $a$. This term can be rewritten as
\begin{align}
 \ee^{\pi \ii \int_{\cM^{4}} s_2 e_2(a) }
=
 \ee^{\pi \ii \int_{D^2} e_2(a) } .
\end{align}
After combining with the first type of particle, the above becomes
\begin{align}
  [\Tr \prod_{S^1}R_{G_b}(a_{ij})] \ee^{\pi \ii \int_{D^2} e_2(a) } .
\end{align}
The term $\ee^{\pi \ii \int_{D^2} e_2(a) }$
introduces $\pm 1$ phase to $R_{G_b}(a_{ij})$ and promotes it
into a representations of $G_f = Z_2 \gext_{e_2} G_b$.
This is why the pointlike excitations are described
by $G_f$ representations.

To summarize, the pointlike excitations are described by $\Rep(G_f)$ when
$k_0=0$ and by $\sRep(G_f)$  when $k_0=1$.  Here $\Rep(G_f)$ is the symmetric
fusion category formed by the representations of $G_f$ where all the
representations are bosons.  $\sRep(G_f)$ is the symmetric fusion category
formed by the representations of $G_f$ where all the representations that
represent the extended $Z_2$ trivially are bosons and the others are fermions.  The
representations that represent the extended $Z_2$ trivially correspond to the
first type of pointlike excitations, which are always bosons regardless the
value of $k_0$.  The representations that represent the extended $Z_2$
non-trivially correspond to the second type of pointlike excitations.  The
second type of pointlike excitations are fermions when $k_0=1$, and bosons
when $k_0=0$.  

Similarly, stringlike excitations are described by worldsheet $W^2$ in
space-time.  The first type of stringlike excitations are created by modifying
the flat condition $a_{ij}a_{jk}a_{ik}^{-1}=\one$ to
\begin{align}
 a_{ij}a_{jk}a_{ik}^{-1} = (\del a)_{ijk} = g
\end{align}
on the triangles that intersect the  worldsheet.  We see that the stringlike
excitations of the first type are labeled by  the group elements.  However, we
can perform a gauge transformation in the region that cover the worldsheet
$W^2$:
$ a_{ij} \to h a_{ij}h^{-1} $.
This changes
\begin{align}
 a_{ij}a_{jk}a_{ik}^{-1} = g \to 
 h a_{ij}a_{jk}a_{ik}^{-1} h^{-1} = hg h^{-1},
\end{align}
\ie changes the string labeled by $g$ to the string labeled by $hgh^{-1}$.
Thus strings labeled by different group elements in the same conjugacy class
are equivalent.  Therefore, stringlike excitations of the first type are
labeled by by the conjugacy classes of $G_b$,  just like a 3+1D gauge theory of
gauge group $G_b$.

The presence of the second type of stringlike excitation modifies the path
integral:
\begin{align}
&\ \ \ 
 Z(\cM^4; \cB(G_b;\Z^f_2),\om_4)
\\
&=
\hskip -1em \sum_{\del a=\one,\dd b=n_3} \hskip -1em
 \ee^{2\pi \ii \int_{W^2}  \frac12 b} \ 
\ee^{2\pi \ii \int_{\cM^{4}} \nu_4(a)
+ \frac{k_0}{2} \Sq^2 b + \frac{1}{2} b e_2(a) }
\nonumber\\
&=
\hskip -1em \sum_{\del a=\one,\dd b=n_3} \hskip -1em
\ee^{2\pi \ii \int_{\cM^{4}} \nu_4(a)
+ \frac{k_0}{2} \Sq^2 b + \frac{1}{2} b e_2(a) + \frac12 b m_2}
,
\nonumber 
\end{align}
where $\Z^f_2$-valued 2-cocycle $m_2$ is the Poincar\'e dual of the worldsheet
$W^2$.

\section{Classify and realize 3+1D EF1 topological orders by
2-gauge theories of 2-gauge-group $\cB(G_b,Z^f_2)$}
\label{clssII}

It was shown that 3+1D AB and EF topological orders 
with emergent bosons and/or fermions
have a unique canonical
boundary.\cite{LW170404221,LW180108530} On the canonical boundary, the boundary
stringlike excitations are labeled by the elements in a finite group.  All
those  boundary string excitations have a unit quantum dimension. For EF1
topological orders with emergent fermions, the canonical boundary also has an
emergent fermionic pointlike excitation with quantum dimension
1.\cite{LW180108530} Those boundary excitations are described by a pointed
unitary fusion 2-category.  Such a pointed unitary fusion 2-category is
classified by a 2-group $\cB(G_b,Z_2^f)$ and a $\R/\Z$-valued 4-cocycle $\om_4$
on the 2-group.  Here $G_b$ is the group that labels the types of boundary
string excitations.  Therefore, \emph{all EF1 3+1D topological orders are
classified by a pair $\cB(G_b,Z_2^f),\bar \om_4$ -- a 2-group and a $\R/\Z$-valued
4-cocycle  on the 2-group}.

To see why pointed fusion 2-categories are classified by the pairs
$(\cB(G_b,Z_2^f), \bar \om_4)$,  we note that the pointed fusion 2-category has
objects labeled by elements in $G_b$, 1-morphisms labeled by elements in $Z_2$
and 2-morphisms corresponding to physical operators. The 2-morphisms are not
all invertible, but for the structural morphisms we only need to consider the
invertible 2-morphisms, thus no generality is lost by restricting 2-morphisms
to $U(1)\simeq \R/\Z$. This way we obtain a 3-group $\cB(G_b,Z_2,\R/\Z)$,
which has
the same classification data as the pointed fusion 2-category. We explain now
in more detail. 

On one hand, by Lemma \ref{lem:fibration}, we have
\begin{align}
  \cB(\R/\Z, 3)\to \cB(G_b,Z_2,\R/\Z) \to \cB(G_b,Z_2),
\end{align}
and $\cB(G_b,Z_2,\R/\Z)$ is classified by the base 2-group
$\cB(G_b,Z_2)$ and an element $\bar \omega_4$ 
in $H^4(\cB(G_b,Z_2),\R/\Z)$. Then the 2-group $\cB(G_b,Z_2)$ is in
turn characterised by $G_b, Z_2, \bar n_3 \in H^3(\cB G_b;\Z_2)$. Thus 3-group
$\cB(G_b,Z_2,\R/\Z)$ is characterised by $(G_b,Z_2,\bar n_3\in
H^3(G_b,Z_2), \bar \omega_4\in H^4(\cB(G_b,Z_2),\R/\Z))$. 

On the other hand, recall the classification data of the pointed fusion
2-category that is listed in \cite{LW180108530}:
\begin{itemize}
  \item Objects $g\in G_b$, 1-morphisms $p_g\in Z_2 \subset \Hom(g,g)$.
\item Interchange law: 2-isomorphisms [$U(1)$ phase factors] $\tilde b(p'_g,q'_h,p_g,q_h)$ that determines the particle
    statistics.
  \item Associator: 1-morphism $n_3(g,h,j): (gh)j\to g(hj)$ in $H^3(\cB G_b;\Z_2)$ and
    2-isomorphisms
    $\tilde n_3(p_g,q_h,r_j)$.
\item Pentagonator: 2-isomorphisms $\nu_4(g,h,j,k)\in C^4(\cB G_b,\R/\Z)$.
\end{itemize}
We thus find an exact correspondence between the above and the
classification data on the higher group side
$(G_b,Z_2,\bar n_3\in H^3(\cB G_b;\Z_2), \bar \omega_4\in H^4(\cB(G_b,Z_2),\R/\Z))$
as below:
$G_b,Z_2,n_3$ are
exactly the same. The 2-group 4-cocycle $\bar \omega_4$ has 3 components $ k_0,\bar e_2,
\bar \nu_4$:
\begin{itemize}
  \item $ k_0$ corresponds to $\tilde b(p'_g,q'_h,p_g,q_h)$ on the 2-category side.
    It has 4 different choices, corresponding to boson, fermion, semion and
    anti-semion statistics respectively. For EF1 topological orders we stick to
    the choice of fermion statistics, which is indicated in our notation by using $Z_2^f$
    instead of $Z_2$.
  \item $\bar e_2$ determines the $Z_2^f$ extension from $G_b$ to $G_f$. Together
    with $ k_0$ it determines the associator 2-morphisms $\tilde n_3(p_g,q_h,r_j)$ on the
    2-category side.
  \item The last component $\bar \nu_4$ is just the pentagonator
    $\nu_4(g,h,j,k)$ on the 2-category side.
  \item Moreover, on both sides they satisfy the same consistent condition
    \eqref{consis}.
\end{itemize}
Since all 3+1D EF1 topological orders are classified by
$\cB(G_b,Z_2^f),\bar \om_4$, and since for each pair $\cB(G_b,Z_2^f),\bar \om_4$ we can
construct a 2-gauge theory to realize a EF1 topological order, we conclude that
exactly soluble 2-gauge theories of 2-gauge-group $\cB(G_b,Z_2^f)$ realize and
classify all 3+1D EF1 topological orders.

\section{Realize  3+1D EF2 topological orders by topological non-linear $\si$-models}

\subsection{Construction of topological non-linear $\si$-models}

In \Ref{KW1458}, it was conjectured that all topological orders with gappable
boundary can be realized by exactly soluble tensor network model defined on
space-time complex.\cite{TV9265,CY9362,GLS0918,WW180109938} In
\Ref{LW180108530}, it was shown that all EF topological orders have a unique
canonical boundary described by a unitary fusion 2-category in
Statement \ref{2cat}.  Motivated by the results in \Ref{CY9362,LW170404221},
here we like to show that all the EF 3+1D bosonic topological orders can be
realized by topological non-linear $\si$-models, a particular type of tensor
network models defined on space-time complex.\cite{GLS0918,KW1458,WW180109938}
The topological non-linear $\si$-models are constructed using the data of
unitary fusion 2-categories described in Statement \ref{2cat}.

Let us remind the readers that the  canonical boundary of a EF topological
order is described by a unitary fusion 2-category $\sA_b^3$.  The boundary
stringlike excitations (the simple objects in $\sA_b^3$)  are labeled by the
elements in $\hat G_b=G_b\gext Z_2^m$.\cite{LW180108530}  All the strings have
a unit quantum dimension and their fusion is described by the group $\hat G_b$:  
\begin{align}
\label{strF}
 g_1 g_2 = g_3,\ \ \  g_1,g_2,g_3 \in \hat G_b.
\end{align}
Also
two strings (two objects) labeled by $g$ and $gm$ (where $g\in \hat G_b$ and
$m$ is the generator of $Z_2^m$) are connected by an 1-morphism $\si_{g,gm}$ of
quantum dimension $\sqrt 2$.  This 1-morphism correspond to an on-string
pointlike excitation.  There is another 1-morphism $f_g$ of quantum dimension
$1$ that connect every string $g$ to itself. The second 1-morphism correspond
to a fermionic pointlike excitation. The fusion of 1-morphisms is given by
\begin{align}
\label{fsiF}
& f_g \otimes f_g = \one,\ \ \ \ \
 f_g \otimes \si_{g,gm} = \si_{g,gm},
\nonumber\\ 
& \si_{g,gm} \otimes \si_{gm,g} = \one \oplus f_g .
\end{align}

We note that the fusion 2-category $\sA_b^3$ has three layers.  The first layer
is formed by objects in a fusion category.  For our case, the simple objects in
fusion ring form a finite group $\hat G_b$ (see \eqn{strF}).  The second layer
is formed by 1-morphisms generated by $\one, f_g, \si_{g,gm}$.  The objects and the
1-morphisms are described by a fusion category (see \eqn{fsiF}).  The third
layer is formed by 2-morphisms, which are complex vector spaces for our case.
The objects plus the 1-morphisms and 2-morphisms are described by the fusion
2-category.  In the first part of this section, we are going to show
that the simple objects and simple morphisms in the
fusion category \eqn{strF} and \eqn{fsiF} (\ie the object and 1-morphism
layers) are described by a simplicial set $\hat \cK(\hat G_b,Z_2^f)$. And from this simplicial set, we can recover the entire fusion category (including semi-simple objects). In the
second part of this section, we will show that the 2-morphism layer is
described by a set of tensors.  So the fusion 2-category is  described by a
topological non-linear $\si$-model with a target complex $\hat \cK(\hat
G_b,Z_2^f)$.

To obtain the bulk topological non-linear $\si$-model that realize the fusion
2-category $\sA_b^3$, let us first ignore the quantum-dimension-$\sqrt 2$
1-morphisms $\si_{g,gm}$.  In this case, the canonical boundary will be
described by a pointed unitary fusion 2-category, \ie by a 2-group $\cB(\hat
G_b,Z_2^f)$ and a $\R/\Z$-valued 4-cocycle $\bar\om_4(\hat{\bar a}, \bar b)$ on
the 2-group, where $\hat{\bar a}$ and $\bar b$ are canonical 1-cochain and
2-cochain of $\cB(\hat G_b,Z_2^f)$.  The tensor network model that realize this
reduced boundary will be a 2-gauge theory of 2-gauge-group $\cB(\hat
G_b,Z_2^f)$.  In other words, the links in the tensor network model have an
index $\hat a_{ij}\in \hat G_b$ which defines $\hat{\bar a}$, and the triangles
in the tensor network model have an index $b_{ijk}\in Z_2^m$ which defines
$\bar b$.  $\hat{\bar a}$ and $\bar b$ satisfy
\begin{align}
 \del \hat{\bar a}= \one, \ \ \ \ 
\dd \bar b = \hat n_3(\hat{\bar a}) ,
\end{align}
where $\hat n_3 \in \cH^3(\hat G_b;\Z_2^f)$.  The corresponding path integral
is given by
\begin{align}
&\ \ \ 
 Z(\cM^{4}) = 
|\hat G_b|^{-N_0} 2^{-N_1}
\hskip-2em
\sum_{\del \hat a=\one,\dd b=\hat n_3(\hat a)} 
\hskip-2em
\ee^{2\pi \ii \int_{\cM^{4}} \om_4(\hat a,b) } 
\end{align}

Now, let us include the 1-morphisms $\si_{g,gm}$ that connect two strings $g$
and $gm$.  But at the moment, we will assume such 1-morphisms to have a unit
quantum dimension and a fusion $\si_{g,gm}\otimes \si_{gm,g}=\one$.  Since the
extra 1-morphism can connect two strings differ by $m$, the flat condition on
$\hat a$ is modified and becomes a quasi-flat condition $ \del \hat{\bar a} \in
Z_2^m$.  In $\cB(\hat G_b,Z_2^f)$, three links $\hat a_{ij}$, $\hat a_{jk}$,
$\hat a_{ki}=(\hat a_{ik})^{-1}$ bound a triangle only when $\hat a_{ij}\hat
a_{jk}\hat a_{ki}=\one$.  Now we add some triangles to the complex $\cB(\hat
G_b,Z_2^f)$ so that three links $\hat a_{ij}$, $\hat a_{jk}$, $\hat a_{ki}$
bound a triangle even when $\hat a_{ij}\hat a_{jk}\hat a_{ki}=m \in Z_2^m$.
Including those extra triangles change the first homotopy group of the target
complex to $\pi_1=\hat G_b/Z_2^m=G_b$.  The new  target complex is denoted as
$\hat \cB(G_b,Z_2^f)$, which is a triangulation of $K(G_b,Z_2^f)$.

Let us compare two triangulations, $\hat \cB(G_b,Z_2^f)$ and $\cB(G_b,Z_2^f)$,
of the same space $K(G_b,Z_2^f)$.  In $\cB(G_b,Z_2^f)$, the links are labeled
by $a_{ij} \in G_b$, while in $\hat \cK(G_b,Z_2^f)$ we double the number of
links, which now are labeled by $\hat a_{ij} \in \hat G_b =Z_2^m \gext G_b$.
The triangles in $ \cB(G_b,Z_2^f)$ are labeled by
$[a_{01},a_{12},a_{02};b_{012}]$ where $a_{01},a_{12},a_{02}$ satisfy
$a_{01}a_{12} (a_{02})^{-1}=\one $. On the other hand, the triangles in $\hat
\cB(G_b,Z_2^f)$ are labeled by $[\hat a_{01},\hat a_{12},\hat a_{02};b_{012}]$
where $\hat a_{01},\hat a_{12},\hat a_{02}$ satisfy $\hat a_{01}\hat a_{12}
(\hat a_{02})^{-1}\in Z_2^m $.  The full structure of $\hat \cK(G_b,Z_2^f)$ is
determined by its canonical 1-cochain $\hat{\bar a}$ and 2-cochain $\bar b$
that satisfy
\begin{align}
\label{amZ2mbn3}
\del \hat{\bar a}  \in Z_2^m, \ \ \ \ \dd \bar b = \hat n_3(\hat{\bar a}).
\end{align}
where $\hat n_3(\hat{\bar a})$ is a 3-cocycle in $\hat \cK (\hat{G}_b,
Z_2^f)$ satisfying 
\begin{align}
 \hat n_3(\hat{\bar a})  &=  n_3( \pi^m (\hat {\bar a})), \ \ \ \ \
\pi^m: \hat{G}_b \to G_b, 
\nonumber\\
n_3(\bar a) &\in H^3(\cB G_b, Z_2^f).  
\end{align}

To have a more rigorous construction of $\hat \cB(G_b,Z_2^f)$, we note that
given a morphism of groups $A_2 \xrightarrow{p_2} \hat G$, $\ker p_2
\xrightarrow{0}  G:= \hat G/\Im p_2$ together with $ G$ action $\alpha$ on
$\ker p_2$ and $n_3 \in H^3(G, \ker p_2^\alpha)$ decide a 2-group $\cB(G, \ker
p_2)$, which as a simplicial set has the following form: $K_n = G^{\times n}
\times (\ker p_2)^{\times (^n_2)}$, where 
\begin{align}
K_1 &= \{ ( a_{01} ) | a_{01} \in \hat G \}, 
\\
K_2 &= \{ ( a_{01}, a_{12},  a_{02};  b_{012}) | a_{01}a_{12}a_{02}^{-1} =\one,
 b_{012} \in \ker p_2 \}, 
\nonumber\\
K_3 &= \{ ( a_{01}, a_{12},
a_{23};  b_{012}, b_{013}, b_{023}, b_{123}) |
 \alpha(a_{01}) b_{123} - b_{023}
\nonumber\\
& 
+b_{013}-b_{012}= n_3(a_{01}, a_{12}, a_{23}) \in \ker p_2
\}, 
\nonumber 
\end{align}
and $K_n$ in general is made up of those $n$-simplices whose 2-faces are
elements of $K_2$ and such that each set of four 2-faces gluing together as a
3-simplex is an element of $K_3$. This is the so-called coskelenton
construction. 

Then we pullback this 2-group structure via the projection map $\hat G
\xrightarrow{\pi^m} G$, we obtain another 2-group. The pullback
simplicial set $\hat K_\bullet$ of $K_\bullet$ through $\hat K_1 \to
K_1$ (both $\hat K_0=K_0=pt$) is
inductively defined as $\hat K_n = K_n \times_{\Hom(\partial \Delta[n],
  K)} \Hom(\partial \Delta[n], K)$. Here $\partial \Delta[n]$ is the
boundary simplicial set of the standard simplicial simplex
$\Delta[n]$. Pullback of a 2-group still satisfies the same Kan
conditions, thus still a 2-group. Then after calculation, we see that
the pullback 2-group as a simplicial set has the following form: $\hat
K_n = \hat G^{\times n} \times A_2^{\times (^n_2)}$,
where 
\begin{align}
\hat K_1 &= \{ ( \hat a_{01} ) | \hat a_{01} \in \hat G \}, 
\\
\hat K_2 &= \{ ( \hat a_{01}, \hat a_{12}, \hat a_{02};  b_{012}) |
\hat a_{01} \hat a_{12} \hat a_{02}^{-1} \in \Im p_2, b_{012} \in \ker p_2 \}, 
\nonumber\\
\hat K_3 &= \{ (  \hat a_{01}, \hat a_{12},
\hat a_{23};  b_{012}, b_{013}, b_{023}, b_{123}) | 
\alpha(\pi^m(\hat a_{01})) b_{123}
\nonumber\\
& - b_{023}+b_{013}-b_{012}= \hat n_3(\hat a_{01}, \hat a_{12}, 
\hat a_{23}) \in \ker p_2
\}, 
\nonumber 
\end{align}
and $\hat K_n$ is similarly defined by coskeleton construction. Here $\hat n_3
= (\pi^m)^* n_3$ is the pullback 3-cocycle. We denote this 2-group by $\hat \cB
(G, \ker p_2)$. Since the pullback construction introduces equivalent 2-groups,
$\hat \cB (G, \ker p_2)$ and $\cB (G, \ker p_2)$ are equivalent 2-groups. To
apply in the above situation, we take $G=G_b$, $A_2=Z^f_2 \times Z^m_2$ and
$p_2 = 0 \times i$ where $i: Z_2^m \to \hat G_b $ is the embedding, thus $\ker
p_2 = Z^f_2$ and $\Im p_2 = Z^m_2$. 

Through the above examples, we see that pointed unitary fusion 2-categories
have a ``geometric'' picture in terms of 2-groups.  The fusion rules in the
2-categories are described by the complex of the 2-groups. The complicated
coherent relations in the 2-categories are described by the cocycle conditions
on the 2-groups side.  In the following, we will develop a ``geometric''
picture, \ie a complex $\hat \cK(G_b;Z_2^f)$, for the unitary fusion 2-category
$\sA_b^3$ that contains non-invertible 1-morphisms.

The complex $\hat \cK(G_b;Z_2^f)$ has one vertex. The links in $\hat
\cK(G_b;Z_2^f)$ are labeled by elements $\hat a_{ij}$ in group $\hat G_b =
Z_2^m\gext_{\rho_2} G_b$, with $\rho_2 \in H^2(\cB G_b;\Z_2)$.  The complex
$\hat \cK(G_b;Z_2^f)$ has the same set of links as $\cB(\hat G_b,Z_2^f)$, but
has a different set of triangles to describe a different set of 1-morphisms.
In $\hat\cK(G_b,Z_2^f)$, three links $\hat a_{ij}$, $\hat a_{jk}$, $\hat
a_{ki}=(\hat a_{ik})^{-1}$ bound a triangle when $\hat a_{ij}\hat a_{jk}\hat
a_{ki}\in Z_2^m$.  When $\hat a_{ij}\hat a_{jk}\hat a_{ki}=\one$, the three
links bound two triangles labeled by $b_{ijk}=0,1$.  When $\hat a_{ij}\hat
a_{jk}\hat a_{ki}=m$, where $m$ generates $Z_2^m$, the three links bound only
one triangle which has a fixed $b_{ijk}=1$.

The tetrahedrons in $\hat \cK(G_b;Z_2^f)$ describe the fusion channels of
1-morphisms \eqn{fsiF}.  Consider a 2-sphere in $\hat \cK(G_b;Z_2^f)$ formed by
four triangles who share their edges.  If all four triangles carry no $m$-flux,
\ie satisfy $\hat a_{ij}\hat a_{jk}\hat a_{ki}=\one$, then the 2-sphere is
filled by a tetrahedron if the label $b_{ijk}$ on the four triangles satisfy
$\sum b_{ijk}\se{2} \bar n_3(a_{ij})$. Here $\bar n_3(\hat a_{ij})$ is a
function that depends on labels $\hat a_{ij}$ of the six links on the 2-sphere.
Note that $\bar n_3(\hat a_{ij})$ is defined only when $\hat a_{ij}\hat
a_{jk}\hat a_{ki}=\one$ for all four triangles.  If two of four triangles carry
$m$-flux, \ie satisfy $\hat a_{ij}\hat a_{jk}\hat a_{ki}=m$, then the 2-sphere
is filled by a tetrahedron regardless the values of the labels $b_{ijk}$ on the
four triangles.  

If all four triangles carry $m$-flux, then the 2-sphere is filled by two
different tetrahedrons, labeled by $c_{0123}=0,1$.  This is because each
triangle with $m$-flux corresponds to the 1-morphism $\si$. The fusion of three
$\si$ is given by $\si\otimes \si \otimes \si =(\one \oplus f) \otimes \si
=2\si$.  The factor 2 means there are two fusion channels, and thus two
different tetrahedrons to fill the  2-sphere.

At higher dimensions, every 3-sphere formed by five tetrahedrons glued along
their 2-faces is filled by a 4-simplex, every 4-spheres formed by six
4-simplexes glued along their 3-faces is filled by a 5-simplex, \etc.  In this
way, we obtain the simplicial set $\hat \cK(G_b;Z_2^f)$ (which is thus
3-coskeleton):
\begin{equation}
\xymatrix{ 
K_0 & 
K_1 \ar@<-1ex>[l]_{d_0, d_1}\ar[l] & 
K_2 \ar@<-1ex>_{d_0, d_1 , d_2}[l] \ar@<1ex>[l] \ar[l] & 
K_3 \ar@<-1ex>[l]_{d_0, ..., d_3} \ar@<1ex>[l]_{\cdot} & 
K_4 \ar@<-1ex>[l]_{d_0, ..., d_4} \ar@<1ex>[l]_{\cdot} \cdots ,
}
\end{equation}
where the simplexes at each dimensions are given by
\begin{align}
\hat K_0 &= \{ pt.\},
\nonumber\\ 
\hat K_1 &= \{ ( \hat a_{01} ) | \hat a_{01} \in \hat G \}, 
\\
\hat K_2 &= \{ ( \hat a_{01}, \hat a_{12}, \hat a_{02};  b_{012}) |
\hat a_{01} \hat a_{12} = \hat a_{02},\ b_{012}=0,1;
\nonumber\\
&\ \ \ \ \
\text{or } \hat a_{01} \hat a_{12} =m \hat a_{02},\ b_{012}=0.
   \}, 
\nonumber\\
\hat K_3 &= \{ (  
\hat a_{01}, \hat a_{12}, \hat a_{23},  
\hat a_{02}, \hat a_{13}, \hat a_{03};  
b_{012}, b_{013}, b_{023}, b_{123};c_{0123}
) 
\nonumber\\
& \hskip -1.3em
| \text{if all }
\del \hat a =\one:\
b_{123} - b_{023}+b_{013}-b_{012}
= \hat n_3(\hat a_{01}, \hat a_{12}, \hat a_{23}),
\nonumber\\ &\ \ \ 
c_{0123}=0 ; \ \
\text{if two } \del \hat a =m :\ c_{0123}=0
.
\}, 
\nonumber 
\end{align}
where $\hat n_3 \in H^3(\cB \hat G_b;\Z_2)$.  The complex $\hat \cK(G_b;Z_2^f)$
describes a fusion category formed by the objects and 1-morphisms in the unitary
fusion 2-category $\sA_b^3$. (The 2-morphisms in $\sA_b^3$ will be discussed in
the later part of this section.)

Since it is a coskeleton construction of a 3-step tower, $\hat \cK(G_b;Z_2^f)$
is certainly a simplicial set. In general, the geometric realization $|Y|$ of
the simplicial set $Y$ is a topological space. By construction, $|Y|$ is given
by $|Y|:= \sqcup Y_i \times \Delta^i/\sim $, where $\sim$ is provided by gluing
along lower dimensional faces provided by the information given by $s$ the
degeneracy maps.  However, $|Y|$ may not be a manifold.  Also, $\hat
\cK(G_b;Z_2^f)$ is not a 2-group any more. First of all, strict $Kan(3, j)!$
are not satisfied, and even non-strict $Kan(4, j)$ are not satisfied.
Nevertheless, $\pi_{\ge 3}(\hat \cK (G_b;Z_2^f))=0$.
Moreover, we still have $\pi_2(\hat \cK (G_b; Z_2^f)) = Z_2^f$ and $\pi_1(\hat
\cK (G_b; Z_2^f))=G_b$. 

Although $\hat \cK(G_b;Z_2^f)$ does not correspond to a 2-group, in the
following, we will show that from the data of $\hat \cK(G_b;Z_2^f)$, one can
recover the fusion category, which is the original fusion 2-category $\sA_b^3$
without the 2-morphism layer. We first let the set of simple objects to be the
links in $\hat \cK(G_b;Z_2^f)$, $C_0:=\hat \cK(G_b;Z_2^f)_1= \hat G_b$. And let
the set of simple 1-morphisms to be the triangles with one side degenerate in
$\hat \cK(G_b;Z_2^f)$. One can picture them as bigons (see Fig.
\ref{fig:bigon}), 
\[
\begin{split}
C_1 := & \{ (1, \hat a_{12}, \hat a_{02}; b_{012} ) \in \hat K_2 \}
\\
=& \{ ( g, g'; b)| g=g', b=0, 1; g'=gm, b=0\}
\end{split}
\] 
  \begin{figure}[tb]
    \centering
    \begin{tikzpicture}[>=latex',mydot/.style={draw,circle,inner sep=1pt},every label/.style={scale=0.9},scale=1.0]
\node[mydot,label=240:\(0\)]                at (0,0)    (p00) {};
\node[mydot,label=90:\(1\)]      at (0,0.5) (p01) {};
\node[mydot,label=-60:\(2\)]     at (2,0)    (p02) {};      
\node[below, scale=0.8] at (0.6, 0.37) (p03) {$b_{012}=b$};

\node[mydot,label=240:\(0\)]                at (4,0)    (p10) {};
\node[mydot,label=90:\(1\)]      at (6,0) (p11) {};
\node[scale=.8] at (5, 0) (p12) {$b \Uparrow$};

      \begin{scope}[<-]
        \draw (p00) --node[left, scale=0.8]{1}  (p01);
        \draw (p00) --node[below,scale=0.8]{$\hat a_{02}=g'$} (p02);
        \draw (p01) --node[above,scale=0.8]{$\hat a_{01}=g$} (p02);
        
\end{scope}
\path[<-]      (p10) edge [bend left] node[scale=0.8,above left]{$g$}  (p11);
\path[<-]      (p10) edge [bend right] node[scale=0.8,below left]{$g'$} (p11);
  
  \end{tikzpicture}
    \caption{Links are simple objects and triangles with degenerate $(0,1)$-sides are simple 1-morphisms.
    \label{fig:bigon} }
  \end{figure}

  \begin{figure}[tb]
    \centering
    \begin{tikzpicture}[>=latex',mydot/.style={draw,circle,inner sep=1pt},every label/.style={scale=1.0},scale=1.0]
\node[mydot,label=240:\(0\)]                at (0,0)    (p00) {};
\node[mydot,label=-180:\(1\)]      at (-0.5,0.5) (p01) {};
\node[mydot,label=90:\(2\)]     at (0,1)    (p02) {};
\node[mydot,label=-60:\(3\)]     at (2,0.5)    (p03) {};      
\node[below, scale=0.8] at (0.4, 0.55) (p04) {$b_{013}$};
\node[below, scale=0.8] at (0.4, 0.9) (p05) {$b_{123}$};

\node[mydot,label=240:\(0\)]                at (4,0.5)    (p10) {};
\node[mydot,label=90:\(1\)]      at (6,0.5) (p11) {};
\node[scale=.8] at (5.3, 0.33) (p12) {$b \Uparrow$}; 
\node[scale=.8] at (5.3, 0.67) (p12) {$b' \Uparrow$};

      \begin{scope}[<-]
        \draw (p00) --node[left, scale=0.8]{1}  (p01);
        \draw (p00) --node[below,scale=0.8]{$\hat a_{02}=g''$} (p03);
        \draw (p01) --node[right,scale=0.8]{$\hat a_{01}=g'$} (p03);
        \draw (p01) --node[left, scale=0.8]{1}  (p02);
         \draw (p00) --node[left, scale=0.8]{1}  (p02);
         \draw (p02) --node[above,scale=0.8]{$\hat a_{02}=g$} (p03);
\end{scope}
\path[<-]      (p10) edge [bend left=60] node[scale=0.8,above left]{$g$}
(p11);
\path[<-]      (p10) edge node[scale=0.8,above left]{$g'$}
(p11);'
\path[<-]      (p10) edge [bend right=60] node[scale=0.8,below left]{$g''$} (p11);
  
  \end{tikzpicture}
    \caption{The composition $\cdot_v$ of 1-morphisms.
    \label{fig:bigon-composition} }
  \end{figure}

Then the composition $\cdot_v$ of 1-morphisms can be read from the information
of $\hat K_3$, which tells which tetrahedrons are allowed, indicated by Fig.
\ref{fig:bigon-composition}.  For example,  we have a unique tetrahedron $(1,
1, g, 1, g, g; 0, b, b+b', b') $ in $\hat K_3$ to fill its $(3,1)$-horn. Then
this implies that $(g, g; b) \cdot_v (g, g; b')= (g, g; b+b')$, here $+$ is the
addition in $Z_2$. Then the only non-unique case is for $(g, gm; 0) \cdot_v
(gm, g; 0)$: there are both $(1, 1, g, 1, gm, g; 0, 0, 0, 0 )$ or $(1, 1, g, 1,
gm, g; 0, 0, 1, 0 )$ to fill the $(3, 1)$-horn. This makes $(g, gm; 0) \cdot_v
(gm, g; 0) = (g, g; 0 \oplus 1)$ a non-simple element. We thus can extend
$\cdot_v$ to an associative product to all semi-simple objects and 1-morphisms.
We call the result category $\sA_b^3$.

\begin{figure}[tb]
    \centering
    \begin{tikzpicture}[>=latex',mydot/.style={draw,circle,inner sep=1pt},every label/.style={scale=0.8},scale=0.8]
\node[mydot,label=240:\(0\)]                at (0,0)    (p00) {};
\node[mydot,label=90:\(1\)]      at (0,0.5) (p01) {};
\node[mydot,label=-60:\(2\)]     at (2,0)    (p02) {};      
\node[below, scale=0.7] at (0.6, 0.37) (p0b) {$b_{012}=b$};
\node[mydot,label=90:\(3\)]     at (2,0.5)    (p03) {};    
\node[mydot,label=-60:\(4\)]     at (4,0)    (p04) {};
\node[below, scale=0.7] at (2.6, 0.37) (p0b') {$b_{234}=b'$};

\node[mydot,label=240:\(0\)]                at (6,0)    (p10) {};
\node[mydot,label=90:\(1\)]      at (8,0) (p11) {};
\node[mydot,label=90:\(2\)]      at (10,0) (p12) {};
\node[scale=.7] at (7, 0) (p1b) {$b \Uparrow$};
\node[scale=.7] at (9, 0) (p1b') {$b' \Uparrow$};

      \begin{scope}[<-]
        \draw (p00) --node[left, scale=0.7]{1}  (p01);
        \draw (p00) --node[below,scale=0.7]{$\hat a_{02}=g_1'$} (p02);
        \draw (p01) --node[above,scale=0.7]{$\hat a_{01}=g_1$} (p02);
\draw (p02) --node[left, scale=0.5]{1}  (p03);
        \draw (p03) --node[above,scale=0.7]{$\hat a_{34}=g_2$} (p04);
        \draw (p02) --node[below,scale=0.7]{$\hat a_{01}=g'_2$} (p04);
        
\end{scope}
\path[<-]      (p10) edge [bend left] node[scale=0.7,above left]{$g_1$}  (p11);
\path[<-]      (p10) edge [bend right] node[scale=0.7,below left]{$g'_1$} (p11);
 \path[<-]      (p11) edge [bend left] node[scale=0.7,above left]{$g_2$}  (p12);
\path[<-]      (p11) edge [bend right] node[scale=0.7,below left]{$g'_2$} (p12);

\node at (2,-1.5) {\( \downarrow\)};


\node[mydot,label=240:\(0\)]                at (0,-3)    (p20) {};
\node[mydot,label=-60:\(2\)]     at (2,-3)    (p22) {};      
\node[below, scale=0.7] at (1, 0.37-3) (p2b) {$b_{012}=b$};
\node[mydot,label=90:\(3\)]     at (2,0.5-3)    (p23) {};    
\node[mydot,label=-60:\(4\)]     at (4,-3)    (p24) {};
\node[below, scale=0.7] at (2.6, 0.37-3) (p2b') {$b_{234}=b'$};
\node[below, scale=0.7] at (2, -0.33-3) (p2b') {$b_{024}=0$};

      \begin{scope}[<-]
        \draw (p20) --node[below,scale=0.7]{$\hat a_{02}=g_1'$} (p22);
        \draw (p20) --node[above,scale=0.7]{$\hat a_{03}=g_1$} (p23);
\draw (p22) --node[left, scale=0.5]{1}  (p23);
        \draw (p23) --node[above,scale=0.7]{$\hat a_{34}=g_2$} (p24);
        \draw (p22) --node[below,scale=0.7]{$\hat a_{01}=g'_2$} (p24);
       
\end{scope}

\path[<-]      (p20) edge [bend right=45] node[scale=0.7,below left]{$g'_1g'_2$} (p24);

    \node at (5,-3) { \(\rightarrow \)}; 


\node[mydot,label=240:\(0\)]                at (6,-3)    (p30) {};
\node[mydot,label=90:\(4\)]      at (10,-3) (p34) {};

\path[<-]      (p30) edge [bend left] node[scale=0.7,above left]{$g_1g_2$}  (p34);
\path[<-]      (p30) edge [bend right] node[scale=0.7,below left]{$g'_1g'_2$} (p34);

  \end{tikzpicture}
    \caption{Fusion of 1-morphisms.
    \label{fig:bigon-h-c} }
  \end{figure}
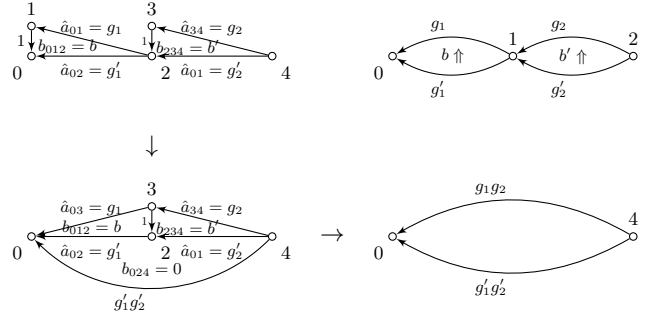

Now we  will read from $\hat K_3$ the fusion product for
$\sA_b^3$, which makes $\sA_b^3$ further into a fusion category. We
only need to take care of fusion of simple objects and simple
1-morphisms, then we can extend the fusion by distribution law to
semi-simple objects and 1-morphisms. The
fusion of simple objects is simply the group multiplication of $\hat
G_b$; the fusion of simple 1-morphisms is again read from
tetrahedrons in $\hat K_3$. If we want to fuse $(g_1, g'_1; b_1)$ and
$(g_2, g'_2; b_2) $, the first step is to transfer the $(0,1)$-side
degenerate triangle $(g_1, g'_1; b_1)
= (1, g_1', g_1; b_1)$ to an $(2,3)$-side degenerate triangle, by
filling the $(3,0)$-horn of the tetrahedron $(0,1,2,3)$ with a unique element 
\[ (1, g'_1, g'_1, g_1, g_1, 1; b_1, 0, b_1, 0) \in \hat K_3.\]
The second step is to fill the $(2,1)$-horn of the triangle $(0,
1, 4)$ without flux with $(g'_1, g'_2, g'_1g'_2; 0)$.  The third step is
to finally fill the $(3,1)$-horn of the tetrahedron $(0, 2, 3, 4)$ and
obtain a triangle $(0, 3, 4)$ with three sides $(g_1, g_2,
g'_1g'_2)$. The fouth step is to transfer this triangle to a trianlge
with sides $(1, g_1 g_2 , g'_1g'_2)$ by filling the $(3, 2)$-horn of a
tetrahedron. The filling can be non-unique only in the third
step. This procedure is illustrated with Fig. \ref{fig:bigon-h-c}.

Following this strategy, the calculation shows that the only
non-unique case
happens when we fuse $(g_1, g_1m; 0) $ and $(g_2, g_2m, 0)$, and $(g_1, g_1m; 0)
\otimes (g_2, g_2m, 0) = (g_1g_2, g_1g_2; 0 \oplus 1)$.  The associator for the fusion product is still given by
$n_3$. Thus we have recovered a fusion 2-category from the simplicial
set $\hat \cK(G_b,Z_2^f)$. 

To obtain the coherence relations (\ie the 2-morphism layer) in the unitary
fusion 2-category $\sA_b^3$, we try to construct topological non-linear
$\si$-models with target complex $\hat \cK(G_b,Z_2^f)$.  To do so, we
assign a complex number to each 4-simplex in $\hat \cK(G_b,Z_2^f)$.  A
4-simplex is labeled by $(\hat a_{ij};b_{ijk},c_{ijkl}| i,j,k,l=0,1,2,3,4)$,
that satisfy
\begin{align}
\label{ban3}
&
\hat a_{ij} \in \hat G_b,\ \ \ b_{ijk}\in \Z_2,\ \ \ c_{ijkl}\in \Z_2;
\nonumber\\
& b_{123} - b_{023}+b_{013}-b_{012}
= \hat n_3(\hat a_{01}, \hat a_{12}, \hat a_{23})
\nonumber\\
& 
\ \ \ \ \ \  \ \ \ \ \ \ \  \ \ \ \ \
\ \ \ \ \ \  \ \ \ \ \ \ \  
\text{ when all four } \del \hat a =\one,
\nonumber\\
& b_{ijk} =0 \text{ when } (\del \bar a)_{ijk} =m .
\nonumber \\
& c_{ijkl} =0 \text{ when one of } \del \bar a =\one .
\end{align}
So we can write such a complex number as
\begin{align}
\tDm {\hat\Om_4} 01234 
\end{align}
which corresponds to the top tensor of the tensor set.  The above number is
non-zero only when $\hat a_{ij},b_{ijk},c_{ijkl}$ satisfy \eqn{ban3}.  We also
assign a positive number $w_0$ to the vertex in $\hat \cK(G_b,Z_2^f)$.  To the
links labeled by $[\hat a_{01}]$ we assign the same positive number $w_1$.  To
the triangle labeled by $[\hat a_{01},\hat a_{12},\hat a_{02};b_{012}]$ we
assign a positive number $w_2(\one)$ or $w_2(m)$ depending on $\hat a_{01} \hat
a_{12} (\hat a_{02})^{-1}= \one$ or $m$.  The path integral that describes the
topological non-linear $\si$-model on space-time with boundary is given by
\begin{align}
& Z(\cM^{4}) =  
\hskip-2em
\sum_{\del \hat a \in Z_2^m,\dd b=\hat n_3(\hat a),c} 
{\prod_i}^\prime w_0
{\prod_{(ij)}}^\prime w_1
{\prod_{(ijk)}}^\prime w_2[\del \hat a)_{ijk}]  \times
\nonumber \\
&
\prod_{(ijklp)} \Big( \tDm {\hat\Om_4} ijklp \Big)^{s_{ijklp}}
,
\end{align}
where $\prod_{(ijklp)}$ is a product over all the $4$-simplices and $s_{ijklp}$
is the orientation of the $4$-simplices (see Fig. \ref{mir}).  Also,
$\prod_{(ijk)}'$ is a product over all the internior triangles, $\prod_{(ij)}'$
is a product over all the internior links, and $\prod_i$ is a product over all
the internior vertices.  

The rank-25 tensor $\hat\Om_4$, as well as the weight tensors $w_0$, $w_1$, and
$w_2$, must satisfy certain conditions in order for the above path integral to
be re-triangulation invariant.  The conditions can be obtained in the following
way: We start with a 5-simplex $(012345)$.  Then, divide the six 4-simplices on
the boundary of the 5-simplex $(012345)$ into two groups.  Then the partition
function on one group of the 4-simplices must equal to the partition function
on the other group of the 4-simplices, after a complex conjugation.

For example, the two groups of the 4-simplices can be $ \big[ (12345), (02345),
(01345) \big] $ and $ \big[ (01245), (01235), (01234) \big] $. This partition
leads to a condition
\begin{align}
\label{cond33}
&\ \ 
\sum_{b_{345}} 
\sum_{
c_{0345}
c_{1345}
c_{2345}
} 
 w_2[(\del \hat a)_{345}]
\nonumber\\ &
 \tDm {\hat\Om_4} 12345 
\nonumber\\
& (\tDm {\hat\Om_4} 02345)^* 
\nonumber\\
& \tDm {\hat\Om_4} 01345
\nonumber\\
=
&\sum_{b_{012}} 
\sum_{
c_{0123}
c_{0124}
c_{0125}
} 
 w_2[(\del \hat a)_{012}]
\nonumber\\ &
 \tDm {\hat\Om_4} 01245 
\nonumber\\
& (\tDm {\hat\Om_4} 01235)^* 
\nonumber\\
& \tDm {\hat\Om_4} 01234
\end{align}
For the partition
$
\big[ 
(12345),
(02345)
\big]
$
and
$
\big[ 
(01345)$, $
(01245)$, $
(01235)$, $
(01234)
\big]
$, we obtain a condition
\begin{align}
\label{cond24}
& \sum_{c_{2345}}
\tDm {\hat\Om_4} 12345 
\nonumber\\
& (\tDm {\hat\Om_4} 02345)^* 
\nonumber\\
&=
w_1
\hskip-15em
\sum_{
\ \ \ \ \  \ \ \ \ \  \ \ \
\ \ \ \ \  \ \ \ \ \  \ \ \
\ \ \ \ \  \ \ \ \ \  \ \ \
\hat a_{01}; b_{012}, b_{013}, b_{014}, b_{015} ;  
c_{0123}
c_{0124}
c_{0125}
c_{0134}
c_{0135}
c_{0145}
}
\hskip-13em
w_2[(\del \hat a)_{012} ]
w_2[(\del \hat a)_{013} ]
w_2[(\del \hat a)_{014} ]
\nonumber\\
&
w_2[(\del \hat a)_{015} ]
 (\tDm {\hat\Om_4} 01345)^*
\nonumber\\
& \tDm {\hat\Om_4} 01245 
\nonumber\\
& (\tDm {\hat\Om_4} 01235)^* 
\nonumber\\
& \tDm {\hat\Om_4} 01234
\end{align}
For the partition
$
\big[ 
(12345)
\big]
$
and
$
\big[ 
(02345)$, $
(01345)$, $
(01245)$, $
(01235)$, $
(01234)
\big]
$, we obtain a condition
\begin{align}
\label{cond15}
 & 
 \tDm {\hat\Om_4} 12345 
\nonumber\\
=&
w_0 
w_1^5
\hskip-2em
\sum_{ \hat a_{01}, \hat a_{02}, \hat a_{03}, \hat a_{04}, \hat a_{05} } 
\sum_{ b_{012}, b_{013}}  
w_2[(\del \hat a)_{012} ]
w_2[(\del \hat a)_{013} ]
\nonumber\\
&
\hskip-21em
\sum_{
\ \ \ \ \ \ \ \ \ \ \ \ 
\ \ \ \ \ \ \ \ \ \ \ \ 
\ \ \ \ \ \ \ \ \ \ \ \ 
\ \ \ \ \ \ \ \ \ \ \ \ 
\ \ \ \ \ \ \ \ \ \ \ \ 
b_{014}, b_{015},b_{045},b_{023};
c_{0123}
c_{0124}
c_{0125}
c_{0134}
c_{0135}
c_{0145}
c_{0234}
c_{0235}
c_{0245}
c_{0345}
 }  
\hskip-20em
w_2[(\del \hat a)_{014} ]
w_2[(\del \hat a)_{015} ]
w_2[(\del \hat a)_{045} ]
w_2[(\del \hat a)_{023} ]
\nonumber\\
&
\hskip-5em
\sum_{\ \ \ \ \ \ \ \ \ \ \ \  b_{024}, b_{025}, b_{034},b_{035} ;
}  
\hskip-4.5em
w_2[(\del \hat a)_{024} ]
w_2[(\del \hat a)_{025} ]
w_2[(\del \hat a)_{034} ]
w_2[(\del \hat a)_{035} ]
\nonumber\\
&\ \ \   \tDm {\hat\Om_4} 02345
\nonumber\\
&\ \ \  (\tDm {\hat\Om_4} 01345)^*
\nonumber\\
&\ \ \   \tDm {\hat\Om_4} 01245 
\nonumber\\
&\ \ \  (\tDm {\hat\Om_4} 01235)^* 
\nonumber\\
&\ \ \  \tDm {\hat\Om_4} 01234
\end{align}
There are many other similar conditions from different partitions.

Each solution of those conditions give us a topological non-linear $\si$-model.
Some of those models have emergent fermions and describe EF topological orders.
We believe that all EF topological orders can be realized this way.

In general, it is very hard to find solutions of those conditions, since that
corresponds to solve billions of non-linear equations with millions of unknown
variables, even for the simplest cases.  One way to make progress is to note
that when restricted to the indices $\hat a$ that satisfy $\del \hat a=\one$,
the tensor $\hat\Om_4$ becomes a $U(1)$-valued 4-cocycle on the 2-group
$\cB(\hat G_b,Z_2^f)$.  This is because some conditions for $\hat\Om_4$, such
as \eqn{cond33}, act within those components of $\hat\Om_4$ whose indices
satisfy $\del \hat a=\one$.  When $\del \hat a=\one$, $w_2(m)$ will not appear
in those conditions.  In this case, if we choose $\hat\Om_4$ to be a
$U(1)$-valued 4-cocycle on the 2-group, the terms in the summation in
\eqn{cond33} will all have the same value. Thus we can replace the summation in
\eqn{cond33} by factors that count the number of the terms in the summation.  
From \eqn{cond33}, we see that those factors cancel out.  In this case, the
condition \eqn{cond33} reduces to the condition for the 4-cocycles on the
2-group.  Thus, the restricted $\hat\Om_4$ must be $U(1)$-valued 4-cocycle on
the 2-group $\cB(\hat G_b,Z_2^f)$, which has a form:
\begin{align}
\label{Dcocy}
& \tDm {\hat\Om_4}01234 \Big|_{\del \hat a=\one,c\text{'s}=0}
\nonumber\\
&=
 \ee^{2\pi \ii \int_{(01234)} \nu_4(\hat a) 
+ \frac{k_0}{2} \Sq^2 b + \frac{1}{2} b e_2(\hat a) }
\end{align}
When $k_0=1$, the tensor $\hat\Om_4$ and the associated topological non-linear
$\si$-model will describe a EF topological order. 
%
%
Starting from the parcial solution \eq{Dcocy} we can use the equations
\eqn{cond33}, \eqn{cond24}, and \eqn{cond15} to find other components of
$\hat\Om_4$ whose indices do not satisfy $\del \hat a=\one$.  

As we have seen that the topological non-linear $\si$-model on the complex
$\hat \cK(G_b,Z_2^f)$ is closely related to the unitary fusion 2-category
$\sA_b^3$ that describes the canonical boundary of a EF topological
order.\cite{LW180108530}  The links in $\hat \cK(G_b,Z_2^f)$ correspond to the
objects in the fusion 2-category.  The 1-morphisms $f_g$ that connect an object
to itself corresponds to triangles with no flux, which are labled by
$\pi_2[\hat \cK(G_b,Z_2^f)]=\Z_2$.  The non-invertible 1-morphisms $\si_{g,gm}$
correspond to triangles with $m$-flux.  If we treat the objects connected by
1-morphisms as equivalent, then the eqquivalent classes of the objects
correspond to $\pi_1[\hat \cK(G_b,Z_2^f)]=G_b$.  The fusion of the objects in
different orders may differ by an 1-morphism which lives in $\pi_2[\hat
\cK(G_b,Z_2^f)]$, It is called an associator. In both \Ref{LW180108530} and
this paper, we use the same symbol $\hat n_3$ to describe the associator.  The
part of the $\hat\Om_4$ tensor, $\hat \nu_4$, also correspond to $\hat \nu_4$
in \Ref{LW180108530} that is another piece of data to describe the unitary
fusion 2-category $\sA_b^3$.  It is this correspondence between topological
non-linear $\si$-models on  $\hat \cK(G_b,Z_2^f)$ and the fusion 2-categories
desecribed  in \Ref{LW180108530} that allows us to conclude that all EF
topological orders are realized by topological non-linear $\si$-models on $\hat
\cK(G_b,Z_2^f)$.

From a consideration of 2-gauge transformations (see \eqn{gauge2a} and
\eqn{gauge2b}), we expect $w_0$ and $w_1$ to contain factors $|\hat G_b|^{-1}$
and $\frac12$ to cancel the volum of the 2-gauge transformations.  If
$w_2(\one)=w_2(m)$ with $m$ being the generator of $Z_2^m$, the solutions should
describe AB or EF1 topological orders.  If $w_2(\one)\neq w_2(m)$, some of those
solutions should describe EF2 topological orders.  In particular, we expect
$w_2(m)$ to be related to the quantum dimension of the non-invertible
1-morphism -- the Majorana zero mode.

\subsection{The canonical boundary of topological non-linear $\si$-models}

In the last section, we constructed topological non-linear $\si$-models using
the data of unitary fusion 2-categories in Statement \ref{2cat}.  In this
section, we like to show that the topological non-linear $\si$-models  have a
canonical boundary described by corresponding unitary fusion 2-category
$\sA_b^3$.

\begin{figure}[t]
\begin{center}
\includegraphics[scale=0.8]{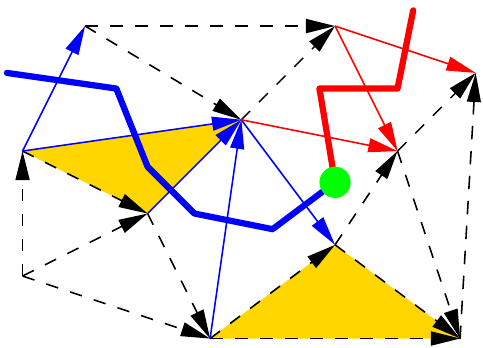}
\end{center}
\caption{
A boundary configuration.
The thin dash-lines corresponds to $\hat a_{ij}=\one$.
The thin colored-lines corresponds to $\hat a_{ij} \neq \one$.
The white triangles corresponds to $b_{ijk}=0$.
The yellow triangles corresponds to $b_{ijk}=1$, which are boundary fermions.
The non-zero $\hat a_{ij}$'s describe boundary strings
on the dual lattice, represented by the thick lines.
The strings with different colors are described by
$g$ and $gm$. The domain wall between two strings
has a Majorana zero mode marked by a green dot.
}
\label{bdry}
\end{figure}

The canonical boundaries of the topological non-linear $\si$-models are very
simple which are given by choosing $\hat a_{ij}=\one$ and $b_{ijk}=0$ on the
boundary.  The states with $\hat a_{ij} \neq \one$ and $b_{ijk} \neq 0$
corresponds excited states with boundary stringlike and pointlike excitations
(see Fig.  \ref{bdry}).

We see that the boundary string are labeled by $\hat a_{ij}$ which is an
element in $\hat G_b$.  They correspond to objects in a unitary fusion
2-category.  $b_{ijk}$ on triangles correspond to 1-morphisms of unit quantum
dimension.  $b_{ijk}=1$ implies the presence of a fermion on the triangle
$(ijk)$.  The condition $\dd b = \hat n_3(\hat a)$ describes how a fermion
worldline can starts or ends at certain configurations of $\hat a$, where $\hat
n_3(\hat a)\neq 0$.  

The Fermi statistics of the particle described by $b_{ijk} \neq 0$ is
determined by the form of the top tensor $\tDm {\hat\Om_4} 01234$ in
\eqn{Dcocy}.  $k_0=1$ will make the particle to be a fermion.

The triangles with $\del \hat a = m$ will carry a Majorana zero mode, provided
that the weight tensor $w_2(\del \hat a)$ satisfies $w_2(\one) \neq w_2(m)$.  If
$w_2(\one) = w_2(m)$, the triangles with $\del \hat a = m$ will not correspond to
a Majorana zero mode.  Those results suggest that the canonical boundaries of
the topological non-linear $\si$-models are described by unitary fusion
2-categories in Statement \ref{2cat}.

To summarize, the topological non-linear $\si$-models are described by the
following data
\begin{align} 
 \hat G_b &=Z_2^m \gext_{\rho_2} G_b,\ \hat n_3(\hat a),\ w_0,\ w_1,\
w_2(\one),\ w_2(m), 
\nonumber\\ & \
\tDm {\hat\Om_4} 01234 .  
\end{align} 
where $\hat n_3(\hat a)$ is defined only when $\del \hat a=\one$. In that case,
it is a $\Z_2$-valued group 3-cocycle for $\hat G_b$: $\hat n_3|_{\del \hat
a=\one} \in \cH^3(\hat G_b;\Z_2)$.  Also, $w_0,w_1,w_2(Z_2^m) ,\hat \Om_4$
satisfy a set of non-linear equations, such as \eqn{cond33}, \eqn{cond24}, and
\eqn{cond15}. If the tensor $\hat \Om_4$ has a form \eq{Dcocy} with $k_0=1$,
then the data describe a EF topological order. Such data also classify the EF
topological orders after quotient out certain equivalence relation.  When $\hat
G_b=Z_2^m \gext_{\rho_2} G_b$ is a non-trivial extension of $G_b$ by $Z_2^m$
and when $w_2(\one) \neq w_2(m)$, the data classify the EF2 topological orders.

Although we have collected many evidences to support the above proposal, many
details still need to be worked out to confirm it.

\section{Summary}

In this paper, we show that higher gauge theories are nothing but familiar
non-linear $\si$-models in the topological-defect-free disordered phase. As a
result, non-linear $\si$-models whose target spaces $K$ satisfy $\pi_1(K)=$
finite group and $\pi_{k>1}(K)= 0$ can realize gauge theories, and non-linear
$\si$-models whose target spaces $K$ satisfy $\pi_1(K),\pi_2(K) = $ finite
group and  $\pi_{k>2}(K)= 0$ can realize 2-gauge theories, \etc.

We discuss in detail how to characterize and classify higher gauge theories,
such as 2-gauge theories.  As an application, we use 2-gauge theories to
realize and classify all 3+1D EF1 topological orders -- 3+1D topological orders
for bosonic systems with emergent fermions, but no Majorana zero modes for
triple string intersections.  We also design topological non-linear
$\si$-models to realize and classify all 3+1D EF2 topological orders -- 3+1D
topological orders for bosonic systems with emergent fermions that have
Majorana zero modes for some triple string intersections.  Since EF topological
orders can be viewed as gauged fermionic SPT state in 3+1D, our result also
give rise to a classification of 3+1D fermionic SPT orders.

To obtain the above results, we developed a ``geometric'' way to view the
unitary fusion 2-category $\sA_b^3$ for the canonical boundary of the EF
topological orders.  We used a special triangulation of a space $K(\hat
G_b,\Z_2^f)$ to described the fusion category formed by the objects and
1-morphisms in $\sA_b^3$.  We used a tensor set defined for the triangulation
to described the 2-morphism layer of 2-category $\sA_b^3$.

We thank Zheng-Cheng Gu, Thomas Schick and Chenjie Wang for helpful discussions.  XGW is
supported by NSF Grant No.  DMR-1506475 and DMS-1664412. CZ is
supported by the German Research Foundation (Deutsche
  Forschungsgemeinschaft (DFG)) through the Institutional
  Strategy of the University of G\"ottingen and DFG ZH 274/1-1.

\appendix

\section{Space-time complex, cochains, and cocycles} 

\label{cochain}

In this paper, we consider models defined on a space-time lattice.  A
space-time lattice is a triangulation of the $d+1$D space-time, which is
denoted as $\cM^{d+1}$.  We will also call the triangulation
$\cM^{d+1}$ as a space-time complex, which is formed by simplices --
the vertices, links, triangles, \etc.  We will use $i,j,\cdots$ to label
vertices of the space-time complex.  The links of the complex (the 1-simplices)
will be labeled by $(i,j),(j,k),\cdots$.  Similarly, the triangles of the
complex  (the 2-simplices)  will be labeled by $(i,j,k),(j,k,l),\cdots$.

\begin{figure}[t]
\begin{center}
\includegraphics[scale=0.5]{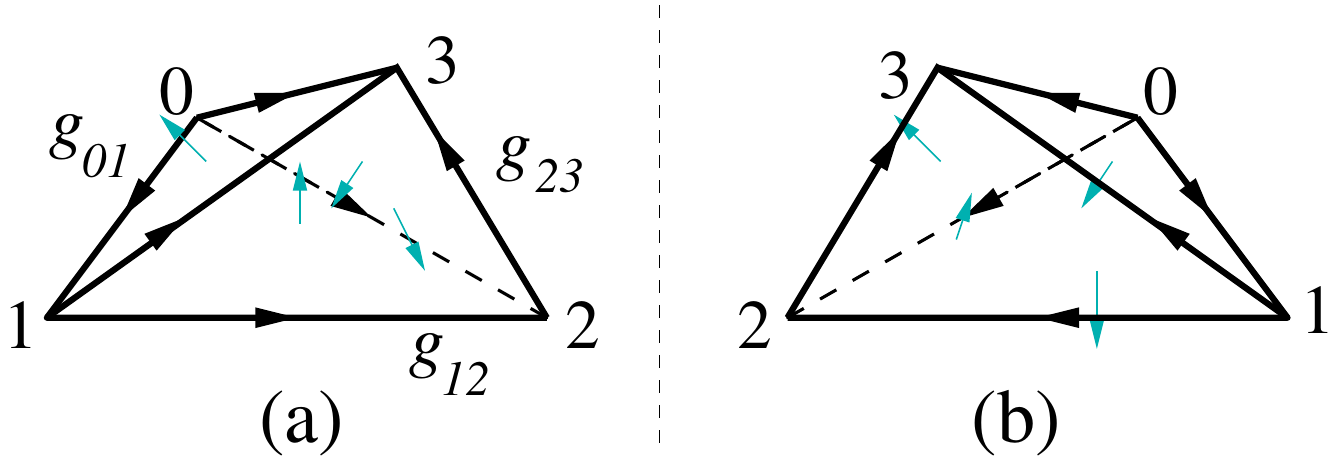} \end{center}
\caption{ (Color online) Two branched simplices with opposite orientations.
(a) A branched simplex with positive orientation and (b) a branched simplex
with negative orientation.  }
\label{mir}
\end{figure}

In order to define a generic lattice theory on the space-time complex
$\cM^{d+1}$ using local tensors $T_{ij\cdots k}$ and
$\om_{d+1}(a^{G_f}_{ij},a^{G_f}_{ik},\cdots)$, it is important to give the
vertices of each simplex a local order.  A nice local scheme to order  the
vertices is given by a branching structure.\cite{C0527,CGL1314,CGL1204} A
branching structure is a choice of orientation of each link in the $d+1$D
complex so that there is no oriented loop on any triangle (see Fig. \ref{mir}).

The branching structure induces a \emph{local order} of the vertices on each
simplex.  The first vertex of a simplex is the vertex with no incoming links,
and the second vertex is the vertex with only one incoming link, \etc.  So the
simplex in  Fig. \ref{mir}a has the following vertex ordering: $0,1,2,3$.

The branching structure also gives the simplex (and its sub-simplices) a
canonical orientation.  Fig. \ref{mir} illustrates two $3$-simplices with
opposite canonical orientations compared with the 3-dimension space in which
they are embedded.  The blue arrows indicate the canonical orientations of the
$2$-simplices.  The black arrows indicate the canonical orientations of the
$1$-simplices.

Given an abelian group $(\M, +)$, an $n$-cochain $f_n$ is an assignment of
values in $\M$ to each $n$-simplex, for example a value $f_{n;i,j,\cdots,k}\in
\M$ is assigned to $n$-simplex $(i,j,\cdots,k)$.  So \emph{a cochain $f_n$ can
be viewed as a bosonic field on the space-time lattice}.

\begin{figure}[tb]
\begin{center}
\includegraphics[scale=0.5]{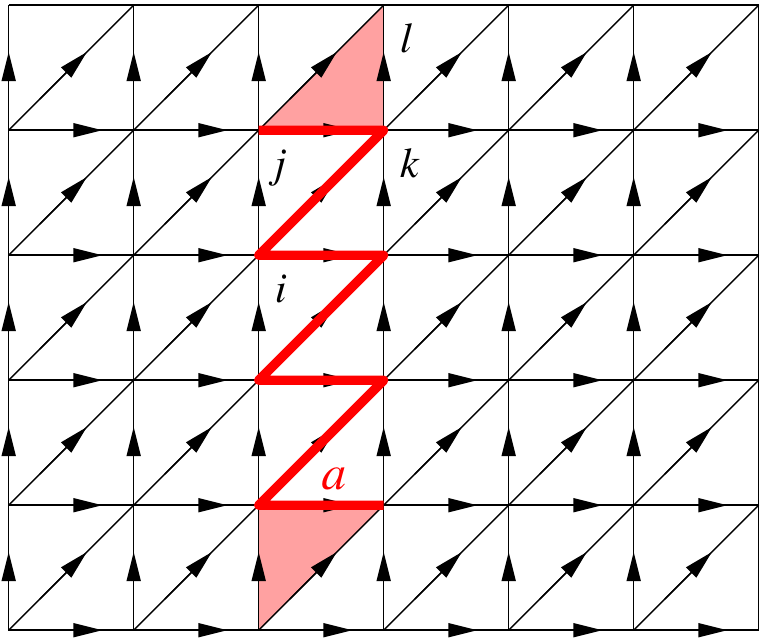} \end{center}
\caption{ (Color online)
A 1-cochain $a$ has a value $1$ on the red links: $ a_{ik}=a_{jk}= 1$ and a
value $0$ on other links: $ a_{ij}=a_{kl}=0 $.  $\dd a$ is non-zero on the
shaded triangles: $(\dd a)_{jkl} = a_{jk} + a_{kl} - a_{jl}$.  For such
1-cohain, we also have $a\smile a=0$.  So when viewed as a $\Z_2$-valued cochain,
$\Bs_2 a \neq a\smile a$ mod 2.
}
\label{dcochain}
\end{figure}

We like to remark that a simplex $(i,j,\cdots,k)$ can have two different
orientations $s_{ij\cdots k}=\pm$. We can use $(i,j,\cdots,k)$ and
$(j,i,\cdots,k)=-(i,j,\cdots,k)$ to denote the same simplex with opposite
orientations.  The value $f_{n;i,j,\cdots,k}$ assigned to the simplex with
opposite  orientations should differ by a sign:
$f_{n;i,j,\cdots,k}=-f_{n;j,i,\cdots,k}$.  So to be more precise $f_n$ is a
linear map $f_n: n\text{-simplex} \to \M$. We can denote the linear map as
$\<f_n, n\text{-simplex}\>$, or
\begin{align}
 \<f_n, (i,j,\cdots,k)\> = f_{n;i,j,\cdots,k} \in \M.
\end{align}
More generally, a \emph{cochain} $f_n$ is a linear map
of $n$-chains:
\begin{align}
	f_n:  n\text{-chains} \to \M,
\end{align}
or (see Fig. \ref{dcochain})
\begin{align}
 \<f_n, n\text{-chain}\> \in \M,
\end{align}
where a \emph{chain} is a composition of simplices. For example, a 2-chain can
be a 2-simplex: $(i,j,k)$, a sum of two 2-simplices: $(i,j,k)+(j,k,l)$, a more
general composition of 2-simplices: $(i,j,k)-2(j,k,l)$, \etc.  The map $f_n$ is
linear respect to such a composition.  For example, if a chain is $m$ copies of
a simplex, then its assigned value will be $m$ times that of the simplex.
$m=-1$ correspond to an opposite orientation.  

We will use $C^n(\cM^{d+1};\M)$ to denote the set of all
$n$-cochains on $\cM^{d+1}$.  $C^n(\cM^{d+1};\M)$ can also be
viewed as a set all $\M$-values fields (or paths) on  $\cM^{d+1}$.  Note
that $C^n(\cM^{d+1};\M)$ is an abelian group under the $+$-operation.

The total space-time lattice $\cM^{d+1}$ correspond to a $(d+1)$-chain.  We
will use the same $\cM^{d+1}$ to denote it.  Viewing $f_{d+1}$ as a linear
map of $(d+1)$-chains, we can define an ``integral'' over $\cM^{d+1}$:
\begin{align}
 \int_{\cM^{d+1}} f_{d+1} \equiv \<f_{d+1},\cM^{d+1}\>.
\end{align}

\begin{figure}[tb]
\begin{center}
\includegraphics[scale=0.5]{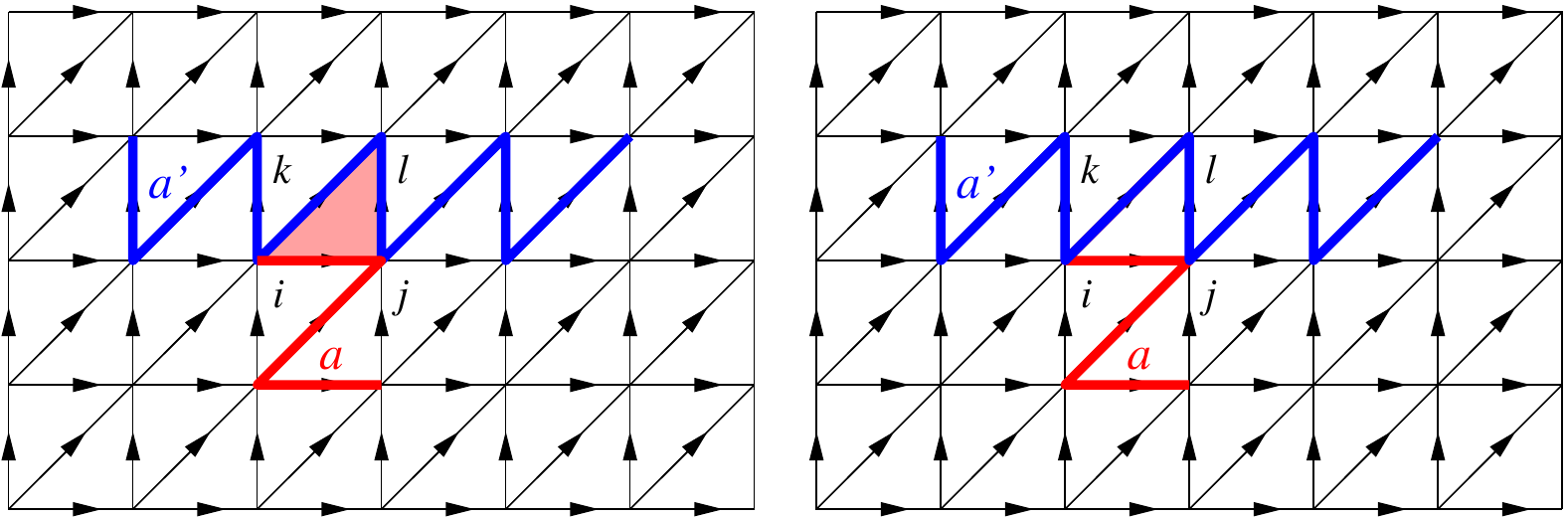} \end{center}
\caption{ (Color online)
A 1-cochain $a$ has a value $1$ on the red links, Another
1-cochain $a'$ has a value $1$ on the blue links.
On the left, $a\smile a'$ is non-zero on the shade triangles:
$(a\smile a')_{ijl}=a_{ij}a'_{jl}=1$.
On the right, $a'\smile a$ is zero on every triangle.
Thus $a\smile a'+a'\smile a$ is not a coboundary.
}
\label{cupcom}
\end{figure}

We can define a derivative operator $\dd$ acting on an $n$-cochain $f_n$, which
give us an $n+1$-cochain (see Fig. \ref{dcochain}):
\begin{align}
&\ \ \ \ \<\dd f_n, (i_0i_1i_2\cdots i_{n+1})\>
\nonumber\\
&=\sum_{m=0}^{n+1} (-)^m 
\<f_n, (i_0i_1i_2\cdots\hat i_m\cdots i_{n+1})\>
\end{align}
where $i_0i_1i_2\cdots \hat i_m \cdots i_{n+1}$ is the sequence
$i_0 i_1 i_2 \cdots i_{n+1}$ with $i_m$ removed, and
$i_0, i_1,i_2 \cdots i_{n+1}$ are the ordered vertices of the $(n+1)$-simplex
$(i_0 i_1 i_2 \cdots i_{n+1})$.

A cochain $f_n \in C^n(\cM^{d+1};\M)$ is called a \emph{cocycle} if $\dd
f_n=0$.   The set of cocycles is denoted as $Z^n(\cM^{d+1};\M)$.  A
cochain $f_n$ is called a \emph{coboundary} if there exist a cochain $f_{n-1}$
such that $\dd f_{n-1}=f_n$.  The set of coboundaries is denoted as
$B^n(\cM^{d+1};\M)$.  Both $Z^n(\cM^{d+1};\M)$ and
$B^n(\cM^{d+1};\M)$ are abelian groups as well.  Since $\dd^2=0$, a
coboundary is always a cocycle: $B^n(\cM^{d+1};\M) \subset
Z^n(\cM^{d+1};\M)$.  We may view two  cocycles differ by a coboundary as
equivalent.  The equivalence classes of cocycles, $[f_n]$, form the so called
cohomology group denoted as
\begin{align}
H^n(\cM^{d+1};\M)=   Z^n(\cM^{d+1};\M)/ B^n(\cM^{d+1};\M),
\end{align}
$H^n(\cM^{d+1};\M)$, as a group quotient of $Z^n(\cM^{d+1};\M)$ by
$B^n(\cM^{d+1};\M)$, is also an abelian group.

For the $\Z_N$-valued cocycle $x_n$, $\dd x_n \se{N} 0$. Thus 
\begin{align}
 \Bs_N x_n \equiv \frac1N \dd x_n 
\end{align}
is a $\Z$-valued cocycle. Here $\Bs_N$ is Bockstrin homomorphism.

From two cochains $f_m$  and $h_n$, we can construct a third cochain
$p_{m+n}$ via the cup product (see Fig. \ref{cupcom}):
\begin{align}
p_{m+n} &= f_m \smile h_n ,
\nonumber\\
\<p_{m+n}, (0 \to {m+n})\> 
&= 
\<f_m, (0 \to m)\> \times
\nonumber\\
&\ \ \ \ 
\<h_n,(m \to {m+n}) \>,
\end{align}
where $i\to j$ is a consecutive sequence from $i$ to $j$: 
\begin{align}
i\to j\equiv i,i+1,\cdots,j-1,j. 
\end{align}

The cup product has the following property 
\begin{align}
\label{cupprop}
 \dd(f_m \smile h_n) &= 
(\dd h_n) \smile f_m 
+ (-)^n h_n \smile (\dd f_m) 
\end{align}
We see that $f_m \smile h_n$ is a cocycle if both $f_m$ and $h_n$ are cocycles.
If both $f_m$ and $h_n$ are  cocycles, then $f_m \smile h_n$ is a coboundary if
one of $f_m$ and $h_n$ is a coboundary.  So the cup product is also an
operation on cohomology groups $\hcup{} : H^m(M^d;\M)\times H^n(M^d;\M) \to
H^{m+n}(M^d;\M)$.  The cup product of two \emph{cocycles} has the following
property (see Fig. \ref{cupcom}) 
\begin{align}
 f_m \smile h_n &= (-)^{mn} h_n \smile f_m + \text{coboundary}
\end{align}

We can also define higher cup product $f_m \hcup{k}
h_n$ which gives rise to a $(m+n-k)$-cochain \cite{S4790}:
\begin{align}
&\ \ \ \
 \<f_m \hcup{k} h_n, (0,1,\cdots,m+n-k)\> 
\nonumber\\
&
 = \hskip -1em \sum_{0\leq i_0<\cdots< i_k \leq n+m-k} \hskip -2em  (-)^p
\<f_m,(0 \to i_0, i_1\to i_2, \cdots)\>\times
\nonumber\\
&
\ \ \ \ \ \ \ \ \ \
\ \ \ \ \ \ \ \ \ \
\<h_n,(i_0\to i_1, i_2\to i_3, \cdots)\>,
\end{align} 
and $f_m \hcup{k} h_n =0$ for  $k>m \text{ or } n$ or $k<0$.
Here
$i\to j$ is the sequence $i,i+1,\cdots,j-1,j$, and
$p$ is the number of permutations to bring the sequence
\begin{align}
 0 \to i_0, i_1\to i_2, \cdots; i_0+1\to i_1-1, i_2+1\to i_3-1,\cdots
\end{align}
to the sequence
\begin{align}
 0 \to m+n-k.
\end{align}
For example
\begin{align}
&
 \<f_m \hcup1 h_n, (0,1,\cdots,m+n-1)\> 
 = \sum_{i=0}^{m-1} (-)^{(m-i)(n+1)}\times
\nonumber\\
&
\<f_m,(0 \to i, i+n\to m+n-1)\>
\<h_n,(i\to i+n)\>.
\end{align} 
We can see that $\hcup0 =\smile$.  
Unlike cup product at $k=0$, the higher cup product of two
cocycles may not be a cocycle. For cochains $f_m, h_n$, we have
\begin{align}
\label{cupkrel}
& \dd( f_m \hcup{k} h_n)=
\dd f_m \hcup{k} h_n +
(-)^{m} f_m \hcup{k} \dd h_n+
\\
& \ \ \
(-)^{m+n-k} f_m \hcup{k-1} h_n +
(-)^{mn+m+n} h_n \hcup{k-1} f_m 
\nonumber 
\end{align}

Let $f_m$ and $h_n$ be cocycles and $c_l$ be a chain, from \eqn{cupkrel} we
can obtain
\begin{align}
\label{cupkrel1}
 & \dd (f_m \hcup{k} h_n) = (-)^{m+n-k} f_m \hcup{k-1} h_n 
\nonumber\\
&
\ \ \ \ \ \ \ \ \ \
 \ \ \ \ \ \ \
+ (-)^{mn+m+n}  h_n \hcup{k-1} f_m,
\nonumber\\
 & \dd (f_m \hcup{k} f_m) = [(-)^k+(-)^m] f_m \hcup{k-1} f_m,
\nonumber\\
& \dd (c_l\hcup{k-1} c_l + c_l\hcup{k} \dd c_l)
= \dd c_l\hcup{k} \dd c_l 
\nonumber\\
&\ \ \ -[(-)^k-(-)^l]
(c_l\hcup{k-2} c_l + c_l\hcup{k-1} \dd c_l) .
\end{align}

From \eqn{cupkrel1}, we see that, for $\Z_2$-valued cocycles $z_n$,
\begin{align}
 \Sq^{n-k}(z_n) \equiv z_n\hcup{k} z_n
\end{align}
is always a cocycle.  Here $\Sq$ is called the Steenrod square.  More generally
$h_n \hcup{k} h_n$ is a cocycle if $n+k =$ odd and $h_n$ is a cocycle.
Usually, the Steenrod square is defined only for $\Z_2$ valued cocycles or
cohomology classes.  Here, we like to define Steenrod square for $\M$-valued
cochains $c_n$:
\begin{align}
\label{Sqdef}
 \Sq^{n-k} c_n \equiv c_n\hcup{k} c_n +  c_n\hcup{k+1} \dd c_n .
\end{align}
From \eqn{cupkrel1}, we see that
\begin{align}
\label{Sqd1}
 \dd \Sq^{k} c_n &= \dd(
c_n\hcup{n-k} c_n +  c_n\hcup{n-k+1} \dd c_n )
\\
&= \Sq^k \dd c_n +(-)^{n}
\begin{cases}
0, & k=\text{odd} \\ 
2  \Sq^{k+1} c_n  & k=\text{even} \\ 
\end{cases}
.
\nonumber 
\end{align}
In particular, when $c_n$ is a $\Z_2$-valued cochain, we have
\begin{align}
\label{Sqd}
  \dd \Sq^{k} c_n \se{2} \Sq^k \dd c_n.
\end{align}

Next, let us consider the action of $\Sq^k$ on the sum of two
 $\M$-valued cochains $c_n$ and $c_n'$:
\begin{align}
\label{Sqplus1}
& \Sq^{k} (c_n+c_n')
 = \Sq^{k} c_n + \Sq^k c_n' +
\nonumber\\
&\ \ \
 c_n \hcup{n-k} c_n' + c_n' \hcup{n-k} c_n 
+ c_n \hcup{n-k+1} \dd c_n' + c_n' \hcup{n-k+1} \dd c_n 
\nonumber\\
&=\Sq^{k} c_n + \Sq^k c_n' 
+[1 + (-)^k]c_n \hcup{n-k} c_n'
\nonumber\\
&\ \ \
-(-)^{n-k} [ - (-)^{n-k} c_n' \hcup{n-k} c_n + (-)^n c_n \hcup{n-k} c_n']
\nonumber\\
&\ \ \
+ c_n \hcup{n-k+1} \dd c_n' + c_n' \hcup{n-k+1} \dd c_n
\nonumber\\
& = 
\Sq^{k} c_n + \Sq^k c_n' 
+[1 + (-)^k]c_n \hcup{n-k} c_n'
\nonumber\\
&
+(-)^{n-k} [ \dd c_n' \hcup{n-k+1} c_n +(-)^n c_n' \hcup{n-k+1} \dd c_n
]
\nonumber\\
&
-(-)^{n-k} 
\dd (c_n'\hcup{n-k+1}c_n) 
+c_n \hcup{n-k+1} \dd c_n'+ c_n' \hcup{n-k+1} \dd c_n
\nonumber\\
&=
\Sq^{k} c_n + \Sq^k c_n'  
+[1 + (-)^k]c_n \hcup{n-k} c_n'
\nonumber \\
&\  \ \
+[1+(-)^{k}]c_n' \hcup{n-k+1} \dd c_n 
-(-)^{n-k} \dd (c_n'\hcup{n-k+1}c_n)
\nonumber\\
&\ \ \
-[(-)^{n-k+1}\dd c_n' \hcup{n-k+1} c_n
- c_n \hcup{n-k+1} \dd c_n']
\nonumber\\
&=
\Sq^{k} c_n + \Sq^k c_n'  
+[1 + (-)^k]c_n \hcup{n-k} c_n'
\nonumber \\
&\  \ \
+[1+(-)^{k}]c_n' \hcup{n-k+1} \dd c_n 
-(-)^{n-k} \dd (c_n'\hcup{n-k+1}c_n)
\nonumber\\
&\ \ \
-\dd (\dd c_n'\hcup{n-k+2} c_n )
+ \dd c_n'\hcup{n-k+2} \dd c_n 
\nonumber\\
&=
\Sq^{k} c_n + \Sq^k c_n'  
+ \dd c_n'\hcup{n-k+2} \dd c_n 
\nonumber \\
&\ \ \
+[1+(-)^{k}][c_n \hcup{n-k} c_n'+ c_n' \hcup{n-k+1} \dd c_n] 
\nonumber\\
&\ \ \
-(-)^{n-k} \dd (c_n'\hcup{n-k+1}c_n)
-\dd (\dd c_n'\hcup{n-k+2} c_n )
.
\end{align}
We see that, if one of the $c_n$ and $c_n'$ is a cocycle,
\begin{align}
\label{Sqplus}
  \Sq^{k} (c_n+c_n') \se{2,\dd} \Sq^{k} c_n + \Sq^k c_n' .
\end{align}
We also see that
\begin{align}
\label{Sqgauge}
&\ \ \ \
 \Sq^{k} (c_n+\dd f_{n-1})
\\
& = \Sq^{k} c_n + \Sq^k \dd f_{n-1} +
[1+(-)^k] \dd f_{n-1}\hcup{n-k} c_n
\nonumber\\
&\ \ \
-(-)^{n-k} \dd (c_n\hcup{n-k+1}\dd f_{n-1})
-\dd (\dd c_n\hcup{n-k+2} \dd f_{n-1} )
\nonumber\\
& = \Sq^{k} c_n 
+ [1+(-)^k] [\dd f_{n-1}\hcup{n-k} c_n +(-)^n \Sq^{k+1}f_{n-1}]
\nonumber\\
&
+\dd [\Sq^k  f_{n-1}
-(-)^{n-k} c_n \hskip -0.5em \hcup{n-k+1} \hskip -0.5em \dd f_{n-1}
-\dd c_n \hskip -0.5em \hcup{n-k+2}  \hskip -0.5em \dd f_{n-1} ]
.
\nonumber 
\end{align}
Using \eqn{Sqplus2}, we can also obtain the following result
if $\dd c_n = $ even
\begin{align}
\label{Sqplus2}
& \ \ \ \
 \Sq^k (c_n+2c_n')
\nonumber\\
& \se{4} \Sq^k c_n+2 \dd (c_n\hcup{n-k+1} c_n') +2 \dd c_n\hcup{n-k+1} c_n'
\nonumber\\
& \se{4} \Sq^k c_n+2 \dd (c_n\hcup{n-k+1} c_n') 
\end{align}

As another application, we note that, for a $\Z_2$ cochain $m_d$ and using
\eqn{cupkrel},
\begin{align}
\label{Sq1Bs}
& \Sq^1(m_{d}) = m_{d}\hcup{d-1} m_{d} + m_{d}\hcup{d} \dd m_{d}
\nonumber\\
&=\frac12 (-)^{d} 
[\dd (m_{d}\hcup{d} m_{d}) 
-\dd m_{d} \hcup{d} m_{d}] 
+\frac12  m_{d} \hcup{d} \dd m_{d} 
\nonumber\\
&=
(-)^{d} \Bs_2 (m_{d}\hcup{d} m_{d}) -(-)^d \Bs_2 m_{d} \hcup{d} m_{d}
+  m_{d} \hcup{d} \Bs_2 m_{d}
\nonumber\\
&=
(-)^{d} \Bs_2 m_{d} 
-2 (-)^d \Bs_2 m_{d} \hcup{d+1} \Bs_2 m_{d}
\nonumber\\
&=
(-)^{d} \Bs_2 m_{d} 
-2 (-)^d \Sq^1 \Bs_2 m_{d} 
\end{align}
where we have used $m_{d} \hcup{d} m_{d}=m_d$.
This way, we obtain a relation between
Steenrod square and Bockstein homomorphism, when
$m_d$ is a $\Z_2$ valued cocycle
\begin{align}
\label{Sq1Bs2}
  \Sq^1(m_{d}) \se{2} \Bs_2 m_{d} .
\end{align}

\section{Lyndon-Hochschild-Serre spectral sequence}
\label{LHS}

The Lyndon-Hochschild-Serre spectral sequence (see \Ref{L4871} page 280,291,
and \Ref{HS5310}) allows us to understand the structure of of the cohomology of
a fiber bundle $F\to X\to B$, $H^*(X;\RZ)$, from $H^*(F;\RZ)$ and
$H^*(B;\RZ)$.  In general, $H^d(X;\M)$, when viewed as an Abelian group,
contains a chain of subgroups
\begin{align}
\label{Lyndon}
\{0\}=H_{d+1}
\subset H_d
\subset \cdots
\subset H_0
=
 H^d(X;\M)
\end{align}
such that $H_l/H_{l+1}$ is a subgroup of a factor
group of $H^l[B,H^{d-l}(F;\M)_{B}]$,
\ie $H^l[B,H^{d-l}(F;\M)_{B}]$
contains a   subgroup $\Ga^k$, such that
\begin{align}
 H_l/H_{l+1} &\subset H^l[B,H^{d-l}(F;\M)_{B}]/\Ga^l,
\nonumber\\
l&=0,\cdots,d.
\end{align}
Note that $\pi_1(B)$  may have a non-trivial action on $\M$ and $\pi_1(B)$ may
have a non-trivial action on $H^{d-l}(F;\M)$ as determined by the structure
$F \to X \to B$.  We add the subscript $B$ to $H^{d-l}(F;\M)$ to
indicate this action.  We also have
\begin{align}
 H_0/H_{1} &\subset H^0[B,H^{d}(F;\M)_{B}],
\nonumber\\
 H_d/H_{d+1}&=H_d = H^d(B;\M)/\Ga^d.
\end{align}
In other words, all the elements in $H^d(X;\M)$ can be one-to-one
labeled by $(x_0,x_1,\cdots,x_d)$ with
\begin{align}
 x_l\in H_l/H_{l+1} \subset H^l[B,H^{d-l}(F;\M)_{B}]/\Ga^l.
\end{align}

Note that here $\M$ can be $\Z,\Z_n,\R,\R/\Z$ \etc.  Let $x_{l,\al}$,
$\al=1,2,\cdots$, be the generators of $H^l/H^{l+1}$. Then we say $x_{i,\al}$
for all $l,\al$ are the generators of $H^d(X;\M)$.  We also call
$H_l/H_{l+1}$, $l=0,\cdots,d$, the generating sub-factor groups of
$H^d(X;\M)$.

The above result implies that we can use $(k_0,k_1,\cdots,k_d)$ with $ k_l\in
H^l[B,H^{d-l}(F;\R/\Z)_{B}] $ to  label all the elements in
$H^d(X;\R/\Z)$. However, such a labeling scheme may not be one-to-one, and it
may happen that only some of $(k_0,k_1,\cdots,k_d)$ correspond to  the
elements in $H^d(X;\R/\Z)$.  But, on the other hand, for every element in
$H^d(X;\R/\Z)$, we can find a $(k_0,k_1,\cdots,k_d)$ that corresponds to it.

For the special case $X=B \times F$, $(k_0,k_1,\cdots,k_d)$ will give us
a one-to-one labeling of the elements in $H^d(B\times F;\R/\Z)$. In fact
\begin{align}
\label{Kunn}
H^d(B & \times F;\R/\Z) = \bigoplus_{l=0}^{d} H^{l}[B,H^{d-l}(F;\R/\Z)]
.
\end{align}

\section{Partition functions for 3+1D pure 2-gauge theory}
\label{Z2gauge}

In this section, we compute the partition function
for the pure 2-gauge theory \eq{Zb4m}
with $n=$ even and $m=$ odd.
Let $C^d(\cM;\M)$ be the
set of $\M$-valued $(d+1)$-cochains on the complex $\cM$, $Z^d(\cM;\M)$ the set of
$(d+1)$-cocycles, and $B^d(\cM;\M)$ the set of $(d+1)$-coboundaries.  When $m=0$, the
partition function is given by the number of $\Z_n$-valued 2-cocycles
$|Z^2(\cM^4;\Z_n)|$, which is $|H^2(\cM^4;\Z_n)|$ times the number of
1-cochains whose derivatives is non-zero.  The number of 1-cochains whose
derivatives is non-zero is the number of 1-cochains
($|C^1(\cM^4;\Z_n)|=n^{N_e}$) divide by $|H^1(\cM^4;\Z_n)|$ and by the number
of number of 0-cochains whose derivatives is non-zero.  The number of
0-cochains whose derivatives is non-zero is the number of 0-cochains
($|C^0(\cM^4;\Z_n)|=n^{N_v}$) divide by $|H^0(\cM^4;\Z_n)|$.  Thus the
partition function is
\begin{align}
&\ \ \ \ Z (\cM^4;\cB(\Z_n,2),0) =  |Z^2(\cM^4;\Z_n)|
\nonumber\\
&= |H^2(\cM^4;\Z_n)| 
\frac{|C^1(\cM^4;\Z_n)|}{|H^1(\cM^4;\Z_n)|} \frac{|H^0(\cM^4;\Z_n)|}{|C^0(\cM^4;\Z_n)|}
\nonumber\\
&= n^{N_e-N_v}
 \frac{|H^2(\cM^4;\Z_n)||H^0(\cM^4;\Z_n)|}{|H^1(\cM^4;\Z_n)|} .
\end{align}
where $N_v$ is the number of vertices and $N_e$ the number of links.
The volume-independent topological partition function is given by
\begin{align}
 Z^\text{top} (\cM^4;\cB(\Z_n,2),0) = 
 \frac{|H^2(\cM^4;\Z_n)||H^0(\cM^4;\Z_n)|}{|H^1(\cM^4;\Z_n)|}
\end{align}
When $m\neq 0$,
The volume-independent topological partition function is given by
\begin{align}
&\ \ \ \
 Z^\text{top} (\cM^4;\cB(\Z_n,2),0) 
\\ & = 
 \frac{|H^0(\cM^4;\Z_n)|}{|H^1(\cM^4;\Z_n)|} 
\sum_{b \in H^2(M^4;\Z_n)}
\ee^{\ii 2\pi \int_{\cM^4} \frac{m}{2n} b^2+\frac{m}2 b\hcup{1}\Bs b }
\nonumber 
\end{align}
where $\sum_{b \in H^2(M^4;\Z_n)} \ee^{\ii 2\pi \int_{\cM^4} 
 \frac{m}{2n} b^2+\frac{m}2 b\hcup{1}\Bs b } $ replaces $|H^2(M^4;\Z_n)|$.

Now, let us compute topological invariants.  On $\cM^4=T^4$, the cohomology
ring $H^*(T^4;\Z_n)$ is generated by $a_I,\ I=1,2,3,4$, where $a_I \in
H^1(T^4;\Z_n)=4\Z_n$.  Using the cohomology ring discussed in \Ref{W161201418},
we can parametrize $b^{\Z_n}$ as
\begin{align}
 b =\al_{IJ} a_Ia_J,\ \ \ \al_{IJ}=-\al_{JI} 
\in \Z_n.
\end{align}
We also have $\Bs b\se{n} 0$.
Thus
\begin{align}
&\ \ \ \
 Z(T^4;\cB(\Z_n,2),m) 
\\ & = 
 \frac{1}{n^3} 
\sum_{\al_{IJ} \in \Z_n}
\ee^{\ii 2\pi \frac{m}{2n} 2 
(
\al_{12}\al_{34} -\al_{13}\al_{24}+ \al_{14}\al_{23}
)
}
\nonumber 
\end{align}
Using $\sum_{\al_1,\al_2\in \Z_n} \ee^{\ii 2\pi \frac{m}{n} \al_1\al_2}=
\<m,n\> n$,
we find that
\begin{align}
 Z^\text{top} (T^4;\cB(\Z_n,2),m)=\<m,n\>^3.
\end{align}

On $\cM^4=S^2\times T^2$, the cohomology ring $H^*(T^2\times S^2;\Z_n)$ is
generated by $a_I,\ I=1,2$ and $b$, where 
$a_I \in H^1(T^2\times S^2;\Z_n)=\Z_n^{\oplus 2}$ and
$b_0 \in H^2(T^2\times S^2;\Z_n)=\Z_n^{\oplus 2}$.
Using the cohomology ring discussed in \Ref{W161201418}, we can
parametrize $b$ as
\begin{align}
 b =\al_1 a_1a_2 +\al_2 b_0 , \ \ \ \al_1,\al_2 \in \Z_n.
\end{align}
Thus
\begin{align}
\label{ZnbT2S2}
&\ \ \ \
 Z^\text{top} (S^2\times T^2;\cB(\Z_n,2),m) 
\\ & = 
 \frac{1}{n} 
\sum_{\al_1,\al_2 \in \Z_n}
\ee^{\ii 2\pi \frac{m}{2n} 2\al_1\al_2}
 =\<m,n\>.
\nonumber 
\end{align}

On $M^4=S^1\times L^3(p)$, we need to use the cohomology ring $H^*(S^1\times
L^3(p);\Z_n)$ calculated in \Ref{W161201418}:
\begin{align}
\label{S1LpqH}
 H^1(S^1\times L^3(p), \Z_n) &= \Z_n\oplus\Z_{\<p,n\>}=\{a_1,a_0\},  
\nonumber\\
 H^2(S^1\times L^3(p), \Z_n) &= \Z_{\<p,n\>}\oplus\Z_{\<p,n\>}=\{a_1a_0,b_0\},  
\nonumber\\
 H^3(S^1\times L^3(p), \Z_n) &= \Z_n\oplus \Z_{\<p,n\>}=\{c_0,a_1b_0\},  
\nonumber\\
 H^4(S^1\times L^3(p), \Z_n) &= \Z_n =\{a_1c_0\}.
\end{align}
where we have also listed the generators. Here $a_1$ comes from $S^1$ and
$a_0,b_0,c_0$ from $L^3(p)$.  
The cohomology ring $H^*(S^1\times L^3(p), \Z_n)$ is given by:
\begin{align}
\label{S1Lpqcup}
a_1^2 &=0,& a_0^2 &= \frac{n^2p(p-1)}{2\<p,n\>^2} b_0,   
\nonumber\\
a_0b_0 &= \frac{n}{\<p,n\>} c_0 , &  b_0^2&=a_0c_0=0.
\end{align}
For $\<n,p\>=1$, $ Z^\text{top} (S^1\times L^3(p);\cB(\Z_n,2),m )=1$.  For
$\<n,p\>\neq 1$, we can parametrize $b$ as
\begin{align}
 b =\al_1 a_0 a_1 +\al_2 b_0  , \ \ \ \al_1,\al_2 \in \Z_{\<n,p\>},
\end{align}
which satisfies $\Bs b = 0$ (see \Ref{W161201418}).
Using $a_0 a_1b_0 =\frac n{\<n,p\>} a_1c_0 $ and $(a_0 a_1)^2=b_0 ^2=0$,
we find that 
\begin{align}
&\ \ \ \
 Z^\text{top} (S^1\times L^3(p);\cB(\Z_n,2),m ) 
\\ & = 
 \frac{1}{\<n,p\>} 
\sum_{\al_1,\al_2 =0}^{\<n,p\>-1}
\ee^{\ii 2\pi \frac{m}{\<n,p\>} \al_1\al_2}
 =\<m,n,p\>.
\nonumber 
\end{align}

On $\cM^4=F^4$, we need to use the cohomology ring $H^*(F^4;\Z_n)$ as described
in \Ref{W161201418}:
\begin{align}
 H^1(F^4;\Z_n)&=\Z_n^{\oplus 2},\ \ 
 H^2(F^4;\Z_n)=\Z_n^{\oplus 2},\ \ 
\nonumber\\
 H^3(F^4;\Z_n)&=\Z_n^{\oplus 2},\ \ 
 H^4(F^4;\Z_n)=\Z_n .
\end{align}
Let $a_1,a_2$ be the
generators of $H^1(F^4;\Z_n)$, $b_1,b_2$ the generators of $H^2(F^4;\Z_n)$,
$c_1,c_2$ be the generators of $H^3(F^4;\Z_n)$, and $v$ be the generator of
$H^4(F^4;\Z_n)$: 
\begin{align}
 H^*(F^4;\Z_n)=\{ a_1,a_2, b_1,b_2, c_1,c_2, v \}.
\end{align}
We find that the non-zero cup products are given by
\begin{align}
 b_1^2=-b_2^2=a_1c_1=a_2c_2=v.
\end{align}
All other cup products vanish.

We can parametrize $b$ as
\begin{align}
 b =\al_1 b_1 +\al_2 b_2 , \ \ \ \al_1,\al_2 \in \Z_n,
\end{align}
where $b_1,b_2$ are generators of $H^2(F^4;\Z_n)$.  Using $b_1^2=-b_2^2=v$,
$b_1b_2=0$, and $\Bs b_1  = \Bs b_2 = 0$ , we find that 
\begin{align} 
\label{ZZnbF4}
 Z^\text{top} (F^4;\cB(\Z_n,2),m) 
 & = 
 \frac{1}{n} 
\sum_{\al_1,\al_2 =0}^{n-1}
\ee^{\ii 2\pi \frac{m}{2n}(\al_1^2-\al_2^2)}
\\
 &= 
\begin{cases}
\<m,n\>, &\text{ if } \frac{mn}{\<m,n\>^2}= \text{ even};\\
0,        &\text{ if } \frac{mn}{\<m,n\>^2}= \text{ odd}.\\
\end{cases}
\nonumber 
\end{align}
The above results, plus some previous results from \Ref{W161201418}, are
summarized in Table \ref{tab:topinv}.

\bibliography{../../bib/wencross,../../bib/all,../../bib/publst} 

\begin{thebibliography}{54}%
\makeatletter
\providecommand \@ifxundefined [1]{%
 \@ifx{#1\undefined}
}%
\providecommand \@ifnum [1]{%
 \ifnum #1\expandafter \@firstoftwo
 \else \expandafter \@secondoftwo
 \fi
}%
\providecommand \@ifx [1]{%
 \ifx #1\expandafter \@firstoftwo
 \else \expandafter \@secondoftwo
 \fi
}%
\providecommand \natexlab [1]{#1}%
\providecommand \enquote  [1]{``#1''}%
\providecommand \bibnamefont  [1]{#1}%
\providecommand \bibfnamefont [1]{#1}%
\providecommand \citenamefont [1]{#1}%
\providecommand \href@noop [0]{\@secondoftwo}%
\providecommand \href [0]{\begingroup \@sanitize@url \@href}%
\providecommand \@href[1]{\@@startlink{#1}\@@href}%
\providecommand \@@href[1]{\endgroup#1\@@endlink}%
\providecommand \@sanitize@url [0]{\catcode `\\12\catcode `\$12\catcode
  `\&12\catcode `\#12\catcode `\^12\catcode `\_12\catcode `\%12\relax}%
\providecommand \@@startlink[1]{}%
\providecommand \@@endlink[0]{}%
\providecommand \url  [0]{\begingroup\@sanitize@url \@url }%
\providecommand \@url [1]{\endgroup\@href {#1}{\urlprefix }}%
\providecommand \urlprefix  [0]{URL }%
\providecommand \Eprint [0]{\href }%
\providecommand \doibase [0]{http://dx.doi.org/}%
\providecommand \selectlanguage [0]{\@gobble}%
\providecommand \bibinfo  [0]{\@secondoftwo}%
\providecommand \bibfield  [0]{\@secondoftwo}%
\providecommand \translation [1]{[#1]}%
\providecommand \BibitemOpen [0]{}%
\providecommand \bibitemStop [0]{}%
\providecommand \bibitemNoStop [0]{.\EOS\space}%
\providecommand \EOS [0]{\spacefactor3000\relax}%
\providecommand \BibitemShut  [1]{\csname bibitem#1\endcsname}%
\let\auto@bib@innerbib\@empty
\bibitem [{\citenamefont {Kane}\ and\ \citenamefont {Mele}(2005)}]{KM0502}%
  \BibitemOpen
  \bibfield  {author} {\bibinfo {author} {\bibfnamefont {C.~L.}\ \bibnamefont
  {Kane}}\ and\ \bibinfo {author} {\bibfnamefont {E.~J.}\ \bibnamefont
  {Mele}},\ }\href@noop {} {\bibfield  {journal} {\bibinfo  {journal} {Phys.
  Rev. Lett.}\ }\textbf {\bibinfo {volume} {95}},\ \bibinfo {pages} {146802}
  (\bibinfo {year} {2005})},\ \Eprint {http://arxiv.org/abs/cond-mat/0506581}
  {cond-mat/0506581} \BibitemShut {NoStop}%
\bibitem [{\citenamefont {{Bernevig}}\ \emph {et~al.}(2006)\citenamefont
  {{Bernevig}}, \citenamefont {{Hughes}},\ and\ \citenamefont
  {{Zhang}}}]{BZ0611399}%
  \BibitemOpen
  \bibfield  {author} {\bibinfo {author} {\bibfnamefont {B.~A.}\ \bibnamefont
  {{Bernevig}}}, \bibinfo {author} {\bibfnamefont {T.~L.}\ \bibnamefont
  {{Hughes}}}, \ and\ \bibinfo {author} {\bibfnamefont {S.-C.}\ \bibnamefont
  {{Zhang}}},\ }\href {\doibase 10.1126/science.1133734} {\bibfield  {journal}
  {\bibinfo  {journal} {Science}\ }\textbf {\bibinfo {volume} {314}},\ \bibinfo
  {pages} {1757} (\bibinfo {year} {2006})},\ \Eprint
  {http://arxiv.org/abs/cond-mat/0611399} {cond-mat/0611399} \BibitemShut
  {NoStop}%
\bibitem [{\citenamefont {Moore}\ and\ \citenamefont {Balents}(2007)}]{MB0706}%
  \BibitemOpen
  \bibfield  {author} {\bibinfo {author} {\bibfnamefont {J.~E.}\ \bibnamefont
  {Moore}}\ and\ \bibinfo {author} {\bibfnamefont {L.}~\bibnamefont
  {Balents}},\ }\href@noop {} {\bibfield  {journal} {\bibinfo  {journal} {Phys.
  Rev. B}\ }\textbf {\bibinfo {volume} {75}},\ \bibinfo {pages} {121306}
  (\bibinfo {year} {2007})},\ \Eprint {http://arxiv.org/abs/cond-mat/0607314}
  {cond-mat/0607314} \BibitemShut {NoStop}%
\bibitem [{\citenamefont {Fu}\ \emph {et~al.}(2007)\citenamefont {Fu},
  \citenamefont {Kane},\ and\ \citenamefont {Mele}}]{FKM0703}%
  \BibitemOpen
  \bibfield  {author} {\bibinfo {author} {\bibfnamefont {L.}~\bibnamefont
  {Fu}}, \bibinfo {author} {\bibfnamefont {C.~L.}\ \bibnamefont {Kane}}, \ and\
  \bibinfo {author} {\bibfnamefont {E.~J.}\ \bibnamefont {Mele}},\ }\href@noop
  {} {\bibfield  {journal} {\bibinfo  {journal} {Phys. Rev. Lett.}\ }\textbf
  {\bibinfo {volume} {98}},\ \bibinfo {pages} {106803} (\bibinfo {year}
  {2007})},\ \Eprint {http://arxiv.org/abs/cond-mat/0607699} {cond-mat/0607699}
  \BibitemShut {NoStop}%
\bibitem [{\citenamefont {Qi}\ \emph {et~al.}(2008)\citenamefont {Qi},
  \citenamefont {Hughes},\ and\ \citenamefont {Zhang}}]{QHZ0824}%
  \BibitemOpen
  \bibfield  {author} {\bibinfo {author} {\bibfnamefont {X.-L.}\ \bibnamefont
  {Qi}}, \bibinfo {author} {\bibfnamefont {T.}~\bibnamefont {Hughes}}, \ and\
  \bibinfo {author} {\bibfnamefont {S.-C.}\ \bibnamefont {Zhang}},\ }\href@noop
  {} {\bibfield  {journal} {\bibinfo  {journal} {Phys. Rev. B}\ }\textbf
  {\bibinfo {volume} {78}},\ \bibinfo {pages} {195424} (\bibinfo {year}
  {2008})},\ \Eprint {http://arxiv.org/abs/arXiv:0802.3537} {arXiv:0802.3537}
  \BibitemShut {NoStop}%
\bibitem [{\citenamefont {Roy}(2006)}]{R0664}%
  \BibitemOpen
  \bibfield  {author} {\bibinfo {author} {\bibfnamefont {R.}~\bibnamefont
  {Roy}},\ }\href@noop {} {\  (\bibinfo {year} {2006})},\ \Eprint
  {http://arxiv.org/abs/cond-mat/0608064} {cond-mat/0608064} \BibitemShut
  {NoStop}%
\bibitem [{\citenamefont {Thouless}\ \emph {et~al.}(1982)\citenamefont
  {Thouless}, \citenamefont {Kohmoto}, \citenamefont {Nightingale},\ and\
  \citenamefont {den Nijs}}]{TKN8205}%
  \BibitemOpen
  \bibfield  {author} {\bibinfo {author} {\bibfnamefont {D.~J.}\ \bibnamefont
  {Thouless}}, \bibinfo {author} {\bibfnamefont {M.}~\bibnamefont {Kohmoto}},
  \bibinfo {author} {\bibfnamefont {M.~P.}\ \bibnamefont {Nightingale}}, \ and\
  \bibinfo {author} {\bibfnamefont {M.}~\bibnamefont {den Nijs}},\ }\href@noop
  {} {\bibfield  {journal} {\bibinfo  {journal} {Phys. Rev. Lett.}\ }\textbf
  {\bibinfo {volume} {49}},\ \bibinfo {pages} {405} (\bibinfo {year}
  {1982})}\BibitemShut {NoStop}%
\bibitem [{\citenamefont {Avron}\ \emph {et~al.}(1983)\citenamefont {Avron},
  \citenamefont {Seiler},\ and\ \citenamefont {Simon}}]{ASS8351}%
  \BibitemOpen
  \bibfield  {author} {\bibinfo {author} {\bibfnamefont {J.}~\bibnamefont
  {Avron}}, \bibinfo {author} {\bibfnamefont {R.}~\bibnamefont {Seiler}}, \
  and\ \bibinfo {author} {\bibfnamefont {B.}~\bibnamefont {Simon}},\
  }\href@noop {} {\bibfield  {journal} {\bibinfo  {journal} {Phys. Rev. Lett.}\
  }\textbf {\bibinfo {volume} {51}},\ \bibinfo {pages} {51} (\bibinfo {year}
  {1983})}\BibitemShut {NoStop}%
\bibitem [{\citenamefont {Avron}\ \emph {et~al.}(1988)\citenamefont {Avron},
  \citenamefont {Sadun}, \citenamefont {Segert},\ and\ \citenamefont
  {Simon}}]{ASS8829}%
  \BibitemOpen
  \bibfield  {author} {\bibinfo {author} {\bibfnamefont {J.~E.}\ \bibnamefont
  {Avron}}, \bibinfo {author} {\bibfnamefont {L.}~\bibnamefont {Sadun}},
  \bibinfo {author} {\bibfnamefont {J.}~\bibnamefont {Segert}}, \ and\ \bibinfo
  {author} {\bibfnamefont {B.}~\bibnamefont {Simon}},\ }\href@noop {}
  {\bibfield  {journal} {\bibinfo  {journal} {Phys. Rev. Lett.}\ }\textbf
  {\bibinfo {volume} {61}},\ \bibinfo {pages} {1329} (\bibinfo {year}
  {1988})}\BibitemShut {NoStop}%
\bibitem [{\citenamefont {Kitaev}(2009)}]{K0986}%
  \BibitemOpen
  \bibfield  {author} {\bibinfo {author} {\bibfnamefont {A.}~\bibnamefont
  {Kitaev}},\ }in\ \href@noop {} {\emph {\bibinfo {booktitle} {Advances in
  Theoretical Physics: Landau Memorial Conference, Chernogolovka, Russia,
  2008}}},\ Vol.\ \bibinfo {volume} {AIP Conf. Proc. No. 1134},\ \bibinfo
  {editor} {edited by\ \bibinfo {editor} {\bibfnamefont {V.}~\bibnamefont
  {Lebedev}}\ and\ \bibinfo {editor} {\bibfnamefont {M.}~\bibnamefont
  {Feigel’man}}}\ (\bibinfo  {publisher} {AIP},\ \bibinfo {address}
  {Melville, NY},\ \bibinfo {year} {2009})\ p.~\bibinfo {pages} {22},\ \Eprint
  {http://arxiv.org/abs/arXiv:0901.2686} {arXiv:0901.2686} \BibitemShut
  {NoStop}%
\bibitem [{\citenamefont {Wen}(1989)}]{Wtop}%
  \BibitemOpen
  \bibfield  {author} {\bibinfo {author} {\bibfnamefont {X.-G.}\ \bibnamefont
  {Wen}},\ }\href@noop {} {\bibfield  {journal} {\bibinfo  {journal} {Phys.
  Rev. B}\ }\textbf {\bibinfo {volume} {40}},\ \bibinfo {pages} {7387}
  (\bibinfo {year} {1989})}\BibitemShut {NoStop}%
\bibitem [{\citenamefont {Wen}(1990)}]{Wrig}%
  \BibitemOpen
  \bibfield  {author} {\bibinfo {author} {\bibfnamefont {X.-G.}\ \bibnamefont
  {Wen}},\ }\href@noop {} {\bibfield  {journal} {\bibinfo  {journal} {Int. J.
  Mod. Phys. B}\ }\textbf {\bibinfo {volume} {4}},\ \bibinfo {pages} {239}
  (\bibinfo {year} {1990})}\BibitemShut {NoStop}%
\bibitem [{\citenamefont {Wen}\ and\ \citenamefont {Niu}(1990)}]{WNtop}%
  \BibitemOpen
  \bibfield  {author} {\bibinfo {author} {\bibfnamefont {X.-G.}\ \bibnamefont
  {Wen}}\ and\ \bibinfo {author} {\bibfnamefont {Q.}~\bibnamefont {Niu}},\
  }\href@noop {} {\bibfield  {journal} {\bibinfo  {journal} {Phys. Rev. B}\
  }\textbf {\bibinfo {volume} {41}},\ \bibinfo {pages} {9377} (\bibinfo {year}
  {1990})}\BibitemShut {NoStop}%
\bibitem [{\citenamefont {Kitaev}\ and\ \citenamefont
  {Preskill}(2006)}]{KP0604}%
  \BibitemOpen
  \bibfield  {author} {\bibinfo {author} {\bibfnamefont {A.}~\bibnamefont
  {Kitaev}}\ and\ \bibinfo {author} {\bibfnamefont {J.}~\bibnamefont
  {Preskill}},\ }\href@noop {} {\bibfield  {journal} {\bibinfo  {journal}
  {Phys. Rev. Lett.}\ }\textbf {\bibinfo {volume} {96}},\ \bibinfo {pages}
  {110404} (\bibinfo {year} {2006})}\BibitemShut {NoStop}%
\bibitem [{\citenamefont {Levin}\ and\ \citenamefont {Wen}(2006)}]{LW0605}%
  \BibitemOpen
  \bibfield  {author} {\bibinfo {author} {\bibfnamefont {M.}~\bibnamefont
  {Levin}}\ and\ \bibinfo {author} {\bibfnamefont {X.-G.}\ \bibnamefont
  {Wen}},\ }\href@noop {} {\bibfield  {journal} {\bibinfo  {journal} {Phys.
  Rev. Lett.}\ }\textbf {\bibinfo {volume} {96}},\ \bibinfo {pages} {110405}
  (\bibinfo {year} {2006})},\ \Eprint {http://arxiv.org/abs/cond-mat/0510613}
  {cond-mat/0510613} \BibitemShut {NoStop}%
\bibitem [{\citenamefont {Chen}\ \emph {et~al.}(2010)\citenamefont {Chen},
  \citenamefont {Gu},\ and\ \citenamefont {Wen}}]{CGW1038}%
  \BibitemOpen
  \bibfield  {author} {\bibinfo {author} {\bibfnamefont {X.}~\bibnamefont
  {Chen}}, \bibinfo {author} {\bibfnamefont {Z.-C.}\ \bibnamefont {Gu}}, \ and\
  \bibinfo {author} {\bibfnamefont {X.-G.}\ \bibnamefont {Wen}},\ }\href@noop
  {} {\bibfield  {journal} {\bibinfo  {journal} {Phys. Rev. B}\ }\textbf
  {\bibinfo {volume} {82}},\ \bibinfo {pages} {155138} (\bibinfo {year}
  {2010})},\ \Eprint {http://arxiv.org/abs/arXiv:1004.3835} {arXiv:1004.3835}
  \BibitemShut {NoStop}%
\bibitem [{\citenamefont {Witten}(1989)}]{W8951}%
  \BibitemOpen
  \bibfield  {author} {\bibinfo {author} {\bibfnamefont {E.}~\bibnamefont
  {Witten}},\ }\href@noop {} {\bibfield  {journal} {\bibinfo  {journal} {Comm.
  Math. Phys.}\ }\textbf {\bibinfo {volume} {121}},\ \bibinfo {pages} {351}
  (\bibinfo {year} {1989})}\BibitemShut {NoStop}%
\bibitem [{\citenamefont {Levin}\ and\ \citenamefont {Wen}(2005)}]{LW0510}%
  \BibitemOpen
  \bibfield  {author} {\bibinfo {author} {\bibfnamefont {M.}~\bibnamefont
  {Levin}}\ and\ \bibinfo {author} {\bibfnamefont {X.-G.}\ \bibnamefont
  {Wen}},\ }\href@noop {} {\bibfield  {journal} {\bibinfo  {journal} {Phys.
  Rev. B}\ }\textbf {\bibinfo {volume} {71}},\ \bibinfo {pages} {045110}
  (\bibinfo {year} {2005})},\ \Eprint {http://arxiv.org/abs/cond-mat/0404617}
  {cond-mat/0404617} \BibitemShut {NoStop}%
\bibitem [{\citenamefont {Kitaev}(2006)}]{K062}%
  \BibitemOpen
  \bibfield  {author} {\bibinfo {author} {\bibfnamefont {A.}~\bibnamefont
  {Kitaev}},\ }\href@noop {} {\bibfield  {journal} {\bibinfo  {journal} {Annals
  of Physics}\ }\textbf {\bibinfo {volume} {321}},\ \bibinfo {pages} {2}
  (\bibinfo {year} {2006})},\ \Eprint {http://arxiv.org/abs/cond-mat/0506438}
  {cond-mat/0506438} \BibitemShut {NoStop}%
\bibitem [{\citenamefont {{Rowell}}\ \emph {et~al.}(2009)\citenamefont
  {{Rowell}}, \citenamefont {{Stong}},\ and\ \citenamefont {{Wang}}}]{RSW0777}%
  \BibitemOpen
  \bibfield  {author} {\bibinfo {author} {\bibfnamefont {E.}~\bibnamefont
  {{Rowell}}}, \bibinfo {author} {\bibfnamefont {R.}~\bibnamefont {{Stong}}}, \
  and\ \bibinfo {author} {\bibfnamefont {Z.}~\bibnamefont {{Wang}}},\
  }\href@noop {} {\bibfield  {journal} {\bibinfo  {journal} {Comm. Math.
  Phys.}\ }\textbf {\bibinfo {volume} {292}},\ \bibinfo {pages} {343} (\bibinfo
  {year} {2009})},\ \Eprint {http://arxiv.org/abs/arXiv:0712.1377}
  {arXiv:0712.1377} \BibitemShut {NoStop}%
\bibitem [{\citenamefont {Chen}\ \emph {et~al.}(2013)\citenamefont {Chen},
  \citenamefont {Gu}, \citenamefont {Liu},\ and\ \citenamefont
  {Wen}}]{CGL1314}%
  \BibitemOpen
  \bibfield  {author} {\bibinfo {author} {\bibfnamefont {X.}~\bibnamefont
  {Chen}}, \bibinfo {author} {\bibfnamefont {Z.-C.}\ \bibnamefont {Gu}},
  \bibinfo {author} {\bibfnamefont {Z.-X.}\ \bibnamefont {Liu}}, \ and\
  \bibinfo {author} {\bibfnamefont {X.-G.}\ \bibnamefont {Wen}},\ }\href@noop
  {} {\bibfield  {journal} {\bibinfo  {journal} {Phys. Rev. B}\ }\textbf
  {\bibinfo {volume} {87}},\ \bibinfo {pages} {155114} (\bibinfo {year}
  {2013})},\ \Eprint {http://arxiv.org/abs/arXiv:1106.4772} {arXiv:1106.4772}
  \BibitemShut {NoStop}%
\bibitem [{\citenamefont {{Barkeshli}}\ \emph {et~al.}(2014)\citenamefont
  {{Barkeshli}}, \citenamefont {{Bonderson}}, \citenamefont {{Cheng}},\ and\
  \citenamefont {{Wang}}}]{BBC1440}%
  \BibitemOpen
  \bibfield  {author} {\bibinfo {author} {\bibfnamefont {M.}~\bibnamefont
  {{Barkeshli}}}, \bibinfo {author} {\bibfnamefont {P.}~\bibnamefont
  {{Bonderson}}}, \bibinfo {author} {\bibfnamefont {M.}~\bibnamefont
  {{Cheng}}}, \ and\ \bibinfo {author} {\bibfnamefont {Z.}~\bibnamefont
  {{Wang}}},\ }\href@noop {} {\  (\bibinfo {year} {2014})},\ \Eprint
  {http://arxiv.org/abs/arXiv:1410.4540} {arXiv:1410.4540} \BibitemShut
  {NoStop}%
\bibitem [{\citenamefont {Kapustin}(2014)}]{K1467}%
  \BibitemOpen
  \bibfield  {author} {\bibinfo {author} {\bibfnamefont {A.}~\bibnamefont
  {Kapustin}},\ }\href@noop {} {\  (\bibinfo {year} {2014})},\ \Eprint
  {http://arxiv.org/abs/arXiv:1403.1467} {arXiv:1403.1467} \BibitemShut
  {NoStop}%
\bibitem [{\citenamefont {Lan}\ \emph {et~al.}(2016)\citenamefont {Lan},
  \citenamefont {Kong},\ and\ \citenamefont {Wen}}]{LW160205946}%
  \BibitemOpen
  \bibfield  {author} {\bibinfo {author} {\bibfnamefont {T.}~\bibnamefont
  {Lan}}, \bibinfo {author} {\bibfnamefont {L.}~\bibnamefont {Kong}}, \ and\
  \bibinfo {author} {\bibfnamefont {X.-G.}\ \bibnamefont {Wen}},\ }\href
  {\doibase 10.1103/PhysRevB.95.235140} {\bibfield  {journal} {\bibinfo
  {journal} {Phys. Rev. B}\ }\textbf {\bibinfo {volume} {95}},\ \bibinfo
  {pages} {235140} (\bibinfo {year} {2016})},\ \Eprint
  {http://arxiv.org/abs/arXiv:1602.05946} {arXiv:1602.05946} \BibitemShut
  {NoStop}%
\bibitem [{\citenamefont {Kong}\ \emph {et~al.}(2017)\citenamefont {Kong},
  \citenamefont {Wen},\ and\ \citenamefont {Zheng}}]{KW170200673}%
  \BibitemOpen
  \bibfield  {author} {\bibinfo {author} {\bibfnamefont {L.}~\bibnamefont
  {Kong}}, \bibinfo {author} {\bibfnamefont {X.-G.}\ \bibnamefont {Wen}}, \
  and\ \bibinfo {author} {\bibfnamefont {H.}~\bibnamefont {Zheng}},\ }\href
  {\doibase 10.1016/j.nuclphysb.2017.06.023} {\bibfield  {journal} {\bibinfo
  {journal} {Nucl. Phys. B}\ }\textbf {\bibinfo {volume} {922}},\ \bibinfo
  {pages} {62} (\bibinfo {year} {2017})},\ \Eprint
  {http://arxiv.org/abs/arXiv:1702.00673} {arXiv:1702.00673} \BibitemShut
  {NoStop}%
\bibitem [{\citenamefont {Kong}\ and\ \citenamefont {Wen}(2014)}]{KW1458}%
  \BibitemOpen
  \bibfield  {author} {\bibinfo {author} {\bibfnamefont {L.}~\bibnamefont
  {Kong}}\ and\ \bibinfo {author} {\bibfnamefont {X.-G.}\ \bibnamefont {Wen}},\
  }\href@noop {} {\  (\bibinfo {year} {2014})},\ \Eprint
  {http://arxiv.org/abs/arXiv:1405.5858} {arXiv:1405.5858} \BibitemShut
  {NoStop}%
\bibitem [{\citenamefont {{Wen}}\ and\ \citenamefont
  {{Wang}}(2018)}]{WW180109938}%
  \BibitemOpen
  \bibfield  {author} {\bibinfo {author} {\bibfnamefont {X.-G.}\ \bibnamefont
  {{Wen}}}\ and\ \bibinfo {author} {\bibfnamefont {Z.}~\bibnamefont {{Wang}}},\
  }\href@noop {} {\  (\bibinfo {year} {2018})},\ \Eprint
  {http://arxiv.org/abs/1801.09938} {arXiv:1801.09938} \BibitemShut {NoStop}%
\bibitem [{\citenamefont {{Lan}}\ and\ \citenamefont
  {{Wen}}(2018)}]{LW180108530}%
  \BibitemOpen
  \bibfield  {author} {\bibinfo {author} {\bibfnamefont {T.}~\bibnamefont
  {{Lan}}}\ and\ \bibinfo {author} {\bibfnamefont {X.-G.}\ \bibnamefont
  {{Wen}}},\ }\href@noop {} {\  (\bibinfo {year} {2018})},\ \Eprint
  {http://arxiv.org/abs/1801.08530} {arXiv:1801.08530} \BibitemShut {NoStop}%
\bibitem [{\citenamefont {Kapustin}\ and\ \citenamefont
  {Thorngren}(2013)}]{KT1321}%
  \BibitemOpen
  \bibfield  {author} {\bibinfo {author} {\bibfnamefont {A.}~\bibnamefont
  {Kapustin}}\ and\ \bibinfo {author} {\bibfnamefont {R.}~\bibnamefont
  {Thorngren}},\ }\href@noop {} {\  (\bibinfo {year} {2013})},\ \Eprint
  {http://arxiv.org/abs/arXiv:1309.4721} {arXiv:1309.4721} \BibitemShut
  {NoStop}%
\bibitem [{\citenamefont {Bullivant}\ \emph
  {et~al.}(2017{\natexlab{a}})\citenamefont {Bullivant}, \citenamefont
  {Calçada}, \citenamefont {Kádár}, \citenamefont {Martin},\ and\
  \citenamefont {Martins}}]{BM160606639}%
  \BibitemOpen
  \bibfield  {author} {\bibinfo {author} {\bibfnamefont {A.}~\bibnamefont
  {Bullivant}}, \bibinfo {author} {\bibfnamefont {M.}~\bibnamefont {Calçada}},
  \bibinfo {author} {\bibfnamefont {Z.}~\bibnamefont {Kádár}}, \bibinfo
  {author} {\bibfnamefont {P.}~\bibnamefont {Martin}}, \ and\ \bibinfo {author}
  {\bibfnamefont {J.~F.}\ \bibnamefont {Martins}},\ }\href {\doibase
  10.1103/PhysRevB.95.155118} {\bibfield  {journal} {\bibinfo  {journal} {Phys.
  Rev.}\ }\textbf {\bibinfo {volume} {B95}},\ \bibinfo {pages} {155118}
  (\bibinfo {year} {2017}{\natexlab{a}})},\ \Eprint
  {http://arxiv.org/abs/1606.06639} {arXiv:1606.06639} \BibitemShut {NoStop}%
\bibitem [{\citenamefont {Bullivant}\ \emph
  {et~al.}(2017{\natexlab{b}})\citenamefont {Bullivant}, \citenamefont
  {Calcada}, \citenamefont {Kádár}, \citenamefont {Martins},\ and\
  \citenamefont {Martin}}]{BM170200868}%
  \BibitemOpen
  \bibfield  {author} {\bibinfo {author} {\bibfnamefont {A.}~\bibnamefont
  {Bullivant}}, \bibinfo {author} {\bibfnamefont {M.}~\bibnamefont {Calcada}},
  \bibinfo {author} {\bibfnamefont {Z.}~\bibnamefont {Kádár}}, \bibinfo
  {author} {\bibfnamefont {J.~F.}\ \bibnamefont {Martins}}, \ and\ \bibinfo
  {author} {\bibfnamefont {P.}~\bibnamefont {Martin}},\ }\href@noop {} {\
  (\bibinfo {year} {2017}{\natexlab{b}})},\ \Eprint
  {http://arxiv.org/abs/1702.00868} {arXiv:1702.00868} \BibitemShut {NoStop}%
\bibitem [{\citenamefont {{Costa de Almeida}}\ \emph
  {et~al.}(2017)\citenamefont {{Costa de Almeida}}, \citenamefont
  {{Ibieta-Jimenez}}, \citenamefont {{Lorca Espiro}},\ and\ \citenamefont
  {{Teotonio-Sobrinho}}}]{CT171104186}%
  \BibitemOpen
  \bibfield  {author} {\bibinfo {author} {\bibfnamefont {R.}~\bibnamefont
  {{Costa de Almeida}}}, \bibinfo {author} {\bibfnamefont {J.~P.}\ \bibnamefont
  {{Ibieta-Jimenez}}}, \bibinfo {author} {\bibfnamefont {J.}~\bibnamefont
  {{Lorca Espiro}}}, \ and\ \bibinfo {author} {\bibfnamefont {P.}~\bibnamefont
  {{Teotonio-Sobrinho}}},\ }\href@noop {} {\  (\bibinfo {year} {2017})},\
  \Eprint {http://arxiv.org/abs/1711.04186} {arXiv:1711.04186} \BibitemShut
  {NoStop}%
\bibitem [{\citenamefont {Parzygnat}(2018)}]{P180201139}%
  \BibitemOpen
  \bibfield  {author} {\bibinfo {author} {\bibfnamefont {A.~J.}\ \bibnamefont
  {Parzygnat}},\ }\emph {\bibinfo {title} {{Two-dimensional algebra in lattice
  gauge theory}}},\ \href
  {https://inspirehep.net/record/1653107/files/arXiv:1802.01139.pdf} {Ph.D.
  thesis},\ \bibinfo  {school} {Connecticut U.} (\bibinfo {year} {2018}),\
  \Eprint {http://arxiv.org/abs/1802.01139} {arXiv:1802.01139} \BibitemShut
  {NoStop}%
\bibitem [{\citenamefont {Delcamp}\ and\ \citenamefont
  {Tiwari}(2018)}]{DT180210104}%
  \BibitemOpen
  \bibfield  {author} {\bibinfo {author} {\bibfnamefont {C.}~\bibnamefont
  {Delcamp}}\ and\ \bibinfo {author} {\bibfnamefont {A.}~\bibnamefont
  {Tiwari}},\ }\href@noop {} {\  (\bibinfo {year} {2018})},\ \Eprint
  {http://arxiv.org/abs/1802.10104} {arXiv:1802.10104} \BibitemShut {NoStop}%
\bibitem [{\citenamefont {Bouzid}\ and\ \citenamefont
  {Tahiri}(2018)}]{BT180300529}%
  \BibitemOpen
  \bibfield  {author} {\bibinfo {author} {\bibfnamefont {B.}~\bibnamefont
  {Bouzid}}\ and\ \bibinfo {author} {\bibfnamefont {M.}~\bibnamefont
  {Tahiri}},\ }\href@noop {} {\  (\bibinfo {year} {2018})},\ \Eprint
  {http://arxiv.org/abs/1803.00529} {arXiv:1803.00529} \BibitemShut {NoStop}%
\bibitem [{\citenamefont {Nikolaus}\ and\ \citenamefont
  {Waldorf}(2018)}]{NW180400677}%
  \BibitemOpen
  \bibfield  {author} {\bibinfo {author} {\bibfnamefont {T.}~\bibnamefont
  {Nikolaus}}\ and\ \bibinfo {author} {\bibfnamefont {K.}~\bibnamefont
  {Waldorf}},\ }\href@noop {} {\  (\bibinfo {year} {2018})},\ \Eprint
  {http://arxiv.org/abs/1804.00677} {arXiv:1804.00677} \BibitemShut {NoStop}%
\bibitem [{\citenamefont {Lau}\ and\ \citenamefont {Dasgupta}(1988)}]{LD8851}%
  \BibitemOpen
  \bibfield  {author} {\bibinfo {author} {\bibfnamefont {M.}~\bibnamefont
  {Lau}}\ and\ \bibinfo {author} {\bibfnamefont {C.}~\bibnamefont {Dasgupta}},\
  }\href {http://stacks.iop.org/0305-4470/21/i=1/a=009} {\bibfield  {journal}
  {\bibinfo  {journal} {Journal of Physics A: Mathematical and General}\
  }\textbf {\bibinfo {volume} {21}},\ \bibinfo {pages} {L51} (\bibinfo {year}
  {1988})}\BibitemShut {NoStop}%
\bibitem [{\citenamefont {Kosterlitz}\ and\ \citenamefont
  {Thouless}(1973)}]{KT7381}%
  \BibitemOpen
  \bibfield  {author} {\bibinfo {author} {\bibfnamefont {J.~M.}\ \bibnamefont
  {Kosterlitz}}\ and\ \bibinfo {author} {\bibfnamefont {D.~J.}\ \bibnamefont
  {Thouless}},\ }\href@noop {} {\bibfield  {journal} {\bibinfo  {journal} {J.
  Phys. C}\ }\textbf {\bibinfo {volume} {6}},\ \bibinfo {pages} {1181}
  (\bibinfo {year} {1973})}\BibitemShut {NoStop}%
\bibitem [{\citenamefont {Lan}\ \emph {et~al.}(2017)\citenamefont {Lan},
  \citenamefont {Kong},\ and\ \citenamefont {Wen}}]{LW170404221}%
  \BibitemOpen
  \bibfield  {author} {\bibinfo {author} {\bibfnamefont {T.}~\bibnamefont
  {Lan}}, \bibinfo {author} {\bibfnamefont {L.}~\bibnamefont {Kong}}, \ and\
  \bibinfo {author} {\bibfnamefont {X.-G.}\ \bibnamefont {Wen}},\ }\href@noop
  {} {\  (\bibinfo {year} {2017})},\ \Eprint
  {http://arxiv.org/abs/arXiv:1704.04221} {arXiv:1704.04221} \BibitemShut
  {NoStop}%
\bibitem [{\citenamefont {Kitaev}(2001)}]{K0131}%
  \BibitemOpen
  \bibfield  {author} {\bibinfo {author} {\bibfnamefont {A.~Y.}\ \bibnamefont
  {Kitaev}},\ }\href {\doibase 10.1070/1063-7869/44/10S/S29} {\bibfield
  {journal} {\bibinfo  {journal} {Phys.-Usp.}\ }\textbf {\bibinfo {volume}
  {44}},\ \bibinfo {pages} {131} (\bibinfo {year} {2001})},\ \Eprint
  {http://arxiv.org/abs/cond-mat/0010440} {cond-mat/0010440} \BibitemShut
  {NoStop}%
\bibitem [{\citenamefont {Lan}\ and\ \citenamefont {Wen}(2014)}]{LW1384}%
  \BibitemOpen
  \bibfield  {author} {\bibinfo {author} {\bibfnamefont {T.}~\bibnamefont
  {Lan}}\ and\ \bibinfo {author} {\bibfnamefont {X.-G.}\ \bibnamefont {Wen}},\
  }\href@noop {} {\bibfield  {journal} {\bibinfo  {journal} {Phys. Rev. B}\
  }\textbf {\bibinfo {volume} {90}},\ \bibinfo {pages} {115119} (\bibinfo
  {year} {2014})},\ \Eprint {http://arxiv.org/abs/arXiv:1311.1784}
  {arXiv:1311.1784} \BibitemShut {NoStop}%
\bibitem [{\citenamefont {Morandi}(1997)}]{Mor97}%
  \BibitemOpen
  \bibfield  {author} {\bibinfo {author} {\bibfnamefont {P.~J.}\ \bibnamefont
  {Morandi}},\ }\href
  {https://web.nmsu.edu/~pamorand/notes/GroupExtensions.pdf} {\bibfield
  {journal} {\bibinfo  {journal} {https://web.nmsu.edu/$\sim$pamorand/notes/}\
  } (\bibinfo {year} {1997})}\BibitemShut {NoStop}%
\bibitem [{\citenamefont {Turaev}\ and\ \citenamefont {Viro}(1992)}]{TV9265}%
  \BibitemOpen
  \bibfield  {author} {\bibinfo {author} {\bibfnamefont {V.~G.}\ \bibnamefont
  {Turaev}}\ and\ \bibinfo {author} {\bibfnamefont {O.~Y.}\ \bibnamefont
  {Viro}},\ }\href@noop {} {\bibfield  {journal} {\bibinfo  {journal}
  {Topology}\ }\textbf {\bibinfo {volume} {31}},\ \bibinfo {pages} {865}
  (\bibinfo {year} {1992})}\BibitemShut {NoStop}%
\bibitem [{\citenamefont {{Gaiotto}}\ \emph {et~al.}(2015)\citenamefont
  {{Gaiotto}}, \citenamefont {{Kapustin}}, \citenamefont {{Seiberg}},\ and\
  \citenamefont {{Willett}}}]{GW14125148}%
  \BibitemOpen
  \bibfield  {author} {\bibinfo {author} {\bibfnamefont {D.}~\bibnamefont
  {{Gaiotto}}}, \bibinfo {author} {\bibfnamefont {A.}~\bibnamefont
  {{Kapustin}}}, \bibinfo {author} {\bibfnamefont {N.}~\bibnamefont
  {{Seiberg}}}, \ and\ \bibinfo {author} {\bibfnamefont {B.}~\bibnamefont
  {{Willett}}},\ }\href {\doibase 10.1007/JHEP02(2015)172} {\bibfield
  {journal} {\bibinfo  {journal} {Journal of High Energy Physics}\ }\textbf
  {\bibinfo {volume} {2}},\ \bibinfo {pages} {172} (\bibinfo {year} {2015})},\
  \Eprint {http://arxiv.org/abs/arXiv:1412.5148} {arXiv:1412.5148} \BibitemShut
  {NoStop}%
\bibitem [{\citenamefont {{Baez}}\ and\ \citenamefont
  {{Lauda}}(2003)}]{BLm0307200}%
  \BibitemOpen
  \bibfield  {author} {\bibinfo {author} {\bibfnamefont {J.~C.}\ \bibnamefont
  {{Baez}}}\ and\ \bibinfo {author} {\bibfnamefont {A.~D.}\ \bibnamefont
  {{Lauda}}},\ }\href@noop {} {\  (\bibinfo {year} {2003})},\ \Eprint
  {http://arxiv.org/abs/math/0307200} {math/0307200} \BibitemShut {NoStop}%
\bibitem [{\citenamefont {Clement}(2002)}]{Clement02}%
  \BibitemOpen
  \bibfield  {author} {\bibinfo {author} {\bibfnamefont {A.}~\bibnamefont
  {Clement}},\ }\href {https://doc.rero.ch/record/482/files/Clement_these.pdf}
  {\bibfield  {journal} {\bibinfo  {journal} {Ph. D. Thesis, University of
  Lausanne}\ } (\bibinfo {year} {2002})}\BibitemShut {NoStop}%
\bibitem [{\citenamefont {Wen}(2017)}]{W161201418}%
  \BibitemOpen
  \bibfield  {author} {\bibinfo {author} {\bibfnamefont {X.-G.}\ \bibnamefont
  {Wen}},\ }\href {\doibase 10.1103/PhysRevB.95.205142} {\bibfield  {journal}
  {\bibinfo  {journal} {Phys. Rev. B}\ }\textbf {\bibinfo {volume} {95}},\
  \bibinfo {pages} {205142} (\bibinfo {year} {2017})},\ \Eprint
  {http://arxiv.org/abs/arXiv:1612.01418} {arXiv:1612.01418} \BibitemShut
  {NoStop}%
\bibitem [{\citenamefont {Crane}\ and\ \citenamefont {Yetter}(1993)}]{CY9362}%
  \BibitemOpen
  \bibfield  {author} {\bibinfo {author} {\bibfnamefont {L.}~\bibnamefont
  {Crane}}\ and\ \bibinfo {author} {\bibfnamefont {D.~N.}\ \bibnamefont
  {Yetter}},\ }\href@noop {} {\  (\bibinfo {year} {1993})},\ \Eprint
  {http://arxiv.org/abs/hep-th/9301062} {hep-th/9301062} \BibitemShut {NoStop}%
\bibitem [{\citenamefont {Gu}\ \emph {et~al.}(2009)\citenamefont {Gu},
  \citenamefont {Levin}, \citenamefont {Swingle},\ and\ \citenamefont
  {Wen}}]{GLS0918}%
  \BibitemOpen
  \bibfield  {author} {\bibinfo {author} {\bibfnamefont {Z.-C.}\ \bibnamefont
  {Gu}}, \bibinfo {author} {\bibfnamefont {M.}~\bibnamefont {Levin}}, \bibinfo
  {author} {\bibfnamefont {B.}~\bibnamefont {Swingle}}, \ and\ \bibinfo
  {author} {\bibfnamefont {X.-G.}\ \bibnamefont {Wen}},\ }\href@noop {}
  {\bibfield  {journal} {\bibinfo  {journal} {Phys. Rev. B}\ }\textbf {\bibinfo
  {volume} {79}},\ \bibinfo {pages} {085118} (\bibinfo {year} {2009})},\
  \Eprint {http://arxiv.org/abs/arXiv:0809.2821} {arXiv:0809.2821} \BibitemShut
  {NoStop}%
\bibitem [{\citenamefont {Costantino}(2005)}]{C0527}%
  \BibitemOpen
  \bibfield  {author} {\bibinfo {author} {\bibfnamefont {F.}~\bibnamefont
  {Costantino}},\ }\href@noop {} {\bibfield  {journal} {\bibinfo  {journal}
  {Math. Z.}\ }\textbf {\bibinfo {volume} {251}},\ \bibinfo {pages} {427}
  (\bibinfo {year} {2005})},\ \Eprint {http://arxiv.org/abs/math/0403014}
  {math/0403014} \BibitemShut {NoStop}%
\bibitem [{\citenamefont {Chen}\ \emph {et~al.}(2012)\citenamefont {Chen},
  \citenamefont {Gu}, \citenamefont {Liu},\ and\ \citenamefont
  {Wen}}]{CGL1204}%
  \BibitemOpen
  \bibfield  {author} {\bibinfo {author} {\bibfnamefont {X.}~\bibnamefont
  {Chen}}, \bibinfo {author} {\bibfnamefont {Z.-C.}\ \bibnamefont {Gu}},
  \bibinfo {author} {\bibfnamefont {Z.-X.}\ \bibnamefont {Liu}}, \ and\
  \bibinfo {author} {\bibfnamefont {X.-G.}\ \bibnamefont {Wen}},\ }\href@noop
  {} {\bibfield  {journal} {\bibinfo  {journal} {Science}\ }\textbf {\bibinfo
  {volume} {338}},\ \bibinfo {pages} {1604} (\bibinfo {year} {2012})},\ \Eprint
  {http://arxiv.org/abs/arXiv:1301.0861} {arXiv:1301.0861} \BibitemShut
  {NoStop}%
\bibitem [{\citenamefont {Steenrod}(1947)}]{S4790}%
  \BibitemOpen
  \bibfield  {author} {\bibinfo {author} {\bibfnamefont {N.}~\bibnamefont
  {Steenrod}},\ }\href {\doibase 10.2307/1969172} {\bibfield  {journal}
  {\bibinfo  {journal} {Annals of Mathematics}\ }\textbf {\bibinfo {volume}
  {48}},\ \bibinfo {pages} {290} (\bibinfo {year} {1947})}\BibitemShut
  {NoStop}%
\bibitem [{\citenamefont {Lyndon}(1948)}]{L4871}%
  \BibitemOpen
  \bibfield  {author} {\bibinfo {author} {\bibfnamefont {R.~C.}\ \bibnamefont
  {Lyndon}},\ }\href {\doibase 10.1215/S0012-7094-48-01528-2} {\bibfield
  {journal} {\bibinfo  {journal} {Duke Mathematical Journal}\ }\textbf
  {\bibinfo {volume} {15}},\ \bibinfo {pages} {271 } (\bibinfo {year}
  {1948})}\BibitemShut {NoStop}%
\bibitem [{\citenamefont {Hochschild}\ and\ \citenamefont
  {Serre}(1953)}]{HS5310}%
  \BibitemOpen
  \bibfield  {author} {\bibinfo {author} {\bibfnamefont {G.}~\bibnamefont
  {Hochschild}}\ and\ \bibinfo {author} {\bibfnamefont {J.-P.}\ \bibnamefont
  {Serre}},\ }\href {\doibase 10.2307/1990851} {\bibfield  {journal} {\bibinfo
  {journal} {Transactions of the American Mathematical Society (American
  Mathematical Society)}\ }\textbf {\bibinfo {volume} {74}},\ \bibinfo {pages}
  {110 } (\bibinfo {year} {1953})}\BibitemShut {NoStop}%
\end{thebibliography}%

\end{document}